\documentclass[a4paper,11pt]{article}

\usepackage{jheppub} 

\usepackage[T1]{fontenc} 
\usepackage{booktabs}

\begin{document}

\title{Monopole and instanton effects in QCD}

\author{Masayasu Hasegawa}
\affiliation{Bogoliubov Laboratory of Theoretical Physics, Joint Institute for Nuclear Research, Dubna, Moscow 141980, Russia}
\emailAdd{hasegawa@theor.jinr.ru}

\abstract{We aim to show the effects of the magnetic monopoles and instantons in quantum chromodynamics (QCD) on observables; therefore, we introduce a monopole and anti-monopole pair in the QCD vacuum of a quenched SU(3) by applying the monopole creation operator to the vacuum. We calculate the eigenvalues and eigenvectors of the overlap Dirac operator that preserves the exact chiral symmetry in lattice gauge theory using these QCD vacua. We then investigate the effects of magnetic monopoles and instantons. First, we confirm the monopole effects as follows: (i) The monopole creation operator makes the monopoles and anti-monopoles in the QCD vacuum. (ii) A monopole and anti-monopole pair creates an instanton or anti-instanton without changing the structure of the QCD vacuum. (iii) The monopole and anti-monopole pairs change only the scale of the spectrum distribution without affecting the spectra of the Dirac operator by comparing the spectra with random matrix theory. Next, we find the instanton effects by increasing the number density of the instantons and anti-instantons as follows: (iv) The decay constants of the pseudoscalar increase. (v) The values of the chiral condensate, which are defined as negative numbers, decrease. (vi) The light quarks and the pseudoscalar mesons become heavy. The catalytic effect on the charged pion is estimated using the numerical results of the pion decay constant and the pion mass. (vii) The decay width of the charged pion becomes wider than the experimental result, and the lifetime of the charged pion becomes shorter than the experimental result. These are the effects of the monopoles and instantons in QCD.}


\arxivnumber{1807.04808}

\maketitle
\flushbottom

\section{Introduction}

Illuminating the mechanism of color confinement is one of the most important research areas in mathematics and physics~\cite{Clay1}. A particle that possesses a single-color charge, for example, a single quark or gluon, has never been observed experimentally. We have only experimentally observed mesons and baryons of color singlets. We still do not know why we cannot observe particles of a single-color charge.

To explain this phenomenon, 'tHooft~\cite{tHooft2} and Mandelstam~\cite{Mandelstam1} provided a convincing description that a magnetic monopole that condenses in the QCD vacuum causes the dual Meissner effect and that color charged particles are confined. A significant number of simulations have been conducted under lattice gauge theory, and sufficient results have been obtained that support this explanation~\cite{Kronfel1, Maedan1, Brandstaeter1, Hioki1, DiGiacomo1, Sasaki1, Bonati2, Sekido2}. Thus, this scenario seems to be widely accepted.

In the Grand Unified Theory (GUT), the existence of a magnetic monopole, the 'tHooft-Polyakov monopole~\cite{tHooft1, Polyakov1} in the early universe, is necessarily derived. The catalytic effect that the presence of magnetic monopoles induces proton decay is theoretically expected; moreover, the close relation between quarks and magnetic monopoles has been mentioned~\cite{Rubakov1, Rubakov2, Wu2, Rubakov3}. The 'tHooft-Polyakov monopole possesses a superheavy mass~\cite{Romanov1}, and it is difficult to directly detect magnetic monopoles to validate the GUT. Experiments that try to observe the proton decay caused by monopole catalysis have been attempted. The catalytic effects, however, have not yet been observed experimentally~\cite{Groom1, Ueno1, Patrizii1}.

The spontaneous breaking of chiral symmetry causes interesting phenomena in the low energy of QCD~\cite{Nambu1, Nambu2, Goldstone1, Goldstone2, Gross1, Kugo1}. Once chiral symmetry spontaneously breaks, a massless pion, which is the NG (Nambu-Goldstone) boson, appears, and the chiral condensate, which is an order parameter of chiral symmetry breaking, obtains non-zero values. The quarks obtain small masses from the non-zero values of the chiral condensate. The pion decay constant is defined as the strength of the coupling constant between the NG boson and the axial-vector current. The pion obtains mass by supposing a partially conserved axial current (PCAC)~\cite{Weinberg1}.

These phenomena are well explained by models concerning the instanton~\cite{Belavi1, Dyakonov6, Shuryak2}. In particular, the models demonstrate that the chiral condensate and the pion decay constant are estimated from the instanton vacuum and that instantons induce the breaking of the chiral symmetry~\cite{Dyakonov1, Dyakonov2, Dyakonov3, Dyakonov4}.

Recently, very interesting experiments that challenge the frontiers of science have been attempted. In condensed matter physics, a research group has generated Dirac monopoles in a Bose-Einstein condensate and observed the monopoles experimentally~\cite{Ray1, Ray2}. These experimental results are also confirmed by simulations based on the model.

In the field of high-energy physics, the "Monopole and Exotics Detector at the LHC (MoEDAL)" experiment has begun. This experiment aims to explore magnetic monopoles and other highly ionizing particles, which are particles beyond the Standard Model, in proton-proton collisions at the Large Hadron Collider (LHC). The search for magnetic monopoles in high-energy collisions has already begun~\cite{Moedal1, Moedal2}.

The purpose of this study is to present indications that the effects of magnetic monopoles and instantons can be detected by experiments to reveal the existence of magnetic monopoles and instantons in the real world. Even if it seems that color confinement and chiral symmetry breaking are not related, we suppose that both phenomena are closely connected to one another through topological objects, i.e., magnetic monopoles and instantons, in the QCD vacuum. The topological objects that inhabit the QCD vacuum play significant roles in the mechanism of color confinement and the breaking of chiral symmetry. First, we demonstrate by conducting simulations of lattice QCD that the monopoles in the low energy of QCD induce the breaking of chiral symmetry through instantons.

In previous studies of lattice QCD, instantons have been found in QCD vacuums~\cite{Ilgenfritz1}, and the relations between the instantons and Abelian monopoles have been studied~\cite{Hart1}. The hadron masses were calculated from the background fields of Abelian monopoles~\cite{Kitahara2}. The fermion zero modes have been derived from the background fields of magnetic monopoles~\cite{Chernodub1, HAoki1}.

In numerical calculations, however, the fermions, which do not preserve the chiral symmetry in lattice gauge theory, are mainly used in the formulation of quarks. Moreover, the quantitative relation between magnetic monopoles and instantons is not clear because monopoles are defined as three-dimensional objects, whereas instantons are defined as four-dimensional objects.

In the present studies, we introduce the monopole and anti-monopole into the QCD vacuum of the quenched SU(3) by applying the monopole creation operator~\cite{Bonati2, DiGH3} to the vacuum. We generate the configurations by varying the values of the magnetic charges of the monopole and anti-monopole. We then calculate the eigenvalues and eigenvectors of the Dirac operator of the overlap fermions using these configurations. The Dirac operator of the overlap fermions, which is defined in lattice gauge theory, preserves the exact chiral symmetry in the continuum limit~\cite{Ginsparg1, Neuberger1, Neuberger2, Lusher1, Chandrasekharan1}. We have attempted to show the quantitative relations among monopoles, instantons, and chiral symmetry breaking, and we have already demonstrated the following results~\cite{DiGH3, DiGH4, DiGHP1, DiGH5}.
\begin{itemize}
\item The monopole creation operator makes only long monopole loops in the QCD vacuum, and the monopole loops become long with increasing values of the magnetic charges.

\item The total number of instantons and anti-instantons is correctly estimated from the topological charges.

\item The monopole of a magnetic charge +1 and the anti-monopole of a magnetic charge -1 make one instanton or one anti-instanton.

\item The additional monopoles and anti-monopoles do not change the vacuum structure and produce only the topological charges.

\item In the study of the maximal Abelian gauge, the total physical length of the monopole loops is in direct proportion to the total number of instantons and anti-instantons.
  
\item The added monopoles and anti-monopoles do not affect the distributions of the eigenvalues of the overlap Dirac operator, and these monopoles change only the scale parameter of the distributions of the eigenvalues. The chiral condensate decreases with increasing values of the magnetic charges (the chiral condensate is defined as a negative value). We obtain these results by comparing the numerical results with the predictions of random matrix theory~\cite{Damgaard1, Damgaard2, Edwards1, Giusti4}.

\item The preliminary results show that the quark masses become heavy by increasing the values of the magnetic charges.
\end{itemize}

It is apparent that the added monopoles and anti-monopoles are closely related to instantons and chiral symmetry breaking. These results, however, have been obtained using configurations with small lattice volumes ($V = 14^{4}$) and one value ($\beta = 6.0000$) of the parameter for the lattice spacing. We have already performed simulations that use a larger lattice volume ($V = 16^{3}\times 32$, $\beta = 6.0000$); however, the numbers of statistical samples are not sufficient.

We have shown in two ways that the values of the chiral condensate, which is defined as having negative values, decrease when varying the magnetic charges of the added monopole and anti-monopole. However, we cannot quantitatively explain this phenomenon.

In this study, we add a monopole and anti-monopole to a larger lattice volume ($V = 18^{3}\times 32$) with a finer lattice spacing ($\beta = 6.0522$) than in our previous studies. The numbers of statistical samples for the observables are sufficiently high. We calculate the low-lying eigenvalues and eigenvectors of the overlap Dirac operator from these configurations~\cite{Giusti6} and estimate the effects of the monopoles and instantons on the observables.

The contents of this article are as follows. In section II, we generate configurations whereby we add the monopole and anti-monopole. To confirm that we successfully added the monopoles and anti-monopoles to the configurations, we calculate the monopole density and the length of the monopole loops from these configurations.

In section 3, we calculate the number of zero modes, the total number of instantons and anti-instantons, and the instanton density using the eigenvalues of the overlap Dirac operator. We show the quantitative relations between monopoles and instantons with the calculations in reference~\cite{DiGH3}. Moreover, we compare the eigenvalues with the predictions in random matrix theory and show that the additional monopoles and anti-monopole do not affect the spectra and change only the scale of the eigenvalue distributions.

In section 4, we make predictions of the decay constants and the chiral condensate based on the models~\cite{Dyakonov1, Dyakonov2, Dyakonov3, Dyakonov4} to quantitatively explain why the decay constants increase and why the values of the chiral condensate decrease.

In section 5, we calculate the pseudoscalar mass, pseudoscalar decay constant, and the chiral condensate from the correlation functions of the operators~\cite{Gimenez1, Giusti3}. We estimate the renormalization constants by non-perturbative calculations~\cite{Bochicchio1, Maiani1, Giusti3, Hernandez2, Wennekers1}. We show that the numerical results correspond to the predictions in section 4.

In section 6, we calculate the normalization factors of the pion and kaon by matching the numerical results with the experimental results~\cite{Gimenez1, Giusti3}. We then re-estimate the decay constants and the chiral condensate by considering the normalization factors. We estimate precisely the instanton effects on the light quark masses and quantitatively explain why the light quark masses increase. We show that the numerical results correspond remarkably to the predictions of the instanton effects on the observables. Finally, we estimate the catalytic effect on the pion decay.

In section 7, we provide a summary and conclusions.

\section{Monopoles}

In this section, we create monopoles and anti-monopoles in configurations with varying magnetic charges and measure the monopole density and the length of the monopole loops to confirm that the monopoles and anti-monopoles are correctly added to the configurations.

\subsection{The monopole creation operator}

In this study, we use the same definition of the monopole creation operator as in reference~\cite{DiGH3}.

\begin{table}[tbp]
  \small 
  \centering
      {\renewcommand{\arraystretch}{1.2}
               \begin{tabular}{|c|c|c|}\hline
                 $D$ & Monopole ($t, \vec{x_{1}}$) & Anti-monopole ($t, \vec{x_{2}}$) \\ \hline
                 Odd & $\left(\frac{32}{2}, \frac{20 + D}{2}, \frac{20 + D}{2}, \frac{19}{2}\right)$ & $\left(\frac{32}{2}, \frac{20 - D}{2}, \frac{20 - D}{2}, \frac{17}{2}\right)$ \\ \hline
                 Even & $\left(\frac{32}{2}, \frac{19 + D}{2}, \frac{19 + D}{2}, \frac{19}{2}\right)$ & $\left(\frac{32}{2}, \frac{19 - D}{2}, \frac{19 - D}{2}, \frac{17}{2}\right)$ \\\hline
               \end{tabular}
           }
            \caption{The locations of the monopole ($t, \vec{x_{1}}$) and anti-monopole ($t, \vec{x_{2}}$). The time $t$ indicates the time slice in which we add the monopole and anti-monopole. The distance between the monopole and anti-monopole is indicated as $D$ (in lattice units). The lattice volume is $V = 18^{3}\times 32$.}\label{tb:location_mon_1}
 \end{table}
We maintain a certain distance $D$ and place the monopole at location $\vec{x_{1}}$ and the anti-monopole at location $\vec{x_{2}}$. We determine the distance $D$ between the monopole and the anti-monopole as $D$ = 9 (1.09 [fm]) by following the method explained in reference~\cite{DiGH3}. We set the time $t = 16$ to create the monopole and anti-monopole in the configurations. Periodic boundary conditions are adopted for each boundary (the space components and the time component) of the lattice. We indicate the locations of the monopole and anti-monopole and the distance in table~\ref{tb:location_mon_1}. 

We vary both the magnetic charges of the monopole from 0 to 6 and the magnetic charges of the anti-monopole from 0 to -6. The magnetic charges are integers. The anti-monopole possesses the opposite charges of the monopole; thus, the total magnetic charges that are added to the configuration is zero. The magnetic charge $m_{c}$ indicates that both the monopole of the magnetic charge $+m_{c}$ and the anti-monopole of the magnetic charge $-m_{c}$ are added.

To check the consistency with the normal configurations, we generate the configurations of the magnetic charge $m_{c} = 0$ and compare the numerical results.
\begin{table}[tbp]
  \small 
  \centering
          {\renewcommand{\arraystretch}{1.1}     
              \begin{tabular}{|c|c|c|c|c|c|c|c|}\hline
                $m_{c}$  & $a^{(1)}$ [fm] & $a^{(2)}$ [fm] & $(n, \alpha_{sm})$ & $T/a$ & $FR (R_{I}/a)$ & $\chi^{2}/d. o. f.$ & $N_{conf}$ \\ \hline
                N. C.  & 8.53(9)$\times10^{-2}$   &    8.98(4)$\times10^{-2}$    & (25, 0.5) & 4 & 1.8 - 8.0 & 1.0/4.0  & 800 \\\hline
0 & 8.52(14)$\times10^{-2}$ &    8.98(6)$\times10^{-2}$     & (30, 0.5) & 5 & 1.8 - 8.0 & 3.5/4.0 & 980 \\\hline
1 & 8.58(12)$\times10^{-2}$ &    9.03(5)$\times10^{-2}$     & (25, 0.5) & 5 & 1.8 - 9.0 & 4.9/5.0 & 1200 \\  \hline
2 & 8.72(8)$\times10^{-2}$   &    9.15(3)$\times10^{-2}$    & (30, 0.5) & 4 & 1.8 - 8.0 & 5.3/4.0  & 980 \\ \hline
3 & 8.75(8)$\times10^{-2}$   &    9.17(3)$\times10^{-2}$    & (25, 0.5) & 4 & 1.8 - 9.0 & 4.6/5.0  & 980 \\ \hline
4 & 8.7(3)$\times10^{-2}$     &    9.03(14)$\times10^{-2}$  & (30, 0.5) & 6 & 1.8 - 9.0 & 6.2/5.0  & 1060 \\\hline
5 & 8.83(18)$\times10^{-2}$ &    9.27(8)$\times10^{-2}$     & (25, 0.5) & 4 & 1.8 - 7.0 & 3.2/3.0 & 1100 \\\hline
6 & 8.66(19)$\times10^{-2}$ &    9.01(7)$\times10^{-2}$     & (25, 0.5) & 5 & 1.8 - 9.0 & 4.3/5.0  & 920 \\\hline
              \end{tabular}
          }
           \caption{The numerical results of the lattice spacing $a^{(1)}$ and $a^{(2)}$. The lattice is $V = 18^{3}\times 32$, $\beta = 6.0522$. N. C. stands for the normal configuration. The number of iterations and the weight factor for the smearing are written as $(n, \alpha_{sm})$. $T/a$ indicates the temporal component of the Wilson loop, which we determine with the lattice spacing. $FR$ indicates the fitting range.}\label{tb:lattice_a_1}
\end{table}

\subsection{The simulation parameters}

We generate the normal configurations and the configurations to which the classical fields of the monopole and anti-monopole are added. General methods, i.e., the heat bath algorithm and the over-relaxation method, are used. The lattice volume and the parameter $\beta$ of the lattice spacing are $V = 18^{3}\times 32$ and $\beta = 6.0522$, respectively.

First, we confirm the effects of the additional monopole and anti-monopole on the scale of the lattice by calculating the lattice spacing. The lattice spacing $a^{(1)}$ is estimated with the Sommer scale $r_{0} = 0.5$ [fm], $\sigma$, and $\alpha$. The parameters of $\sigma$ and $\alpha$ are obtained by fitting the function
\begin{equation}
V(R) = V_{0} - \frac{\alpha}{R} + \sigma R
\end{equation}
to the numerical results of the static potential $V(R)$, which is computed from Wilson loops. The lattice spacing $a^{(2)}$ is determined using $\sqrt{\sigma} = 440$ [MeV]. To reduce the effects of excited states, we perform the smearing~\cite{Ape1} to the gauge links of the spatial components. Moreover, we improve the spatial component $R$ of the Wilson loop to $R_{I}$ using the Green function~\cite{Necco1, Necco2}. The numerical results of the lattice spacing and the smearing parameters are shown in table~\ref{tb:lattice_a_1}.

Table~\ref{tb:lattice_a_1} shows that the additional monopoles and anti-monopoles do not affect the lattice spacing, and the numerical results are reasonably consistent with the analytic results, which are calculated from formula~\cite{Necco1}. Hereafter, we use the value of the lattice spacing $a$ = 8.5274$\times10^{-2}$ [fm] and the Sommer scale $r_{0}$ = 0.5 [fm].

\subsection{The monopole density and the length of the monopole loops}

In this subsection, to clearly show that we add the monopole and anti-monopole to the configurations, we iteratively diagonalize the SU(3) matrix under the condition of the maximal Abelian gauge by using the simulated annealing algorithm. We perform 20 iterations to prevent the Gribov copies from influencing the numerical results. We then derive the Abelian monopole that holds the $U(1)\times U(1)$ symmetry from the Abelian link variables by performing the Abelian projection to the SU(3) matrix~\cite{tHooft3}.

The monopole current $k_{\mu}^{i}$ in SU(3)~\cite{DeGrand1, Kronfel1, Kitahara2} is defined on the dual site $^{*}n$ such that it satisfies the condition $\sum_{i}k_{\mu}^{i} (^{*}n) = 0$ as follows:
\begin{equation}
k_{\mu}^{i} (^{*}n) \equiv - \epsilon_{\mu\nu\rho\sigma}\nabla_{\nu}n_{\rho\sigma}^{i}(n + \hat{\mu})
\end{equation}
The index $i$ indicates the color, and $n_{\rho\sigma}^{i}$ is defined as the number of Dirac strings that pierce through a plaquette on a plane defined by the directions $\rho$ and $\sigma$. We adopt the normalization factor from reference~\cite{Bornyakov4}.

The monopole current satisfies the current conservation law $\nabla_{\mu}^{*}k_{\mu}^{i} (^{*}n) = 0$. Therefore, the monopole currents form the loops. The derivatives $\nabla_{\mu}$ and $\nabla_{\mu}^{*}$ indicate the forward and backward derivatives on the lattice, respectively.
The following is a definition of the monopole density $\rho_{m}$ as a three-dimensional object~\cite{Bornyakov4}:
\begin{equation}
\rho_{m} = \frac{1}{12V}\sum_{i, \mu}\sum_{^{*}n}|k_{\mu}^{i}(^{*}n)|/a^{3} \ \ [\mbox{GeV}^{3}]
\end{equation}
We count the numbers of the absolute values of the monopole currents that form the closed loops $C$~\cite{Bode1} and define the length of the closed loops $L_{m}$ as a one-dimensional object as follows:
\begin{equation}
L_{m} \equiv \frac{a}{12}\sum_{i, \mu}\sum_{^{*}n \in C}|k_{\mu}^{i}(^{*}n)| \ \ [\mbox{fm}]
\end{equation}
\begin{figure}[tbp]
  \centering
    \includegraphics[width=150mm]{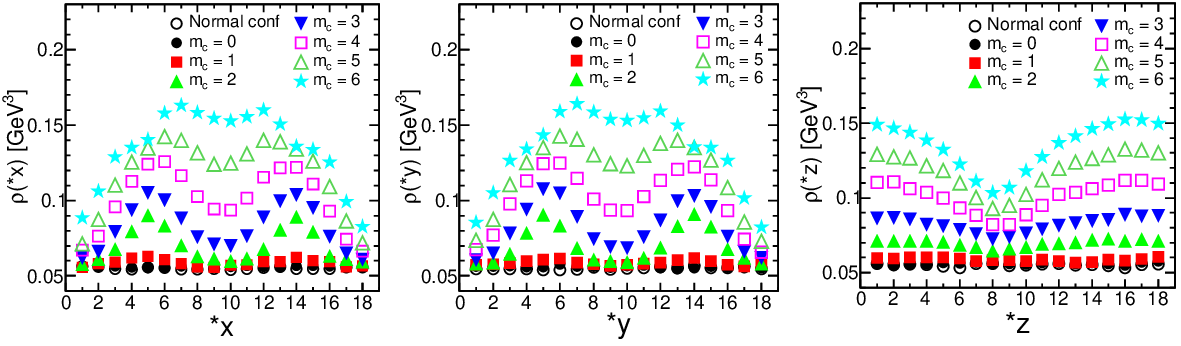}
\setlength\abovecaptionskip{-3.0pt}
\caption{The monopole density on the dual-site $(^{*}x,^{*}y,^{*}z)$. The normalization factor $V$ is $l_{s}^{2}\times l_{t}$.}\label{fig:mono_density_xyz}
\end{figure}

First, we calculate the monopole density on a dual-site using the normal configuration and the configurations of the additional monopole and anti-monopole to confirm whether the monopole and anti-monopole are appropriately added to the configurations. Figure~\ref{fig:mono_density_xyz} shows that the additional monopole and anti-monopole diffuse in the spatial lattice after increasing the magnetic charges $m_{c}$. As indicated in table~\ref{tb:mon_dens_1}, the monopole density $\rho_{m}$ increases with increasing magnetic charge $m_{c}$. Incidentally, we calculate the monopole density $\rho_{m}^{nd}$ without diagonalizing the configurations, and we list the computed results in the same table~\ref{tb:mon_dens_1}. The computed results show that the monopole density $\rho_{m}^{nd}$ does not vary even if we increase the magnetic charges $m_{c}$.
\begin{table}[tbp]
  \centering
  \small 
  \begin{tabular}{|c|c|c|c|c|c|c|}\hline
    $m_{c}$ & $\rho_{m}$ & $\rho_{m}^{nd}$ & $L_{m}^{T}$ & $L_{m}^{L}$ & $L_{m}^{S}$  & $N_{conf}$ \\ 
    & {\footnotesize [GeV$^{3}$]} & {\footnotesize [GeV$^{3}$]}  & {\footnotesize [fm]} &  {\footnotesize [fm]} &  {\footnotesize [fm]} &  \\ \hline
    {\footnotesize Normal Conf} &  0.0551(3)  & 5.2998(6) &  70.7(4)  &  28.4(5)  &  42.3(5) & 100 \\    \hline
    0 &  0.0561(3)  & 5.2992(7) &  72.0(4)  &  29.8(6)  &  42.3(6) & 100 \\\hline
    1 &  0.0587(3)  & 5.2993(6) &  75.4(4)  &  30.2(7)  &  45.2(6) & 100 \\ \hline
    2 &  0.0698(3)  & 5.2998(7) &  89.7(4)  &  47.1(7)  &  42.6(6) & 100  \\\hline
    3 &  0.0820(4)  & 5.3017(6) &  105.3(5) &  65.0(6)  &  40.3(5) & 100 \\\hline
    4 &  0.1007(4)  & 5.3024(6) &  129.4(5) &  89.1(5)  &  40.3(3) & 100 \\\hline
    5 &  0.1182(4)  & 5.3034(7) &  151.9(5) &  112.0(6) &  39.9(3) & 100 \\\hline
    6 &  0.1348(5)  & 5.3062(6) &  173.2(6) &  131.9(6) &  41.2(4) & 100 \\\hline
  \end{tabular}
  \caption{The computed results of the monopole densities $\rho_{m}$ and $\rho_{m}^{nd}$ and the lengths of the monopole loops $L_{m}^{T}$, $L_{m}^{L}$, and $L_{m}^{S}$.}\label{tb:mon_dens_1}
\end{table}
\begin{figure}[tbp]
  \centering
  \includegraphics[width=83mm]{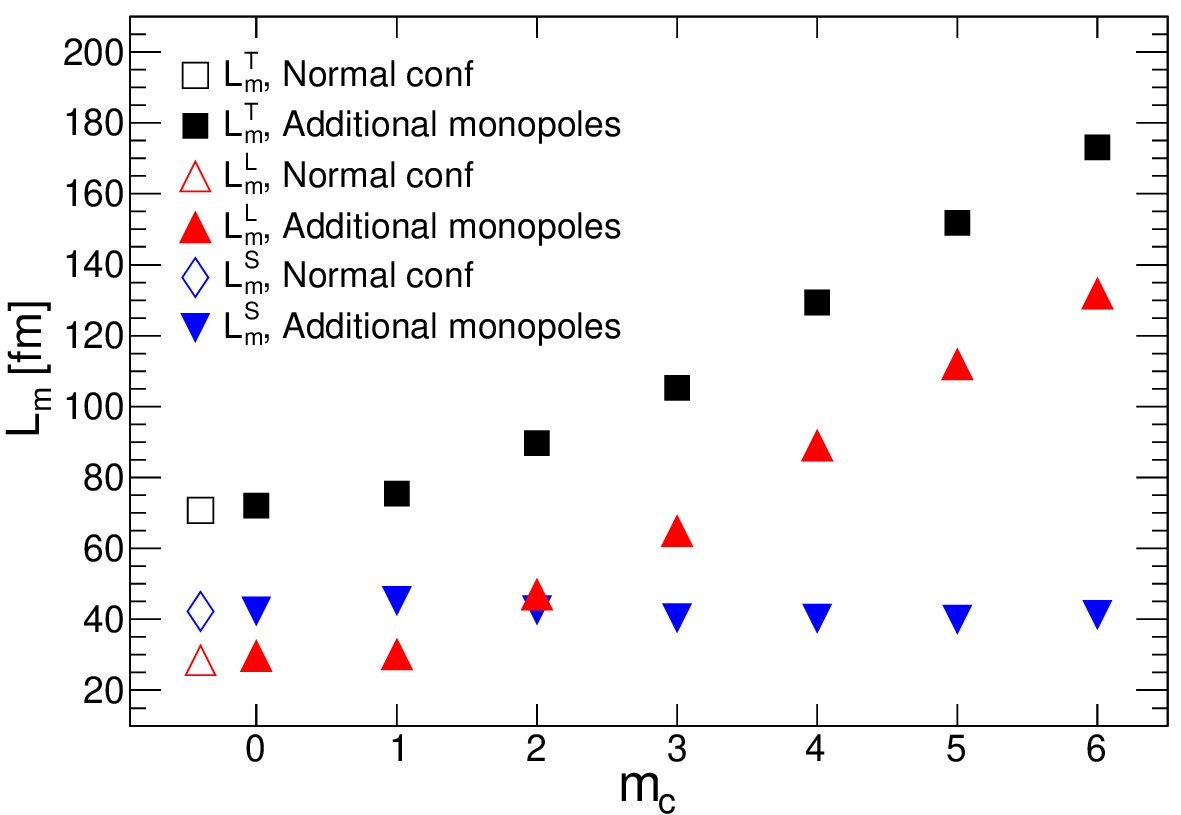}
  \setlength\abovecaptionskip{-3.0pt}
  \caption{The physical length of monopole loops $L_{m}$ versus the magnetic charges $m_{c}$. $L_{m}^{T}$, $L_{m}^{L}$, and $L_{m}^{S}$, which indicate the total length of the loops, the longest loops, and the shortest loops, respectively.}\label{fig:mono_loops_vs_mc}
\end{figure}

Next, we measure the length of the monopole loops. We define the lengths of the monopole loops as $L_{m}^{T}$, $L_{m}^{L}$, and $L_{m}^{S}$, which indicate the total length of the loops, the longest loops, and the shortest loops, respectively. The shortest loops are defined as the remainder after the longest loops are subtracted from the total length. The computed results are provided in table~\ref{tb:mon_dens_1}.

As shown in figure~\ref{fig:mono_loops_vs_mc}, the length of the longest loop $L_{m}^{L}$ linearly increases with increasing magnetic charge $m_{c}$; however, the length of the shortest loops $L_{m}^{S}$ does not change. This shows that the monopole creation operator produces only the long monopole loops in the configurations.

Hereafter, we do not diagonalize the SU(3) matrix under a particular gauge condition, and we do not apply the Abelian projection to the SU(3) matrix.


\section{Monopole effects}

In this section, we briefly explain the Dirac operator of the overlap fermions. We calculate the eigenvalues and eigenvectors of the overlap Dirac operator. The total number of instantons and anti-instantons in the configurations are estimated. We show the quantitative relation between instantons and monopoles by comparing them with our predictions. We compare the eigenvalues with the predictions of random matrix theory and show the monopole effects.

\subsection{Overlap fermions}

The operator $D$ denotes the Dirac operator of the overlap fermions that satisfy chiral symmetry~\cite{Ginsparg1, Lusher1, Neuberger1, Neuberger2}. The Dirac operator is defined by the Hermitian Wilson Dirac operator $H_{W}$ as follows:
\begin{equation}
D(\rho) = \frac{\rho}{a} \left( 1 + \frac{\gamma_{5}H_{W}(\rho)}{\sqrt{H_{W}(\rho)^{\dagger}H_{W}(\rho)}}\right)
\end{equation}
The Hermitian Wilson Dirac operator $H_{W}$ is
\begin{equation}
H_{W}(\rho) =  \gamma_{5}\left(D_{W} - \frac{\rho}{a}\right).
\end{equation}
The parameter $\rho$ is a real-valued mass parameter. We set $\rho = 1.4$~\cite{Hernandez1}. The massless Wilson Dirac operator $D_{W}$ is defined as~(\ref{eq:def_wilson1}).
The overlap Dirac operator is approximated by using the sign function and is derived as follows:
\begin{equation}
D(\rho) = \frac{\rho}{a} \left[ 1 +  \gamma_{5} \mbox{sign}(H_{W}(\rho)) \right]
\end{equation}

In this study, we use the numerical methods explained in reference~\cite{Giusti6}. We solve the eigenvalue problems $D|\psi_{i}\rangle = \lambda_{i}|\psi_{i}\rangle\label{eq:solv_eigen1}$ by using the subroutines ({\small ARPACK}) and retain 100 pairs of the low-lying eigenvalues and eigenvectors for one configuration. The index $i$ indicates the number of pairs. We do not use the smearing method or the cooling method to calculate the Dirac operator.

\subsection{Monopole effects on instantons and topological charges}

There are fermion zero modes in the spectra of the eigenvalues of the overlap Dirac operator. The number of zero modes of the positive chirality is $n_{+}$, and the number of zero modes of the negative chirality is $n_{-}$. The topological charge is defined as $Q = n_{+} - n_{-}$, and the topological susceptibility $\frac{\langle Q^{2} \rangle}{V}$ is calculated from the topological charges.

As mentioned in the previous study~\cite{DiGH3}, however, we have never simultaneously detected the zero modes of the positive chirality and the zero modes of the negative chirality from the same configuration. The zero modes that we observe in our simulations are the topological charges. The number of zero modes, which we observe in our simulations, is the absolute value of the topological charge $N_{Z} = |Q|$. The total number of instantons and anti-instantons $N_{I}$ in the lattice volume $V$ is analytically computed from the square of the topological charges $\langle Q^{2} \rangle$ of the lattice volume $V$ as follows~\cite{Edwards3, DiGH3}:
\begin{equation}
N_{I} = \langle Q^{2} \rangle\label{eq:num_ins}
\end{equation}
The value $\langle \mathcal{O} \rangle$ indicates the average value given by the sum of the samples divided by the number of configurations.

The total number of instantons and anti-instantons of the normal configuration $N_{I}$, which is calculated from formula~(\ref{eq:num_ins}) and the numerical result of the topological charges, is $N_{I} = 9.7(5)$. The number density of the instantons and anti-instantons in the physical volume $V_{phys}$ = 9.8582 [fm$^{4}$] ($V = 18^{3}\times32$, $\beta = 6.0522$) is $\frac{N_{I}}{V} = 1.48 (7) \times 10^{-3} \ [\mbox{GeV}^{4}]$.
\begin{table}[tbp]
  \small 
  \centering
          {\renewcommand{\arraystretch}{1.3}
              \begin{tabular}{|c|c|c|c|c|c|c|c|}\hline
                $m_{c}$ & $N_{Z}^{Pre}$ & $N_{Z}$ & $N_{I}^{Pre}$  &  $N_{I}$ & $\frac{N_{I}^{Pre}}{V}${\footnotesize[GeV$^{4}$]} & $\frac{N_{I}}{V}${\footnotesize[GeV$^{4}$]} & $N_{conf}$\\  \hline
                {\footnotesize Normal conf} & 2.5748 & 2.48(7) & 10.414 & 9.7(5) &  1.6000$\times10^{-3}$    & 1.48(7)$\times10^{-3}$ & 800\\    \hline          
                0 & 2.5748 & 2.66(7) & 10.414 & 10.8(5) & 1.6000$\times10^{-3}$    & 1.66(8)$\times10^{-3}$   & 800  \\    \hline          
                1 & 2.6975 & 2.65(7) & 11.414 & 11.3(6) & 1.7536$\times10^{-3}$    & 1.73(9)$\times10^{-3}$   & 838  \\   \hline           
                2 & 2.8144 & 2.91(8) & 12.414 & 13.6(7) & 1.9073$\times10^{-3}$    & 2.09(11)$\times10^{-3}$  & 810  \\   \hline           
                3 & 2.9265 & 3.03(9) & 13.414 & 15.0(8) & 2.0609$\times10^{-3}$    & 2.31(12)$\times10^{-3}$  & 800  \\   \hline           
                4 & 3.0343 & 3.14(8) & 14.414 & 15.7(8) & 2.2146$\times10^{-3}$    & 2.42(12)$\times10^{-3}$  & 868 \\   \hline           
                5 & 3.1383 & 3.23(9) & 15.414 & 16.5(8) & 2.3682$\times10^{-3}$   & 2.54(13)$\times10^{-3}$  & 810 \\   \hline           
                6 & 3.2388 & 3.29(9) & 16.414 & 17.7(9) & 2.5219$\times10^{-3}$    & 2.72(14)$\times10^{-3}$  & 870 \\   \hline                    
              \end{tabular}         
          }
            \caption{The results of the number of zero modes $N_ {Z}$, the total number of instantons and anti-instantons $N_ {I}$, and the instanton density $\frac{N_{I}}{V}$. The superscript $Pre$ indicates the predicted values.}\label{tb:zero_instantons}
\end{table}
\begin{figure}[tbp]
  \centering
    \includegraphics[width=83mm]{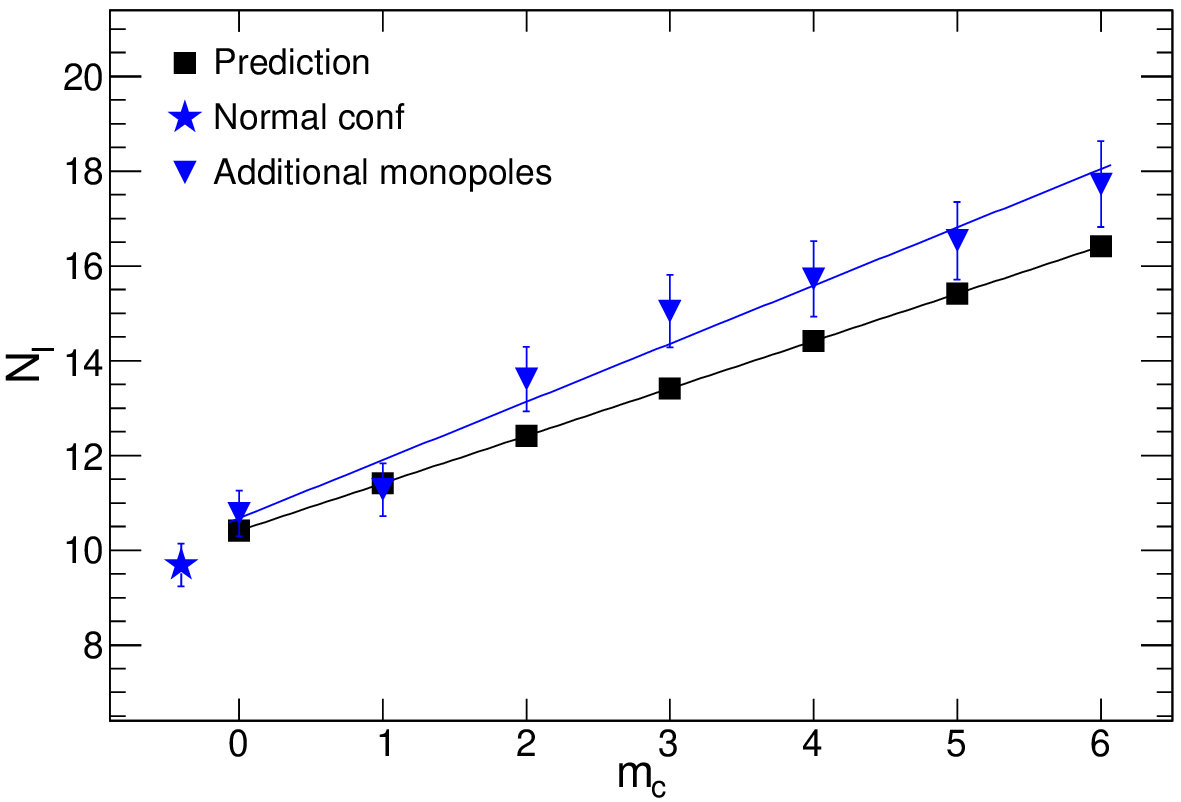}
\setlength\abovecaptionskip{-3.0pt}
\caption{The total number of instantons and anti-instantons $N_{I}$ versus the magnetic charges $m_{c}$. The blue and black lines indicate the fitting results.}\label{fig:num_ins_pred}
\end{figure}

The number density $\rho_{I}$ of the instantons (or anti-instantons) computed in the instanton liquid model~\cite{Shuryak1_1} is $\rho_{I} = 8 \times 10^{-4} \  \  [\mbox{GeV}^{4}]$. We suppose CP invariance; thus, the number density of the instantons and anti-instantons in the volume $V$ is
\begin{equation}
 2\rho_{I} = \frac{N_{I}}{V} = 1.6 \times 10^{-3} \  \  [\mbox{GeV}^{4}]\label{eq:inst_dens_1}.
\end{equation}
The total number of instantons and anti-instantons $N_{I}^{nor}$ in the physical volume $V_{phys}$ = 9.8582 [fm$^{4}$] ($V = 18^{3}\times32$, $\beta = 6.0522$) of the normal configuration is estimated as follows:
\begin{equation}
N_{I}^{nor} =  10.4138\label{eq:inst_dens_2}
\end{equation}
These results show that we can properly calculate the total number of instantons and anti-instantons $N_{I}$ in the physical volume $V_{phys}$ from the topological charges $Q$ using formula~(\ref{eq:num_ins}).

The total number of the instantons and anti-instantons of the magnetic charges $m_{c}$ is predicted as follows:
\begin{align}
  N_{I}^{Pre} = m_{c} + N_{I}^{nor}\label{eq:inst_dens_pre_1}
\end{align}

Moreover, we can analytically predict the numbers of zero modes $N_{Z}^{Pre}$, which are detected in our simulations, using the result~(\ref{eq:inst_dens_2}). The analytic formulas are given in appendix B of reference~\cite{DiGH3} (we provide the analytic formulas for magnetic charges $m_{c} =$ 5 and 6 in appendix~\ref{sec:number_zero}).

We list the results of the number of zero modes $N_{Z}$ that we observed, the total number of instantons and anti-instantons $N_{I}$, and instanton density $\frac{N_{I}}{V}$, as shown in table~\ref{tb:zero_instantons}. The predictions generated with the formulas in appendix B of reference~\cite{DiGH3}, appendix~\ref{sec:number_zero}, and~(\ref{eq:inst_dens_pre_1}) are indicated with the superscript $Pre$ in the same table. The numerical results are consistent with the predictions as shown in table~\ref{tb:zero_instantons}.

To evaluate how many monopoles create instantons and anti-instantons in the configurations, we fit the linear function $N_{I} = Am_{c} + B$ to the prediction and numerical results of $N_{I}$, as shown in figure~\ref{fig:num_ins_pred}. The fitting results are $A = 1.23(13)$, $B = 10.7(4)$, and $\chi^{2}/ d. o. f. = 2.9/5.0$. The fitting result of the intercept $B$ is consistent with the total number of instantons and anti-instantons of the normal configuration $N_{I} = 9.7(5)$ and the result~(\ref{eq:inst_dens_2}). The slope of the numerical result $A$ is approximately 1 of the slope of the prediction~(\ref{eq:inst_dens_pre_1}). Therefore, the monopole of a magnetic charge +1 and the anti-monopole of a magnetic charge -1 make one instanton or one anti-instanton.
\begin{figure}[tbp]
  \centering
    \includegraphics[width=145mm]{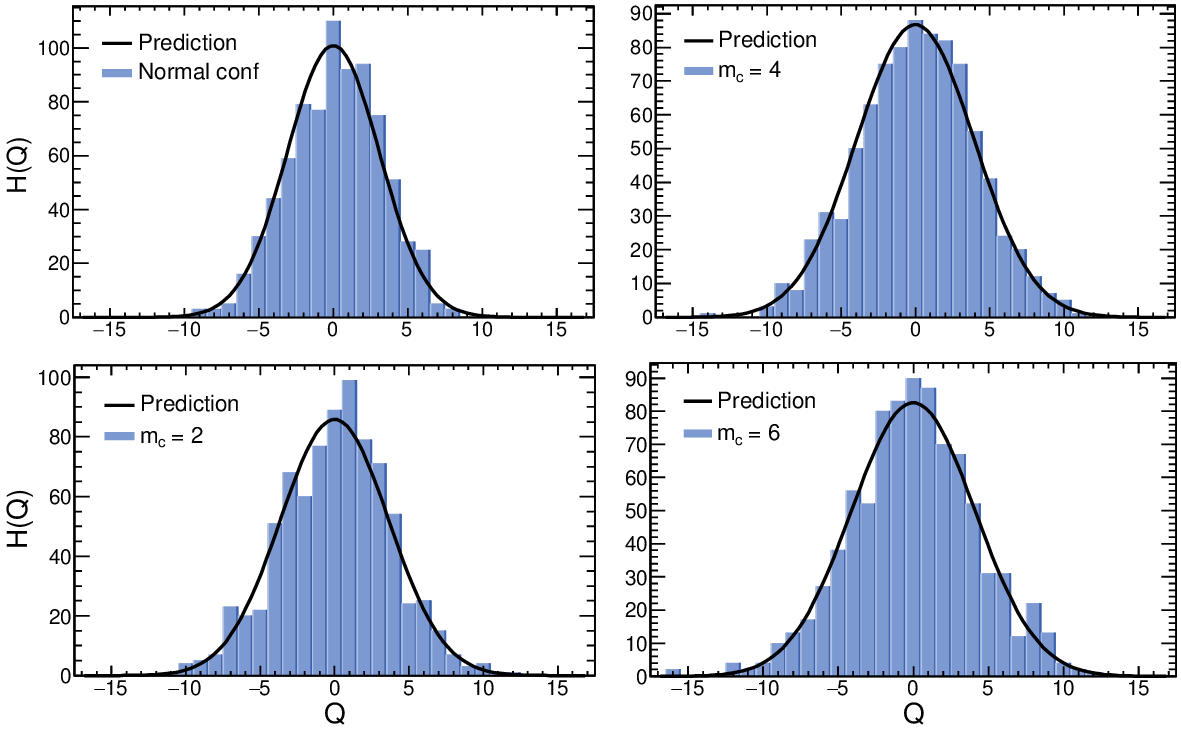}
\setlength\abovecaptionskip{-3.0pt}
\caption{Comparisons of the histogram $H(Q)$ of the topological charges $Q$ of the normal configurations (upper left) and the magnetic charges $m_{c}$ = 2 (lower left), 4 (upper right), and 6 (lower right). The black lines indicate the fitting results according to the distribution functions for each magnetic charge.}\label{fig:qtop_fit}
\end{figure}
\begin{table}[tbp]
  \small
   \centering
          {\renewcommand{\arraystretch}{1.1}
              \begin{tabular}{|c|c|c|c|c|}\hline
                $m_{c}$ & $\langle \delta^{2} \rangle$ & $\mathcal{O}(V^{-1})$  & $\chi^{2}/d. o. f.$ & $N_{conf}$ \\ \hline
                {\footnotesize Normal conf} & 9.6(5) & -2(4)$\times10^{-2}$ & 18.7/17.0 & 800 \\ \hline
                0   & 10.1(5) & -3(3)$\times10^{-2}$  & 28.32/19.0 & 800 \\ \hline
                1   & 10.1(6) & -1(3)$\times10^{-2}$& 12.1/19.0 & 838 \\ \hline
                2   & 11.2(8) & -3(3)$\times10^{-2}$ & 27.7/22.0 & 810 \\ \hline
                3   & 11.7(9) & -3(3)$\times10^{-2}$ & 23.6/22.0 & 800 \\ \hline
                4   & 11.5(8) & -1(3)$\times10^{-2}$ & 12.4/21.0 & 868\\    \hline            
                5   & 10.9(1.0) & -3(3)$\times10^{-2}$ & 27.8/22.0 & 810 \\ \hline
                6   & 10.6(9) & -3(3)$\times10^{-2}$  & 24.1/24.0 & 870\\ \hline
              \end{tabular}
          }
           \caption{The fitting results of $\langle \delta^{2} \rangle$ and correction term $\mathcal{O}(V^{-1})$ for each magnetic charge.}\label{tb:fitting_top_q_prediction}
\end{table}

The distribution of the topological charges computed using the overlap Dirac operator in the quenched QCD becomes the following Gaussian distribution~\cite{Giusti4, DeDebbio1}:
\begin{equation} 
  P(Q) = \frac{\mathrm{e}^{-\frac{Q^{2}}{2\langle \delta^{2}\rangle}}}{\sqrt{2\pi\langle \delta^{2}\rangle}}\left[ 1 + \mathcal{O}(V^{-1})\right].\label{gauss_1}
\end{equation}
We made the distribution function of the topological charges for each magnetic charge $m_{c}$ = 0 - 4 with formula (39) in reference~\cite{DiGH3}. We provide the distribution functions~(\ref{eq:dis_func_mc5}) - (\ref{eq:dis_func_mc6}) for the magnetic charges $m_{c}$ = 5 - 6 in appendix~\ref{sec:dis_top_func}. The distribution functions comprise Gaussian distributions with the same fitting parameter $\langle\delta^{2}\rangle$ and correction term $\mathcal{O}(V^{-1})$ as the distribution function~(\ref{gauss_1}). We fit these distribution functions to the distributions of the topological charges as shown in figure~\ref{fig:qtop_fit}. Table~\ref{tb:fitting_top_q_prediction} indicates that the fitting results of $\langle \delta^{2} \rangle$ of the configurations with the additional monopoles and anti-monopoles are reasonably consistent with the fitting result of the normal configuration. The correction terms $\mathcal{O}(V^{-1})$ are zero, and the values of $\chi^{2}/d. o. f.$ range from 0.6 to 1.5. Therefore, these results clearly indicate that the monopole creation operator adds the topological charges to the configurations without affecting the vacuum structure. We can properly predict the increases of the topological charges.

Finally, these results are consistent with previous results~\cite{DiGH3}.

\subsection{Comparisons with random matrix theory}\label{sec:RMT}


In this subsection, we increase the number of normal configurations to $N_{conf} = 1144$ and the number of configurations of the magnetic charges $m_{c} = 5$ to $N_{conf} = 1566$ to precisely compare with RMT.

We first present the distributions of the nearest-neighbor spacing to study the effects of the additional monopoles and anti-monopoles on the short-range fluctuations of the low-lying eigenvalues. The nearest-neighbor spacing $s$ is given by $s_{i}^{n} = \xi_{i + 1}^{n} - \xi_{i}^{n}$. The superscript $n$ is the configuration number, and the subscript $i$ is the eigenvalue number. The unfolded eigenvalues $\xi$ are obtained in the following way~\cite{Guhr1}. We compute the eigenvalues $\tilde{\lambda}$ of the improved overlap Dirac operator $\tilde{D}(\rho)$. The improved overlap Dirac operator $\tilde{D}(\rho)$ is defined from the massless overlap Dirac operator $D(\rho)$ as follows~\cite{Capitani}:
\begin{equation}
\tilde{D}(\rho) =  \left( 1 - \frac{a}{2\rho} D(\rho) \right)^{-1}D(\rho).\label{imp_op1}
\end{equation}
The eigenvalues $\tilde{\lambda}$ are projected onto the imaginary axis to be near the continuum limit. These eigenvalues are pure imaginary numbers, and all eigenvalues come in positive and negative pairs $\pm i\tilde{\lambda}_{i}^{n}$.

We put the nonzero and positive eigenvalues in ascending order $\tilde{\lambda}_{1}^{n} < \cdots < \tilde{\lambda}_{i}^{n} < \cdots < \tilde{\lambda}_{k}^{n}$ and fit them by the following polynomial of the degree $d = 3$ for each configuration.
\begin{equation}
 N_{pol}(\tilde{\lambda}^{n}) = \sum_{d = 0}^{3}a_{d}^{n}\tilde{\lambda}_{d}^{n}
\end{equation}
The unfolded eigenvalue is obtained by $\xi_{i}^{n} = N_{pol}(\tilde{\lambda}_{i}^{n})$.

The distribution of the nearest-neighbor spacing falls into the three different ensemble classes that obey the symmetries that are universally given in the GRMT. 
The three ensembles are the Gaussian orthogonal ensemble (GOE), the Gaussian unitary ensemble (GUE), and the Gaussian symplectic ensemble (GSE)~\cite{Wigner1,Dyson1}. The distributions of the nearest-neighbor spacing of the GUE is given in the GRMT as follows~\cite{Guhr2}:
\begin{equation}
P(s) = \frac{32}{\pi^{2}}s^{2}\exp\left(-\frac{4s^{2}}{\pi}\right)\label{eq:dis_uni}
\end{equation}
\begin{figure}[tbp]
 \centering
   \includegraphics[width=150mm]{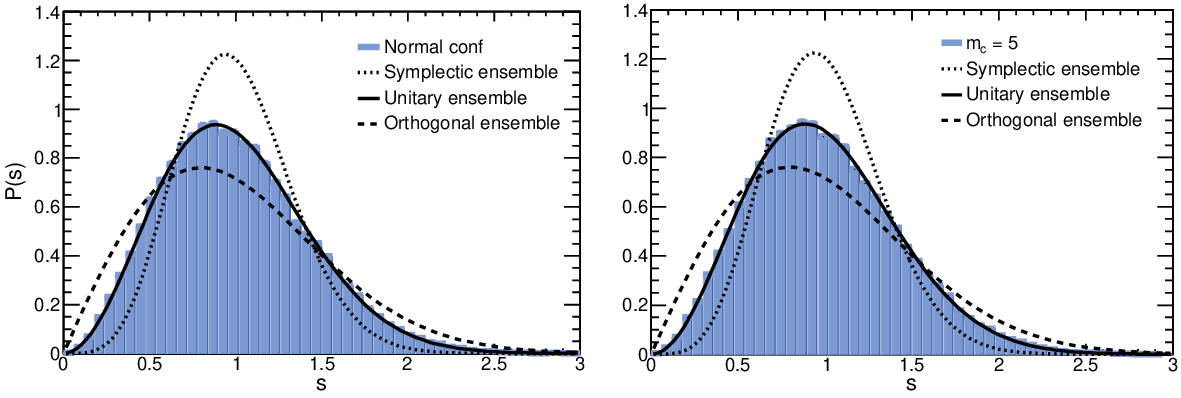}
\setlength\abovecaptionskip{-3.0pt}
\caption{The distributions $P(s)$ of the nearest-neighbor spacing $s$ of the normal configuration (left) and the configuration of $m_{c} = 5$ (right). The distributions of the GOE, the GUE, and the GSE in the GRMT are represented by the dashed lines, the full lines, and the dotted lines, respectively.}\label{fig:near_or_mc5}
\end{figure}
Figure~\ref{fig:near_or_mc5} shows that the distributions of the nearest-neighbor spacing that are calculated with the normal configurations and the configurations of the magnetic charges $m_{c} = 5$ agree perfectly with the distribution of the GUE in the GRMT. The additional monopoles and anti-monopoles do not affect the short-range fluctuations of the low-lying eigenvalues.

Next, to probe the effects of the additional monopoles and anti-monopoles on the spectrum of a long interval of the length $L$, we calculate the spectral rigidity $\Sigma(L)$, which is introduced by Dyson and Mehta~\cite{Dyson2}. The spectral rigidity of the interval $[\alpha, \alpha + L]$ is calculated as follows~\cite{Bohigas1}:
\begin{align}
  & \tilde{\xi}_{i}^{n} = \xi_{i}^{n} - \left(\alpha - \frac{L}{2} \right)\\ 
  & \overline{\langle\Delta_{3} (L) \rangle}  = \frac{k^{2}}{16} - \frac{1}{L^{2}}\left(\sum_{i = 1}^{k}\tilde{\xi}_{i}\right)^{2}\nonumber \\
  & \hspace{18mm} + \frac{3k}{2L^{2}}\left(\sum_{i = 1}^{k}\tilde{\xi}_{i}^{2}\right)  - \frac{3}{L^{4}}\left(\sum_{i = 1}^{k}\tilde{\xi}_{i}^{2}\right)^{2}  +\frac{1}{L}\left[\sum_{i = 1}^{k}(k - 2i + 1)\tilde{\xi}_{i}\right]\label{eq:Delta3_rmt}
\end{align}
The configuration number $n$ in equation~(\ref{eq:Delta3_rmt}) is eliminated. We set the starting point $\alpha$ on the unfolded scale from 1 to 13 for the normal configuration and from 1 to 10 for the configuration of $m_{c} = 5$, and we calculate the spectral average. We then compute the ensemble average over the configurations. In the computations, almost all positive eigenvalues are used. Here, the spectral average and the ensemble average are denoted as $\overline{\langle \cdots \rangle}$.
\begin{figure}[tbp]
 \centering
   \includegraphics[width=150mm]{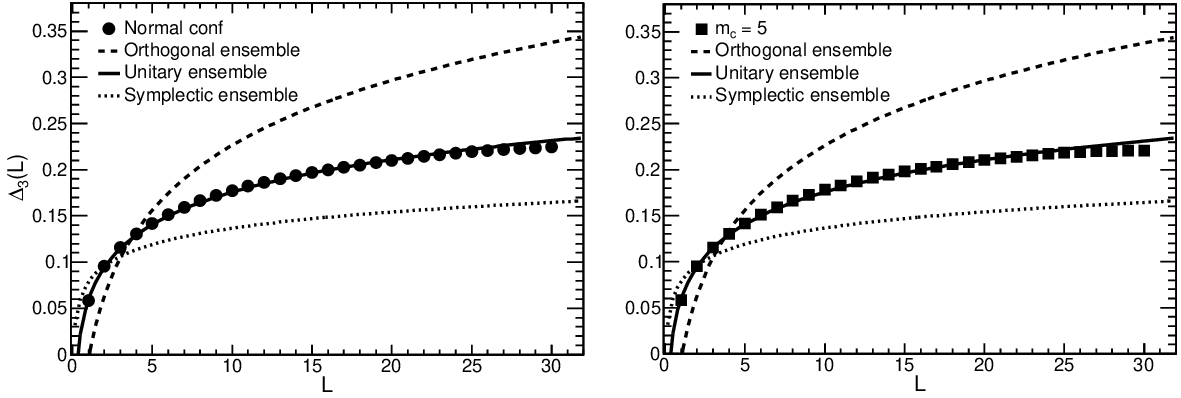}
\setlength\abovecaptionskip{-3.0pt}
\caption{The spectral rigidity $\Delta_{3}(L)$ of the normal configuration (left) and the configuration of $m_{c} = 5$ (right). The results of the GOE, the GUE, and the GSE in the GRMT are represented by the dashed line, the full line, and the dotted line, respectively.}\label{fig:Delta3_or_mc5}
\end{figure}

In the GRMT, the spectral rigidity of the GUE is predicted as follows~\cite{Guhr2}:
\begin{equation}
  \Delta_{3} (L) = \frac{1}{2\pi^{2}}\left[ \ln (2\pi L) + \gamma - \frac{5}{4} \right], \  \gamma = 0.5772.\label{eq:Delta3_rmt_predic}
\end{equation}
Figure~\ref{fig:Delta3_or_mc5} shows that the numerical results of the spectral rigidity are remarkably consistent with the function~(\ref{eq:Delta3_rmt_predic}). Therefore, the additional monopoles and anti-monopoles do not affect the spectrum of the long interval of length $L$.

\begin{figure}[tbp]
 \centering
   \includegraphics[width=150mm]{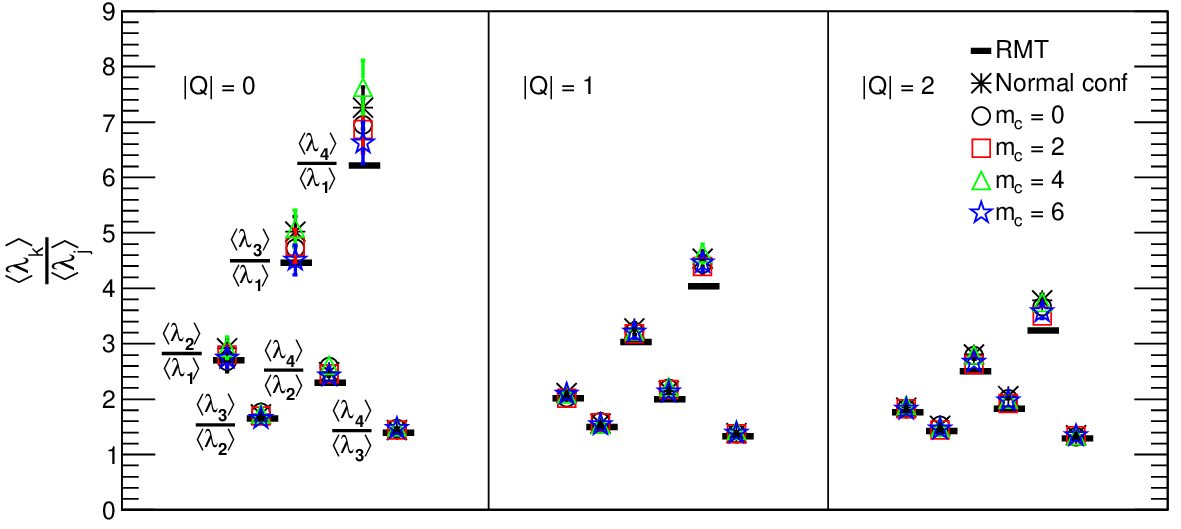}
\setlength\abovecaptionskip{-3.0pt}
\caption{The comparisons of the ratios of the eigenvalues $\tilde{\lambda}_{k}^{|Q|}$ with the prediction of the chRMT for each topological charge sector $|Q| = 0, 1, 2$. The normal configuration and the configurations of the magnetic charges $m_{c} = 0, 2, 4, 6$ are used. The black horizontal lines and the colored symbols indicate the prediction and the numerical results, respectively.}\label{fig:rmt_test}
\end{figure}

The distribution $P(z)$ of the scaled eigenvalues $z$ of the Dirac operator in the $\epsilon$-regime of QCD, which is classed according to three ensembles and each topological charge sector $|Q|$, is universally predicted in chiral random matrix theory (chRMT)~\cite{Nishigaki1,Dammgard1}. The scaled $k$-th eigenvalue of the chRMT $z_{k}^{|Q|}$ and the $k$-th eigenvalue of the Dirac operator $\lambda_{k}^{|Q|}$ have the following relation.
\begin{equation}
  z_{k}^{|Q|} = \Sigma V \lambda_{k}^{|Q|}\label{eq:rmt_eigen}
\end{equation}
The scale parameter $\Sigma$ is a free parameter that is determined from data. To remove the uncertainty that comes from the free parameter, we calculate the ratio of the eigenvalues $\frac{\langle\tilde{\lambda}_{k}\rangle^{|Q|}}{\langle\tilde{\lambda}_{j}\rangle^{|Q|}}$~\cite{Giusti4} and list the results in table~\ref{tb:rmt_test_tb}. $\langle\cdots\rangle$ indicates the ensemble average. We then compare the results with the prediction of the chRMT as shown in figure~\ref{fig:rmt_test}. The table and figure indicate that the ratios of the eigenvalues are consistent with the prediction of the chRMT even if we increase the magnetic charges to $m_{c} = 6$. The additional monopoles and anti-monopoles do not affect the low-lying eigenvalues of the overlap Dirac operator.
\begin{table}[tbp]
  \centering
  \begin{scriptsize}
              \begin{tabular}{|c|c|c|c|c|c|c|c|c|c|c|} \hline
               $|Q|$&$k/j$&N. C.&$m_{c}=0$&$m_{c}=1$&$m_{c}=2$&$m_{c}=3$&$m_{c}=4$&$m_{c}=5$&$m_{c}=6$&RMT\\ \hline
                0&2/1&2.92(17)&2.69(19)&2.79(19)&2.80(18)&3.0(2)&2.9(2)&2.67(16)&2.74(17)&2.70\\\cline{2-11}  
                &3/1&5.0(3)&4.7(3)&4.8(3)&4.8(3)&5.2(4)&5.1(3)&4.7(3)&4.5(3)&4.46\\\cline{2-11}  
                &4/1&7.3(4)&7.0(5)&6.9(4)&6.8(4)&7.5(5)&7.6(5)&6.8(4)&6.6(4)&6.22\\\cline{2-11}  
                &3/2&1.72(6)&1.76(8)&1.73(8)&1.70(7)&1.71(8)&1.73(8)&1.75(7)&1.65(7)&1.65\\\cline{2-11}  
                &4/2&2.49(8)&2.59(10)&2.48(11)&2.44(9)&2.46(10)&2.61(11)&2.53(9)&2.42(9)&2.30\\\cline{2-11}  
                &4/3&1.44(4)&1.47(5)&1.44(5)&1.43(5)&1.44(5)&1.50(6)&1.45(4)&1.47(5)&1.40\\\hline  
                1&2/1&2.11(8)&2.04(8)&2.13(8)&2.02(8)&2.09(9)&2.09(8)&2.05(6)&2.09(8)&2.02\\\cline{2-11}  
                &3/1&3.27(11)&3.23(12)&3.29(12)&3.18(12)&3.32(13)&3.23(12)&3.19(8)&3.23(12)&3.03\\\cline{2-11}  
                &4/1&4.53(15)&4.48(17)&4.62(17)&4.39(16)&4.55(17)&4.63(17)&4.37(11)&4.47(16)&4.04\\\cline{2-11}  
                &3/2&1.55(4)&1.58(4)&1.54(4)&1.57(5)&1.59(5)&1.54(4)&1.56(3)&1.54(4)&1.50\\\cline{2-11}  
                &4/2&2.15(5)&2.19(6)&2.17(6)&2.17(6)&2.17(6)&2.21(6)&2.13(4)&2.14(6)&2.00\\\cline{2-11}  
                &4/3&1.39(3)&1.38(3)&1.40(3)&1.38(3)&1.37(4)&1.43(3)&1.37(3)&1.39(3)&1.33\\\hline
                2&2/1&1.85(6)&1.83(7)&1.89(6)&1.84(7)&1.86(7)&1.88(7)&1.89(5)&1.81(7)&1.76\\\cline{2-11}  
                &3/1&2.80(8)&2.79(10)&2.75(8)&2.64(10)&2.76(11)&2.79(10)&2.81(7)&2.66(9)&2.50\\\cline{2-11}  
                &4/1&3.79(11)&3.67(12)&3.75(11)&3.52(13)&3.65(14)&3.75(13)&3.74(9)&3.57(12)&3.24\\\cline{2-11}  
                &3/2&1.52(4)&1.53(4)&1.46(4)&1.44(4)&1.49(4)&1.49(4)&1.48(3)&1.47(4)&1.42\\\cline{2-11}  
                &4/2&2.05(5)&2.01(6)&1.99(5)&1.92(5)&1.97(5)&1.99(5)&1.98(4)&1.97(5)&1.83\\\cline{2-11}  
                &4/3&1.35(3)&1.32(3)&1.36(3)&1.34(3)&1.32(3)&1.34(3)&1.33(2)&1.34(3)&1.29\\\hline
              \end{tabular}
              \end{scriptsize}
  \caption{The numerical results of $\frac{\langle\tilde{\lambda}_{k}\rangle^{|Q|}}{\langle\tilde{\lambda}_{j}\rangle^{|Q|}}$ and the prediction. The normal configurations (N. C.) and the configurations of the magnetic charges $m_{c}$ from 0 to 6 are used. The results of the comparison with RMT are presented in Table 3~\cite{Giusti4}.}\label{tb:rmt_test_tb}
\end{table}

The chRMT provides the following distribution functions $P_{k}^{|Q|}(z)$ of the scaled first eigenvalues $z$ for each topological charge sector $|Q|$ of the GUE~\cite{Nishigaki1,Dammgard1}.
\begin{align}
P_{k=1}^{|Q| = 0}(z) & = \frac{z}{2}\exp\left(-\frac{z^{2}}{4}\right),\label{eq:first_rmt_1}\\
P_{k=1}^{|Q| = 1}(z) & = \frac{z}{2}\exp\left(-\frac{z^{2}}{4}\right)I_{2}(z),\label{eq:first_rmt_2}\\
P_{k=1}^{|Q| = 2}(z) & = \frac{z}{2}\exp\left(-\frac{z^{2}}{4}\right)\left[I_{2}^{2}(z) - I_{1}(z)I_{3}(z)\right],\label{eq:first_rmt_3}\\
P_{k=1}^{|Q| = 3}(z) & = \frac{z}{2}\exp\left(-\frac{z^2}{4}\right)\left[I_{2}^{3}(z) - 2I_{1}(z)I_{2}(z)I_{3}(z)\right.\nonumber\\
  &\left. \hspace{30mm}+ I_{0}(z)I_{3}^{2}(z) +  I_{1}^{2}(z)I_{4}(z) - I_{0}(z)I_{2}(z)I_{4}(z)\right].\label{eq:first_rmt_4}
\end{align}
\begin{table}[tbp]
  \small
   \centering
          {\renewcommand{\arraystretch}{1.1}
              \begin{tabular}{|c|c|c|c|c|c|}\hline
                $m_{c}$ & $|Q|$ & $N_{conf}$ & $\Sigma^{a}$ [GeV$^{3}$]  & $\Sigma^{b}$ [GeV$^{3}$] & $\chi^{2}/d. o. f.$ \\ \hline
                {\footnotesize Normal conf} & 0 & 157 & 2.08(10)$\times10^{-2}$ & 2.04(10)$\times10^{-2}$ & 12.9/13.0 \\ \cline{2-6} 
                & 1 &  249  & 1.98(6)$\times10^{-2}$ & 2.01(8)$\times10^{-2}$ & 42.6/16.0 \\ \cline{2-6} 
                & 2 &  245  & 2.11(5)$\times10^{-2}$ & 2.19(6)$\times10^{-2}$ & 41.3/16.0 \\ \cline{2-6} 
                & 3 &  192  & 2.02(5)$\times10^{-2}$ & 2.04(5)$\times10^{-2}$ & 36.4/15.0 \\ \hline
                5 & 0 & 142 & 2.64(14)$\times10^{-2}$ & 2.73(14)$\times10^{-2}$ & 12.7/15.0 \\  \cline{2-6}  
                & 1 & 291  & 2.56(6)$\times10^{-2}$ & 2.51(5)$\times10^{-2}$ & 17.5/18.0 \\  \cline{2-6} 
                & 2 & 275  & 2.81(6)$\times10^{-2}$ & 2.80(6)$\times10^{-2}$ & 23.6/19.0 \\  \cline{2-6} 
                & 3 & 244  & 2.55(5)$\times10^{-2}$ & 2.58(6)$\times10^{-2}$ & 40.6/23.0\\ \hline
              \end{tabular}
          }
           \caption{The results of the scale parameters $\Sigma^{a}$ and $\Sigma^{b}$ of each topological charge sectors $|Q|$ and $k = 1$.}\label{tb:fitting_sigma}
\end{table}
\begin{table}[tbp]
    \begin{footnotesize}
    \centering
        {\renewcommand{\arraystretch}{1.1}
          \begin{tabular}{|c|c|c|c|c|c|c|c|c|c|}\hline
            $k$ & $|Q|$ & {\scriptsize Normal conf} & $m_{c}=0$ & $m_{c}=1$ & $m_{c}=2$ & $m_{c}=3$ & $m_{c}=4$ & $m_{c}=5$ & $m_{c}=6$\\ 
            & & $\times10^{-2}$  & $\times10^{-2}$ & $\times10^{-2}$ & $\times10^{-2}$ & $\times10^{-2}$ & $\times10^{-2}$ & $\times10^{-2}$ & $\times10^{-2}$ \\\hline
            1  & 0 &  2.08(10) & 1.98(13) & 2.17(12) & 2.35(14) & 2.71(18) & 2.89(17) & 2.64(14) & 2.63(14)\\ \cline{2-10}
            & 1 &  1.98(6) &  1.99(7) &  2.14(7) &  2.36(8) &  2.51(8) &  2.61(9) &  2.56(6) &  2.60(8) \\\cline{2-10}
            & 2 &  2.11(5) &  2.01(6) &  2.16(6) &  2.19(7) &  2.50(8) &  2.63(8) &  2.81(6) &  2.65(8) \\ \hline
            2  & 0 &  1.92(5) &  1.99(7) &  2.10(8) &  2.27(6) &  2.42(9) &  2.67(10) & 2.67(8) &  2.60(9) \\ \cline{2-10}
            & 1 &  1.90(4) &  1.96(4) &  2.02(4) &  2.36(5) &  2.43(6) &  2.52(5) &  2.53(4) &  2.51(5) \\ \cline{2-10}
            & 2 &  2.01(4) &  1.94(4) &  2.02(4) &  2.10(4) &  2.37(5) &  2.46(5) &  2.62(4) &  2.58(6) \\ \hline
            3  & 0 &  1.85(4) &  1.87(5) &  2.00(5) &  2.21(6) &  2.34(6) &  2.54(7) &  2.52(6) &  2.60(7) \\ \cline{2-10}
            & 1 &  1.83(3) &  1.86(3) &  1.97(3) &  2.25(4) &  2.29(5) &  2.45(4) &  2.43(4) &  2.44(4) \\ \cline{2-10}
            & 2 &  1.88(3) &  1.80(3) &  1.96(3) &  2.08(4) &  2.27(4) &  2.35(4) &  2.50(4) &  2.49(5) \\ \hline
            4  & 0 &  1.78(3) &  1.77(4) &  1.95(4) &  2.15(5) &  2.27(5) &  2.36(6) &  2.42(5) &  2.47(5) \\ \cline{2-10}
            & 1 &  1.76(2) &  1.79(3) &  1.87(3) &  2.17(3) &  2.23(4) &  2.28(3) &  2.37(3) &  2.35(4) \\ \cline{2-10}
            & 2 &  1.80(2) &  1.77(3) &  1.87(3) &  2.01(3) &  2.22(3) &  2.27(3) &  2.43(3) &  2.41(4) \\ \hline
            Ave. & Ave. & 1.91(5)  & 1.90(6)  &2.02(6)   & 2.21(6)  & 2.38(8)  & 2.50(8)  & 2.54(6)  & 2.53(7) \\ \hline 
          \end{tabular}          
        }
\end{footnotesize}
        \caption{The results of the scale parameter $\Sigma^{a}$ in the [GeV$^{3}$] unit. Ave. indicates the average value calculated using all the results of the $k$-th eigenvalues and the topological charge sectors $|Q|$.}\label{tb:sigma_all_a}
\end{table}
We determine the scale parameter $\Sigma$ in the following two ways~\cite{Edwards1,Giusti4}. First, $\Sigma^{a}$ is calculated analytically using the numerical results of the first eigenvalues of each topological sector, the equations~(\ref{eq:rmt_eigen}), and the distribution functions~(\ref{eq:first_rmt_1})-(\ref{eq:first_rmt_4}). Second, four fitting functions of one free parameter $\Sigma^{b}$ are made using the distribution functions~(\ref{eq:first_rmt_1})-(\ref{eq:first_rmt_4}). We fit them to the histograms of the first eigenvalues of each topological sector and determine the free parameter $\Sigma^{b}$. The histograms are normalized to unity.
\begin{figure}[tbp]
 \centering
   \includegraphics[width=150mm]{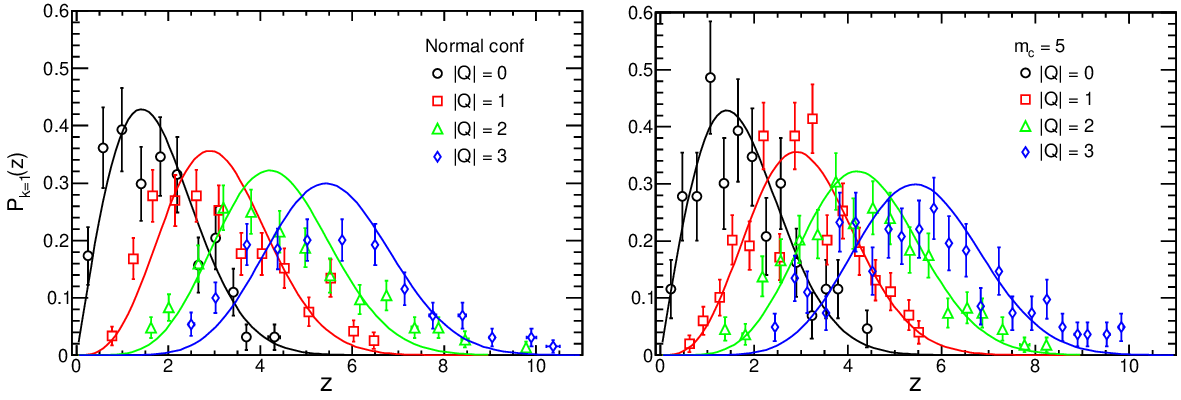}
\setlength\abovecaptionskip{-3.0pt}
\caption{The distributions of the first eigenvalues of each topological charge sector $|Q| = 0, 1, 2, 3$. The normal configurations (left) and the configurations of the magnetic charge $m_{c} = 5$ (right) are used. The colored symbols demonstrate the distributions that are re-scaled with the scale parameter $\Sigma^{b}$. The colored lines indicate the distributions of the chRMT.}\label{fig:k1_q0_1_2_3}
\end{figure}

\begin{figure}[tbp]
 \centering
   \includegraphics[width=75mm]{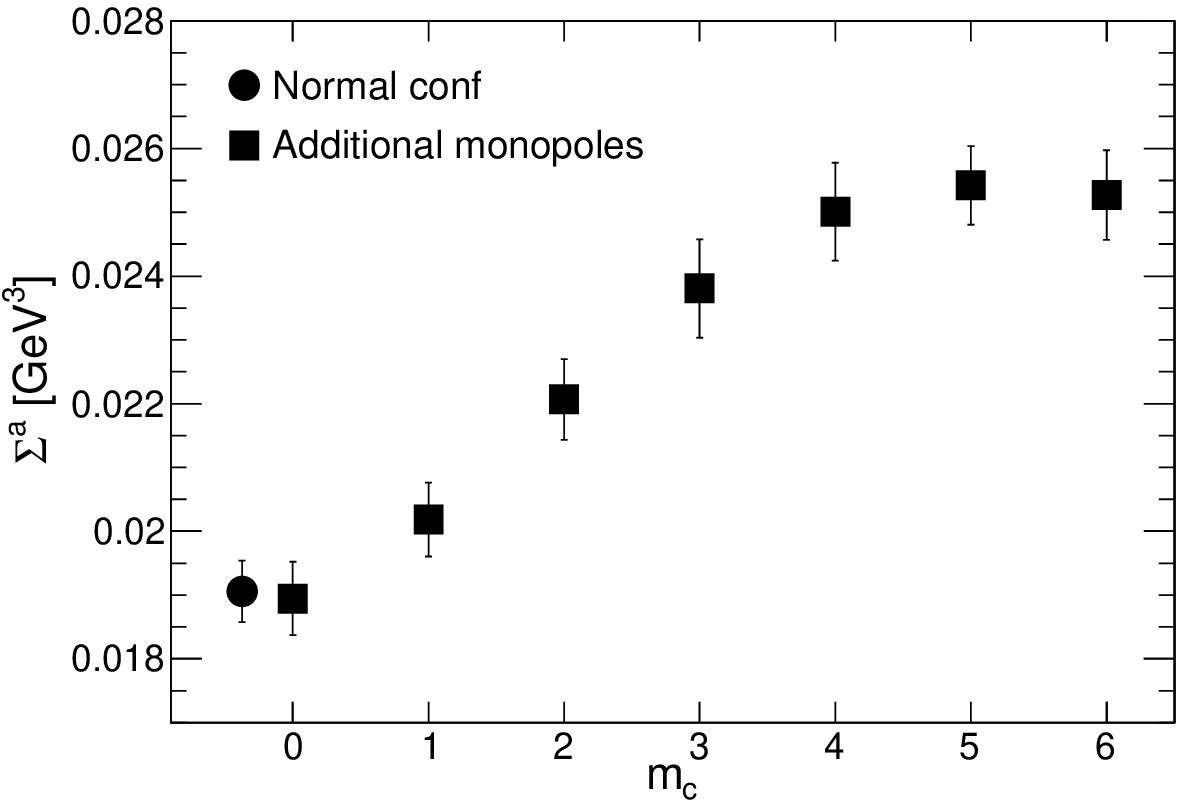}
\setlength\abovecaptionskip{-3.0pt}
\caption{The scale parameter $\Sigma^{a}$ versus the magnetic charges $m_{c}$. The results of $\Sigma^{a}$ are the average values in table~\ref{tb:sigma_all_a}.}\label{fig:mc_scale_para}
\end{figure}

The results of $\Sigma^{a}$ and $\Sigma^{b}$ of the normal configuration and the configuration of the magnetic charge $m_{c} = 5$ are presented in table~\ref{tb:fitting_sigma}. The results of $\Sigma^{a}$ that uses all configurations are displayed in table~\ref{tb:sigma_all_a}. Table~\ref{tb:fitting_sigma} shows that the results of the $\Sigma^{a}$ and $\Sigma^{b}$ of each topological charge sector are consistent with one another. We re-scale the first eigenvalues $\tilde{\lambda}_{k}^{|Q|}$ using the fitting results of $\Sigma^{b}$ of each topological charge sector and compare them with the distribution functions of the chRMT as shown in figure~\ref{fig:k1_q0_1_2_3}. The errors are estimated with the jackknife method. Finally, figure~\ref{fig:mc_scale_para} clearly shows that the scale parameter $\Sigma^{a}$ linearly increases when the magnetic charges $m_{c}$ increase.
The chiral condensate is estimated from this scale parameter~\cite{Wennekers1,DiGHP1}. Therefore, this figure indicates that the values of the chiral condensate linearly decrease when the magnetic charges $m_{c}$ increase. We explain the reason for this in the sections below.

These results demonstrate that the additional monopoles and anti-monopoles do not affect the eigenvalues of the overlap Dirac operator, and they change only the scale of the distribution of the eigenvalues. This scale parameter linearly increases with increasing values of the magnetic charges $m_{c}$. These results correspond exactly to the results in the previous study~\cite{DiGHP1}.


\section{Predictions of the chiral condensate and the decay constants}\label{sec:pion_decay1}

In previous studies~\cite{DiGH2, DiGHP1, DiGH5}, we have shown that the values of the chiral condensate, which are defined as negative values, decrease with increasing values of the magnetic charge $m_{c}$. We found that the decay constants slightly increase with increasing values of the magnetic charge $m_{c}$. However, we cannot explain these results.

In this section, we make predictions to quantitatively explain the decreases in the chiral condensate and increases in the decay constants based on the models concerning the instanton.

\subsection{The predictions of the chiral condensate}\label{sec:chiral_cond_pre1}

The chiral condensate is calculated from the phenomenological models concerning the instanton~\cite{Dyakonov1, Dyakonov2, Dyakonov3, Dyakonov6, Shuryak2}. To quantitatively compare the numerical results in the sections below, we show the following consequence of the chiral condensate as calculated from the model of the instanton vacuum~\cite{Dyakonov3}.
\begin{align}
  \langle{\bar\psi} \psi \rangle = -\frac{1}{{\bar\rho}}\left( \frac{\pi N_{c}}{13.2}\right)^{\frac{1}{2}} \left( \frac{N_{I}}{V}\right)^{\frac{1}{2}} = - (272.7 \  [\mbox{MeV}])^{3}\label{eq:chiral_ins1}
\end{align}
Here, we use the number density of the instantons and anti-instantons~(\ref{eq:inst_dens_1}). $N_{c}$ represents the number of colors. The average size of the instanton~\cite{Shuryak1_1} is
\begin{equation}
\frac{1}{\bar{\rho}} = 6.00 \times10^{2} \ \  [\mbox{MeV}]\label{eq:ins_size}.
\end{equation}
As an important consequence of the models, the value of the chiral condensate decreases in direct proportion to the square root of the number density of the instantons and anti-instantons.

Next, we estimate the chiral condensate in the chiral limit ($m_{q} \rightarrow 0$) using the Gell-Mann-Oakes-Renner (GMOR) relation~\cite{Gellmann1} and the experimental results as follows:
\begin{align}
  \langle{\bar\psi} \psi \rangle  = - \lim_{\bar{m}_{q} \rightarrow 0}\frac{(m_{\pi}F_{\pi})^{2}}{2\bar{m}_{q}} = - (274^{+18}_{-8} \  [\mbox{MeV}])^{3}\label{eq:gmor_res_1}
\end{align}
Here, we suppose that the partially conserved axial current (PCAC) relation holds. We use the following result of the decay constant in the chiral limit as calculated according to the chiral perturbation theory~\cite{Colangelo3}:
\begin{equation}
  F_{0}^{\chi PT} = \lim_{m_{q} \rightarrow 0} F_{PS} = 86.2 (5) \ \ \ [\mbox{MeV}]\label{eq:chi_pt_f0}
\end{equation}
The experimental result of the average mass of the light quarks~\cite{PDG_2017} is
\begin{equation}
  {\bar m}_{q}^{Exp.} = \frac{m_{u} + m_{d}}{2} = 3.5^{+0.7}_{-0.3} \ \ \ [\mbox{MeV}]~\label{eq:exp_q}.
\end{equation}
The experimental result of the pion mass~\cite{PDG_2017} is
\begin{equation}
   m_{\pi^{\pm}}^{Exp.}  = 139.57061(24) \ \ \ [\mbox{MeV}]~\label{eq:exp_mpi}.
\end{equation}

In the studies of lattice QCD that use the overlap Dirac operator, the renormalization group invariant (RGI) scalar condensate $\langle{\bar\psi} \psi \rangle^{\overline{MS}}$ into the $\overline{MS}$-scheme at 2 [GeV] is computed from the scale parameter $\Sigma$ in random matrix theory~\cite{Wennekers1}. We reported the following result of the RGI chiral condensate~\cite{DiGHP1} into the $\overline{MS}$-scheme at 2 [GeV] using the same methods as reference~\cite{Wennekers1}. Moreover, the renormalized chiral condensate is estimated with the GMOR relation and the correlation functions of the operators into the $\overline{MS}$-scheme at 2 [GeV]~\cite{Giusti3}. The finding of the chiral condensate~(\ref{eq:chiral_ins1}) computed from the phenomenological model corresponds to these results. This clearly shows that the chiral condensate can be properly calculated from the number density of the instantons and anti-instantons.
\begin{table}[tbp]
  \small
  \centering
  \begin{tabular}{|c|c|c|c|c|c|c|} \hline
    $m_{c}$  & {\footnotesize$\left(\frac{N_{I}^{Pre}}{V}\right)^{\frac{1}{2}}$} & {\footnotesize$\left(\frac{N_{I}}{V}\right)^{\frac{1}{2}}$}   & {\footnotesize$\left(\frac{N_{I}^{Pre}}{V}\right)^{\frac{1}{4}}$} & {\footnotesize$\left(\frac{N_{I}}{V}\right)^{\frac{1}{4}}$} & $\langle \bar{\psi}\psi\rangle^{Pre}$ & $F_{0}^{Pre}$ \\
    & {\footnotesize[GeV$^{2}$]} & {\footnotesize[GeV$^{2}$]} & {\footnotesize[MeV]} &  {\footnotesize[MeV]}  & {\footnotesize[GeV$^{3}$]} & {\footnotesize[MeV]} \\ \hline
    {\footnotesize Normal conf}& 4.0000$\times10^{-2}$ & 3.85(9)$\times10^{-2}$  &200.00 & 196(2) &  -2.0280$\times10^{-2}$  & 85.366 \\ \hline
    0  & 4.0000$\times10^{-2}$ & 4.07(9)$\times10^{-2}$  &200.00 & 202(2) &  -2.0280$\times10^{-2}$  & 85.366\\ \hline
    1  & 4.1877$\times10^{-2}$ & 4.16(10)$\times10^{-2}$ &204.64 & 204(3) &  -2.1231$\times10^{-2}$  & 87.345\\ \hline
    2  & 4.3672$\times10^{-2}$ & 4.57(12)$\times10^{-2}$ &208.98 & 214(3) &  -2.2142$\times10^{-2}$  & 89.199\\ \hline
    3  & 4.5397$\times10^{-2}$ & 4.81(12)$\times10^{-2}$ &213.07 & 219(3) &  -2.3016$\times10^{-2}$  & 90.943\\ \hline
    4  & 4.7059$\times10^{-2}$ & 4.92(12)$\times10^{-2}$ &216.93 & 222(3) &  -2.3859$\times10^{-2}$  & 92.593\\ \hline
    5  & 4.8664$\times10^{-2}$ & 5.04(12)$\times10^{-2}$ &220.60 & 224(3) &  -2.4672$\times10^{-2}$  & 94.159\\ \hline
    6  & 5.0218$\times10^{-2}$ & 5.22(13)$\times10^{-2}$ &224.09 & 228(3) &  -2.5460$\times10^{-2}$  & 95.650\\ \hline
  \end{tabular}
  \caption{The numerical results of {\footnotesize$\left(\frac{N_{I}}{V}\right)^{\frac{1}{2}}$} and {\footnotesize$\left(\frac{N_{I}}{V}\right)^{\frac{1}{4}}$}, and their predictions {\footnotesize$\left(\frac{N_{I}^{Pre}}{V}\right)^{\frac{1}{2}}$} and {\footnotesize$\left(\frac{N_{I}^{Pre}}{V}\right)^{\frac{1}{4}}$}. The predictions of the chiral condensates $\langle \bar{\psi}\psi\rangle^{Pre}$ and the decay constants $F_{0}^{Pre}$.}\label{tb:instantons_chiral_fpi}
\end{table}

To quantitatively explain why the values of the chiral condensate decrease with increasing values of the magnetic charges $m_{c}$, we derive the following relational expression between the chiral condensate and the magnetic charges $m_{c}$ using formula~(\ref{eq:chiral_ins1})
\begin{align}
  &    \langle{\bar\psi} \psi \rangle^{Pre} = -\frac{1}{{\bar\rho}}\left( \frac{\pi N_{c}}{13.2} \right)^{\frac{1}{2}}\left( \frac{N_{I}^{Pre}}{V}\right)^{\frac{1}{2}}.\label{eq:chiral_ins_prediction1}
\end{align}
The total number of instantons and anti-instantons $N_{I}^{Pre}$ is~(\ref{eq:inst_dens_pre_1}). This prediction indicates that the value of the chiral condensate decreases in direct proportion to the square root of the number density of the instanton and anti-instantons. We calculate the prediction of the chiral condensates ${\langle{\bar\psi}\psi\rangle^{Pre}}$ by substituting the results of $\left(\frac{N_{I}^{Pre}}{V}\right)^{\frac{1}{2}}$ for formula~(\ref{eq:chiral_ins_prediction1}), and we list them in table~\ref{tb:instantons_chiral_fpi}.

\subsection{The predictions of the decay constants}\label{sec:decay_cons_pre}

The decay constant of the pseudoscalar in the chiral limit $F_{0}$, which is calculated using the configurations with the additional monopoles and anti-monopoles, is derived from the number density of the instantons and anti-instantons~(\ref{eq:inst_dens_pre_1}), the GMOR relation~(\ref{eq:gmor_res_1}), and the prediction of the chiral condensate~(\ref{eq:chiral_ins_prediction1}) as follows:
\begin{equation}
  F_{0}^{Pre} = \frac{1}{m_{\pi}}\left( \frac{2\bar{m}_{q}}{{\bar\rho}} \right)^{\frac{1}{2}} \left( \frac{\pi N_{c}}{13.2} \right)^{\frac{1}{4}} \left( \frac{N_{I}^{Pre}}{V} \right)^{\frac{1}{4}}\label{eq:pion_decay}
\end{equation}

The decay constant of the pseudoscalar in the chiral limit $F_{0}^{Pre}$ of the normal configuration ($m_{c} = 0$) is
\begin{equation}
 F_{0}^{Pre} = 85^{+9}_{-4} \ [\mbox{MeV}]\label{eq:f0_pred}.
\end{equation}
Here, we use formula~(\ref{eq:pion_decay}) and results~(\ref{eq:inst_dens_1}),~(\ref{eq:ins_size}),~(\ref{eq:exp_q}), and~(\ref{eq:exp_mpi}). The finding is clearly consistent with result~(\ref{eq:chi_pt_f0}) of the chiral perturbation theory. Therefore, we can properly predict the decay constant of the pseudoscalar in the chiral limit using formula~(\ref{eq:pion_decay}). The large errors of~(\ref{eq:f0_pred}), however, come from the experimental outcome of the average mass of the light quarks. For convenience, we do not consider the errors of the experimental results when comparing the prediction with the numerical results.

We substitute the results of the instanton densities $\left(\frac{N_{I}^{Pre}}{V}\right)^{\frac{1}{4}}$ for formula~(\ref{eq:pion_decay}) and calculate the prediction $F_{0}^{Pre}$. We list the computed results of $F_{0}^{Pre}$ in table~\ref{tb:instantons_chiral_fpi}.


\section{The PCAC relation, decay constants, and chiral condensate}

In this section, we calculate the correlation functions of the operators and estimate the renormalized decay constants, the mass of the pseudoscalar meson, and the renormalized chiral condensate. We inspect the increases in the decay constants and the decreases in the values of the chiral condensate by comparing the predictions with the numerical results.

\subsection{The correlation functions}

We calculate the correlation functions of the operators using the pairs of the eigenvalues $\lambda_{i}$ and eigenvectors $\psi_{i}$ of the massless overlap Dirac operator $D$. We use the technique in~\cite{Giusti2, DeGrand2} to calculate the quark propagators. The advantages of this technique are that we do not need to solve the eigenvalue problems of the massive overlap Dirac operator for each bare quark mass, and the excited terms of the correlation functions are removed. The validity of the results has already been shown in~\cite{Giusti2, DeGrand2}.

The quark propagator is defined from the spectral decomposition in the non-relativistic limit, similar to a quantum theory, as follows:
\begin{equation}
G(\vec{y}, y^{0}; \vec{x}, x^{0}) \equiv \sum_{i}\frac{\psi_{i}(\vec{x}, x^{0}) \psi_{i}^{\dagger}(\vec{y}, y^{0})}{\lambda_{i}^{mass}}\label{eq:qpro}
\end{equation}
The eigenvalues $\lambda_{i}^{mass}$ of the massive overlap Dirac operator $D (\bar{m}_{q})$ are calculated from the eigenvalues $\lambda_{i}$ of the massless overlap Dirac operator $D$ as follows:
\begin{equation}
\lambda_{i}^{mass} =\left(1-\frac{a\bar{m}_{q}}{2\rho} \right)\lambda_{i} + \bar{m}_{q}\label{eq:mass_d}
\end{equation}
The massive overlap Dirac operator $D (\bar{m}_{q})$~\cite{Neuberger1, Neuberger2, Niedermayer1} is defined as
\begin{equation}
  D (\bar{m}_{q}) = \left(1 - \frac{a\bar{m}_{q}}{2\rho} \right) D + \bar{m}_{q}\label{eq:ovmass}
\end{equation}
The parameter $\bar{m}_{q}$ is the bare quark mass. In this study, we set the masses of the light quarks $\bar{m}_{ud}$ and $\bar{m}_{sud}$ that compose the pion and kaon, respectively, as follows:
\begin{itemize}
\item Pion
\begin{equation}
  \bar{m}_{ud} \equiv \frac{m_{u} + m_{d}}{2}\label{eq:light_quark_masses1}
\end{equation}
\item Kaon
\begin{equation}
  \bar{m}_{sud} \equiv \frac{m_{s} + \bar{m}_{ud}}{2}\label{eq:light_quark_masses2}
\end{equation}
\end{itemize}

The quark bilinear operators of the scalar $\mathcal{O}_{S}$ and the pseudoscalar $\mathcal{O}_{PS}$ are defined as
\begin{align}
  & \mathcal{O}_{S} = \bar{\psi}_{1}\left( 1 - \frac{a}{2\rho}D\right)\psi_{2}, \ \ \mathcal{O}_{S}^{C} = \bar{\psi}_{2}\left( 1 - \frac{a}{2\rho}D\right)\psi_{1}\\
  & \mathcal{O}_{PS} = \bar{\psi}_{1}\gamma_{5}\left( 1 - \frac{a}{2\rho}D \right) \psi_{2}, \ \ \mathcal{O}_{PS}^{C} = \bar{\psi}_{2}\gamma_{5}\left( 1 - \frac{a}{2\rho}D \right) \psi_{1}.
\end{align}
The operator of the axial vector current $\mathcal{A}_{\mu}$ is defined as follows:
\begin{equation}
 \mathcal{A}_{\mu}= \bar{\psi}_{1}\gamma_{\mu}\gamma_{5}\left( 1 - \frac{a}{2\rho}D \right)\psi_{2}, \ \ \mathcal{A}_{\mu}^{C} = \bar{\psi}_{2}\gamma_{\mu}\gamma_{5}\left( 1 - \frac{a}{2\rho}D \right)\psi_{1}
\end{equation}
The superscript $C$ denotes the Hermitian transpose of the operator. The factor $\left(1 - \frac{a}{2\rho}\lambda_{j}\right)$ in the expressions of the quark bilinear operators comes from the definition of the fermion field $\psi$ in the overlap notation
\begin{equation}
  \psi_{a}(\vec{x}, x_{0}) \rightarrow \left(1 - \frac{a}{2\rho}D\right)\psi_{a}(\vec{x}, x_{0}), \  \ (a = 1, 2).
\end{equation}
The anti-particle of the fermion in the overlap notation is
\begin{equation}
  \bar{\psi}_{a}(\vec{x}, x_{0}) \rightarrow \bar{\psi}_{a}(\vec{x}, x_{0}), \ \ (a = 1, 2).
\end{equation}
We use the notations and definitions of reference~\cite{Niedermayer1}.

The correlation function of the scalar density is
\begin{equation}
  C_{SS}(\Delta t) = \frac{a^{3}}{V} \sum_{\vec{x}_{1}}\sum_{\vec{x}_{2}, \ t}\langle \mathcal{O}_{S}^{C}(\vec{x}_{2}, t) \mathcal{O}_{S}(\vec{x}_{1}, t + \Delta t)\rangle\label{eq:s_corre}.
\end{equation}
Similarly, the correlation function of the pseudoscalar density is
\begin{equation}
  C_{PS}(\Delta t) = \frac{a^{3}}{V} \sum_{\vec{x}_{1}}\sum_{\vec{x}_{2}, \ t}\langle \mathcal{O}_{PS}^{C}(\vec{x}_{2}, t) \mathcal{O}_{PS}(\vec{x}_{1}, t + \Delta t)\rangle\label{eq:ps_corre}.
  \end{equation}

We compute the correlation function between the partial derivative of the axial vector current and the pseudoscalar density as follows~\cite{Bochicchio1, Maiani1}:
\begin{equation}
aC_{AP}(\Delta t) = \frac{a^{4}}{V} \sum_{\vec{x}_{1}}\sum_{\vec{x}_{2}, \ t} \left\langle \left[\nabla_{0}^{*}\mathcal{A}_{0}^{C}(\vec{x}_{2}, t) \right] \mathcal{O}_{PS}(\vec{x}_{1}, t + \Delta t)\right\rangle 
\end{equation}
The partial derivative acts only on the axial vector current $A_{\mu}$ as follows:
\begin{equation}
  a\nabla_{0}^{*}A_{0}(\vec{x}, x^{0}) \equiv \frac{A_{0}(\vec{x}, x^{0} + 1) - A_{0}(\vec{x}, x^{0} - 1)}{2}.
\end{equation}

To reduce errors, we calculate the correlation functions between all spatial sites $\vec{x}$ and $\vec{y}$; moreover, we take the sum of the temporal sites $x^{0}$~\cite{DeGrand2}.

In the study of quenched QCD, the number of zero modes is not suppressed due to the lattice artifact of the finite volume. Such zero modes undesirably affect the PCAC relation near the chiral limit~\cite{Blum1, Giusti3}. In particular, we want to precisely evaluate the effects of monopoles and instantons on the physical quantities near the chiral limit. To remove the undesirable effect near the chiral limit, we subtract the scalar correlator $C_{SS}$ from the pseudoscalar correlator $C_{PS}$. The definition of the correlation function~\cite{Giusti3, Blum1} is the following:
\begin{equation}
  C_{PS-SS}(\Delta t) \equiv C_{PS}(\Delta t) - C_{SS}(\Delta t)\label{eq:corre_1}
\end{equation}

We vary the bare quark mass in the range $ 1.296\times10^{-2} \leq a\bar{m}_{q} \leq 6.482\times10^{-2}$ in the lattice unit, which corresponds to the range of $ 30 \ [\mbox{MeV}] \leq \bar{m}_{q} \leq 150 \ [\mbox{MeV}]$ in physical units. We calculate the correlation function~(\ref{eq:corre_1}) using the normal configurations and the configurations with the additional monopoles and anti-monopoles. The numbers of configurations that we use for the calculations of the correlation functions are listed in table~\ref{tb:zero_instantons}. We set a lower limit to the bare quark mass so that the relation $m_{PS}L_{s} \geq 2.4$, which is derived from the limit $m_{\pi}L \gg 1$ of the $p$-expansion~\cite{Colangelo3}, is satisfied. $L_{s}$ indicates the spatial length of the lattice in this study.

We suppose that the correlation function $C_{PS-SS}$ can be approximated by the following function~\cite{Gimenez1}:
\begin{equation}
  C_{PS-SS}(t) = \frac{a^{4}G_{PS-SS}}{am_{PS}} \exp\left( {-\frac{m_{PS}}{2}T}\right) \cosh\left[m_{PS} \left(\frac{T}{2} - t \right) \right]\label{eq:fiting_corre}.
\end{equation}
We fit this function to the numerical results, obtain the coefficient $a^{4}G_{PS-SS}$ and the pseudoscalar mass $am_{PS}$, and evaluate the decay constants and the chiral condensate. We set the fitting range so that the fitting value of $\chi^{2}/d. o. f.$ is approximately 1. The fitting results of the coefficient $a^{4}G_{PS-SS}$ and the pseudoscalar mass $am_{PS}$ are given in tables~\ref{tb:numerical_results_or_mc0},~\ref{tb:numerical_results_mc1_2},~\ref{tb:numerical_results_mc3_4}, and~\ref{tb:numerical_results_mc5_6} in appendix~\ref{sec:corre_func_fit_res_1}.

Moreover, to calculate the renormalization constant for the axial vector $Z_{A}$, we calculate the ratio~\cite{Gimenez1} of the correlation functions of $C_{AP}$ and $C_{PS}$, which is defined as follows:
\begin{equation}
  a\rho(\Delta t) \equiv \frac{aC_{AP}(\Delta t)}{2C_{PS}(\Delta t)}\label{eq:corre_2}
\end{equation}
We suppose that the parameter $a\rho(\Delta t)$ becomes constant~\cite{Giusti3}. We fit the constant function $a\rho (\Delta t) = aC$ to the numerical results of the ratio~(\ref{eq:corre_2}). The fitting results of $a\rho(\Delta t)$ are given in table~\ref{tb:numerical_results_arho_all} in appendix~\ref{sec:corre_func_fit_res_1}. The fitting range is $13 \leq t/a \leq 19$. The values of $\chi^{2}/ d. o. f.$ are very large because the errors of the ratio $a\rho (\Delta t)$ are very small. The numbers of configurations that we use for the computations are provided in table~\ref{tb:zero_instantons}.

\subsection{The PCAC relation}

We analyze the effects of the additional monopoles and anti-monopoles on the PCAC relation by comparing the results calculated using the normal configurations and the configurations with the additional monopoles and anti-monopoles. We suppose that the PCAC relation~\cite{Weinberg1} holds between the square of the pseudoscalar mass $m_{PS}^{2}$ and the bare quark mass $\bar{m}_{q}$ as follows:
\begin{equation}
  m_{PS}^{2} = A\bar{m}_{q}\label{eq:pcac_def}
\end{equation}
In this expression, the coefficient $A$ is a constant number that includes the factor 2 derived from the equations $2\bar{m}_{q} = {m_{i} + m_{j}}$. The subscripts $i, j$ indicate the flavors of the quarks. The bare quark mass $\bar{m}_{q}$ is defined as~(\ref{eq:light_quark_masses1}) and~(\ref{eq:light_quark_masses2}).
\begin{figure}[tbp]
  \begin{minipage}{0.48\hsize}
    \centering
    \includegraphics[keepaspectratio, width=70mm]{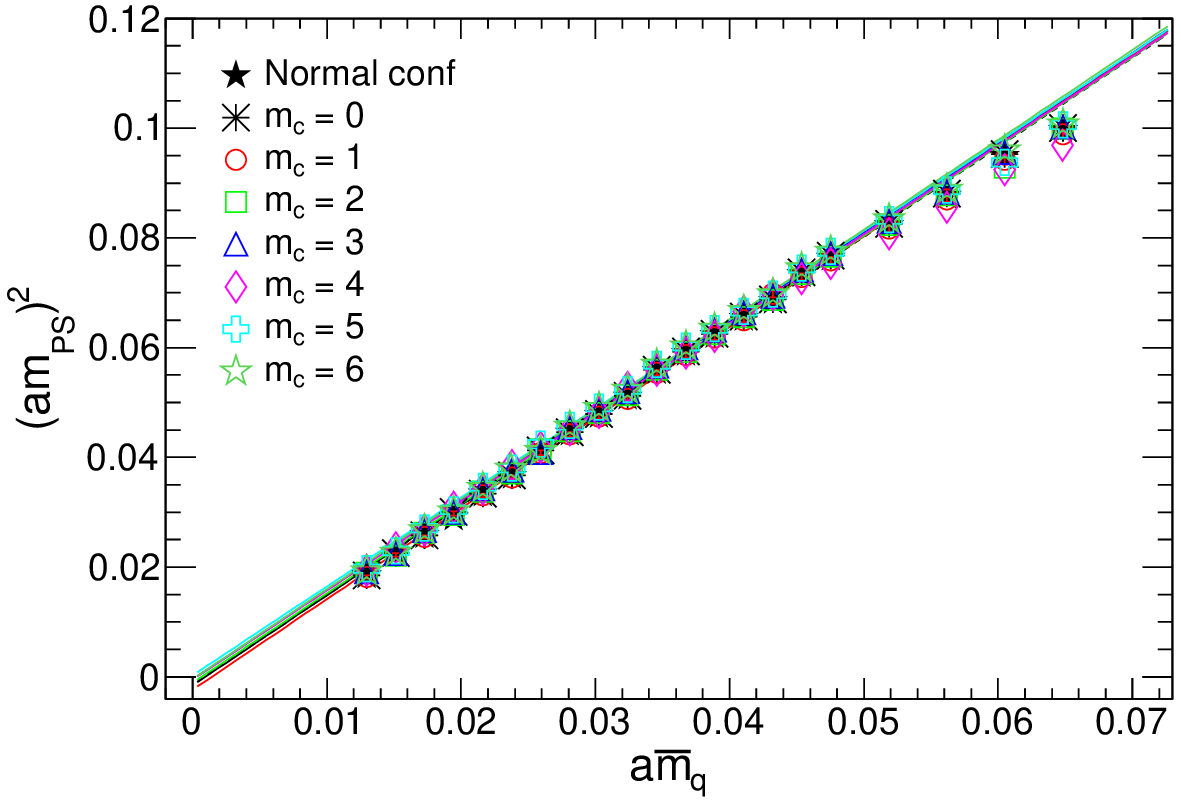}
    \hfill
    \setlength\abovecaptionskip{-3.0pt}
       \caption{The PCAC relation. The colored symbols indicate the numerical results, and the colored lines indicate the fitting results.}\label{fig:pcac_fitting_all}
  \end{minipage}
  \hspace{4mm}
  \begin{minipage}{0.48\hsize}
    \centering
    \vspace{-4 mm}
    \includegraphics[keepaspectratio, width=70mm]{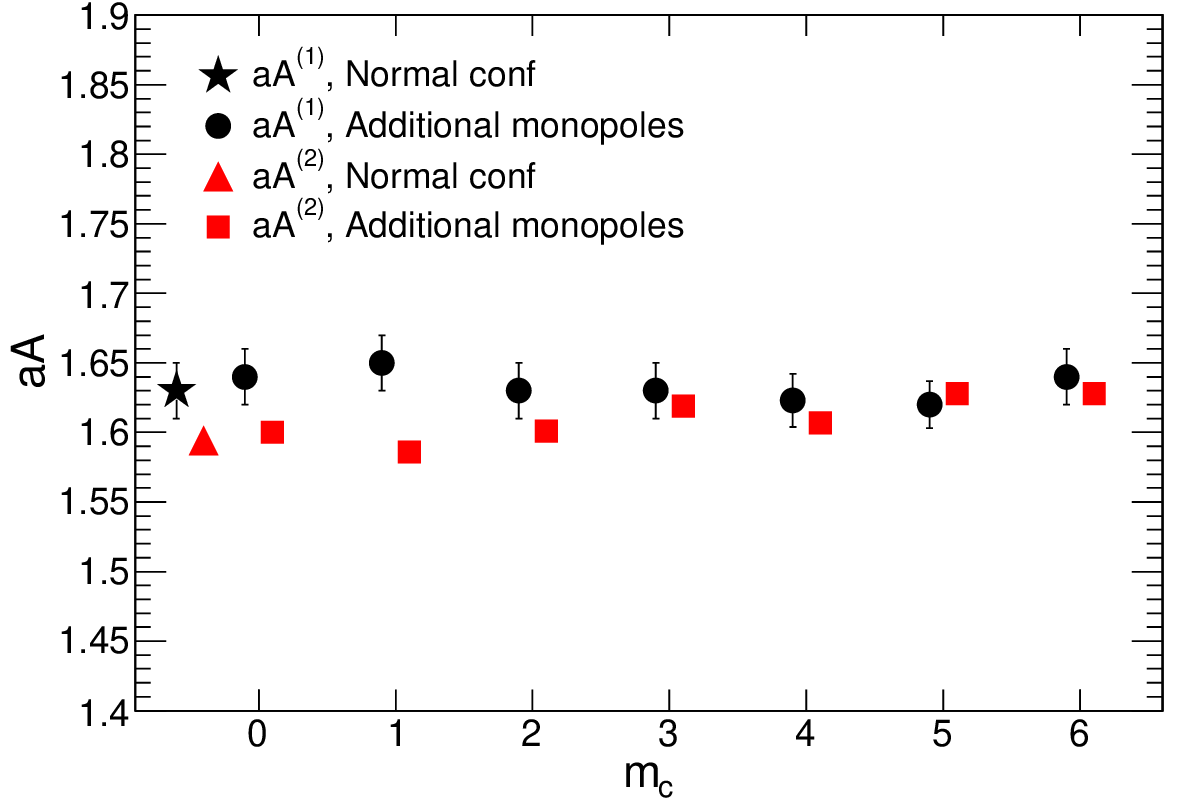}
    \setlength\abovecaptionskip{-3.0pt}
     \caption{Comparisons of the fitting results of the slopes $aA^{(1)}$ and $aA^{(2)}$.}\label{fig:pcac_slope1}
  \end{minipage}
\end{figure}
\begin{table}[tbp]
  \small
  \centering
  \begin{tabular}{|c|c|c|c|c|c|c|c|}\hline
    $m_{c}$ & $aA^{(1)}$ & $a^{2}B$ & $aA^{(2)}$ & $\hat{Z}_{S}$ &$FR(a\bar{m}_{q})$& $\chi^{2}/d. o. f.$ & $\chi^{2}/d. o. f.$  \\
    & & $\times10^{-3}$ &  & &$\times10^{-2}$& $(A^{(1)}, B)$ & $(A^{(2)})$ \\\hline
    {\footnotesize Normal conf}   & 1.63(2) & -1.4(7) & 1.594(4) &  0.93(3) & 2.5 - 4.8 & 9.0/9.0 & 12.7/10.0  \\\hline
    0 & 1.64(2)   & -1.6(8) & 1.600(4) &  0.93(3) & 2.5 - 4.8 & 9.4/9.0 & 13.5/10.0  \\\hline
    1 & 1.65(2)   & -2.4(8) & 1.586(4) &  0.93(3) & 2.5 - 4.6 & 7.9/8.0 & 15.7/9.0   \\\hline
    2 & 1.63(2)   & -1.1(9) & 1.601(4) &  0.93(3) & 2.8 - 4.8 & 8.0/8.0 & 9.5/9.0    \\\hline
    3 & 1.63(2)   & -0.5(9) & 1.619(4) &  0.92(3) & 2.8 - 4.8 & 8.2/8.0 & 8.5/9.0    \\\hline
    4 & 1.623(19) & -0.5(6) & 1.607(4) &  0.92(3) & 2.1 - 4.4 & 9.3/9.0 & 9.9/10.0   \\\hline
    5 & 1.620(17) & -0.3(5) & 1.628(4) &  0.91(3) & 2.5 - 4.6 & 8.0/8.0 & 8.1/9.0    \\\hline
    6 & 1.64(2)   & -0.4(8) & 1.628(4) &  0.91(3) & 2.8 - 4.8 & 8.4/8.0 & 8.7/9.0    \\\hline
  \end{tabular}
  \caption{The fitting results of the slopes $aA^{(1)}$ and $aA^{(2)}$ and the intercept $a^{2}B$. The numerical results of the renormalization constant $\hat{Z}_{S}$.}\label{tb:pcac_1_3}
\end{table}

The chiral perturbation theory predicts that the logarithmic divergence near the chiral limit appears in the correlation between the square of the pseudoscalar mass and the bare quark mass~\cite{Gasser1}. Therefore, we investigate the logarithmic divergence in the range of the bare quark mass 10 [MeV] $\leq \bar{m}_{q} \leq$ 150 [MeV]; however, we have not observed the chiral logarithms. Therefore, we fit a linear function
\begin{equation}
(am_{PS})^{2} = aA^{(1)}a\bar{m}_{q} + a^{2}B\label{eq:pcac_fit_1}
\end{equation}
to the numerical results of the square of the pseudoscalar mass $(am_{PS})^{2}$, as shown in figure~\ref{fig:pcac_fitting_all}. The fitting ranges are determined such that the values of $\chi^{2}/d. o. f.$ are approximately 1. The fitting results of the slope $aA^{(1)}$, the intercept $a^{2}B$, and the values of $\chi^{2}/d. o. f.$ are presented in table~\ref{tb:pcac_1_3}. The fitting results of the intercept $a^{2}B$ are almost zero. To reduce the errors that come from the number of free parameters of the fitting, we fit function
\begin{equation}
  (am_{PS})^{2} = aA^{(2)}a\bar{m}_{q}
\end{equation}
to the numerical results. The fitting results of the slope $aA^{(2)}$ and the values of $\chi^{2}/d. o. f.$ are listed in table~\ref{tb:pcac_1_3}. Figure~\ref{fig:pcac_slope1} shows that the additional monopoles and anti-monopoles do not affect the values of the slopes $A^{(1)}$ and $A^{(2)}$. In the sections below, we calculate the renormalization constant $\hat{Z}_{S}$ for the scalar density and the light quark masses using the fitting results of the slope $A^{(2)}$.

As a consequence of this subsection, the fitting results of the slope and intercept indicate that the additional monopoles and anti-monopoles do not affect the PCAC relation. This result indicates that even if the average masses of the light quarks become heavy by increasing the values of the magnetic charges $m_{c}$ of the additional monopole and anti-monopole, formula~(\ref{eq:pion_decay}) is unaffected because the PCAC relation holds.

\subsection{The renormalization constants $\hat{Z}_{S}$ and $Z_{A}$}\label{sec:zs_za}

First, we determine the renormalization constant $\hat{Z}_{S}$ for the scalar density by the non-perturbative calculations~\cite{Hernandez2, Wennekers1}. There is the relation~\cite{Alexandrou1} between the renormalization constant $Z_{m}$ for the bare quark mass $\bar{m}_{q}$ of the massive overlap Dirac operator~(\ref{eq:ovmass}) and the renormalization constant $\hat{Z}_{S}$ for the bare scalar density as follows:
\begin{equation}
  \hat{Z}_{S} = \frac{1}{Z_{m}}\label{eq:z_s_and_zm}
\end{equation}

We calculate the bare quark mass $\bar{m}_{q}r_{0}$ at the reference mass $(m_{PS}r_{0})_{ref.}^{2}$ = 1.5736~\cite{Hernandez2} of the kaon using the fitting results of the slope $A^{(2)}$ in table~\ref{tb:pcac_1_3}. Here, we convert the scale in the lattice unit $a$ into a physical scale using the Sommer scale $r_{0} = 0.5$ [fm]. We then compute the renormalization constant $\hat{Z}_{S}$ by substituting the computed results of the bare quark mass for the following formula:
\begin{equation}
  \hat{Z}_{S} (g_{0}) = \frac{1}{Z_{m} (g_{0})} = \left.\frac{(\bar{m}_{q}r_{0}) (g_{0})}{U_{M}}\right|_{(m_{PS}r_{0})_{ref.}^{2}}.
\end{equation}
The bare quark mass $\bar{m}_{q}r_{0}$ and the renormalization constants $\hat{Z}_{S}$ and $Z_{m}$ rely on the bare coupling $g_{0}$. The factor $U_{M}$ is the renormalization group-invariant quark mass. We use the result $U_{M} = 0.181 (6)$ from reference~\cite{Hernandez2}. The results of $\hat{Z}_{S}$ are displayed in table~\ref{tb:pcac_1_3}.

To confirm our calculations, we set the value of the parameter $\beta = 6.0000$ for the lattice spacing to be the same as another research group in the literature~\cite{Wennekers1} 
and calculate the renormalization constant $\hat{Z}_{S}$ using the normal configurations. Our result is $\hat{Z}_{S} = 0.95(3)$. This result is approximately 10$\%$ smaller than the result of the other group~\cite{Wennekers1}. We suppose that this is because we remove the excited states of the correlation functions.
\begin{figure}[tbp]
\centering
    \includegraphics[width=83mm]{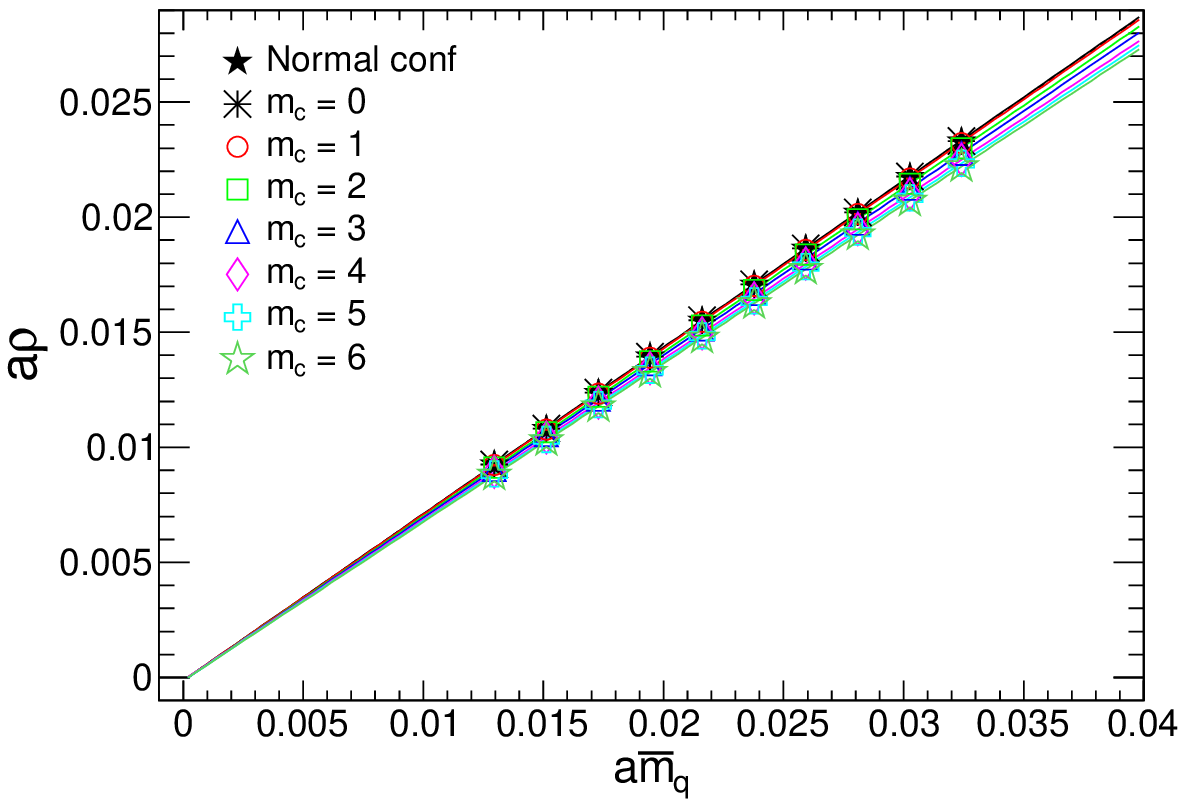}
\setlength\abovecaptionskip{-3.0pt}
\caption{The ratio $a\rho$ of the correlation functions versus the bare quark mass $a\bar{m}_{q}$. The colored lines indicate the fitting results.}\label{fig:arho_amq_1}
\end{figure}
\begin{table}[tbp]
  \small
  \centering
  \begin{tabular}{|c|c|c|c|c|c|} \hline
    $m_{c}$ & $A$ & $aB\times10^{-4}$ & $Z_{A}$ & $FR (a\bar{m}_{q}) \times10^{-2}$ & $\chi^{2}/d. o. f.$  \\\hline
    {\footnotesize Normal conf} & 0.7235(3) & -1.40(5) & 1.3822(5) & 1.2 - 3.1 & 6.6/7.0  \\\hline
    0 & 0.7244(2) & -1.38(5) &  1.3805(5) & 1.2 - 3.1 & 7.6/7.0  \\\hline
    1 & 0.7215(3) & -1.35(5) &  1.3860(5) & 1.2 - 3.1 & 5.6/7.0  \\\hline
    2 & 0.7144(2) & -1.51(5) &  1.3997(5) & 1.2 - 3.1 & 7.4/7.0  \\\hline
    3 & 0.7076(3) & -1.46(5) &  1.4132(5) & 1.2 - 3.1 & 6.2/7.0  \\\hline
    4 & 0.6984(2) & -1.31(5) &  1.4319(5) & 1.2 - 3.1 & 5.9/7.0  \\\hline
    5 & 0.6938(2) & -1.42(5) &  1.4413(5) & 1.2 - 3.1 & 7.9/7.0  \\\hline
    6 & 0.6895(2) & -1.43(5) &  1.4502(5) & 1.2 - 3.1 & 6.8/7.0  \\ \hline
  \end{tabular}
  \caption{The fitting results of slope $A$ and intercept $aB$ obtained by fitting the function $a\rho = Aa\bar{m}_{q} + aB$. The numerical results of the renormalization constants $Z_{A}$.}\label{tb:rho_1}
\end{table}

Next, we calculate the renormalization constant $Z_{A}$ for the axial vector current using the following relation~\cite{Giusti3}:
\begin{equation}
a\rho = \frac{1}{Z_{A}}a\bar{m}_{q}.
\end{equation}
The numerical results of the ratio $a\rho$ of the correlation functions are listed in table~\ref{tb:numerical_results_arho_all} in appendix~\ref{sec:corre_func_fit_res_1}. We fit the linear function
\begin{equation}
a\rho = A a\bar{m}_{q} + aB
\end{equation}
to the numerical results of $a\rho$, as shown in figure~\ref{fig:arho_amq_1}. The fitting ranges are determined such that the values of $\chi^{2}/d. o. f.$ are approximately 1. The fitting results of slope $A$, intercept $aB$, and $\chi^{2}/d. o. f.$ are presented in table~\ref{tb:rho_1}. 

Finally, the renormalization constant $Z_{A}$ is calculated by taking the inverse of the fitting result of slope $A$, and the computed results are shown in table~\ref{tb:rho_1}. The values of the renormalization constant $Z_{A}$ slightly increase with increasing magnetic charge $m_{c}$. We suppose that this results from the effect of the discretization.

We compare our numerical result of $Z_{A}$, which is calculated using the normal configurations ($V = 16^{3}\times32$, $\beta = 6.0000$), with the computed results of other groups. Our result is $Z_{A} = 1.4247(4)$. This finding is approximately 8$\%$ smaller than the results of other groups~\cite{Wennekers1, Giusti3_2}. We assume the same rationale as the computed result of $\hat{Z}_{S}$.

\subsection{The decay constant of the pseudoscalar $F_{PS}$}\label{sec:fps_comp}

The decay constant of the pseudoscalar $F_{PS}$ is defined as follows~\cite{Giusti3}:
\begin{equation}
  aF_{PS} = \frac{2a\bar{m}_{q}\sqrt{a^{4}G_{PS-SS}}}{(am_{PS})^{2}}\label{eq:numerical_fpi_1}
\end{equation}
In this notation, the pion decay constant is $F_{\pi}$ = 93 [MeV]. We calculate the decay constant $aF_{PS}$ with the fitting results of the coefficient $a^{4}G_{PS-SS}$ and pseudoscalar mass $am_{PS}$ at the bare quark mass $a\bar{m}_{q}$. The results of the decay constant $aF_{PS}$, which are calculated using the normal configurations and the configurations with the additional monopoles and anti-monopoles, are shown in tables~\ref{tb:numerical_results_or_mc0},~\ref{tb:numerical_results_mc1_2},~\ref{tb:numerical_results_mc3_4}, and~\ref{tb:numerical_results_mc5_6} in appendix~\ref{sec:corre_func_fit_res_1}.
\begin{figure}[tbp]
  \begin{minipage}{0.48\hsize}
    \centering
    \includegraphics[keepaspectratio, width=70mm]{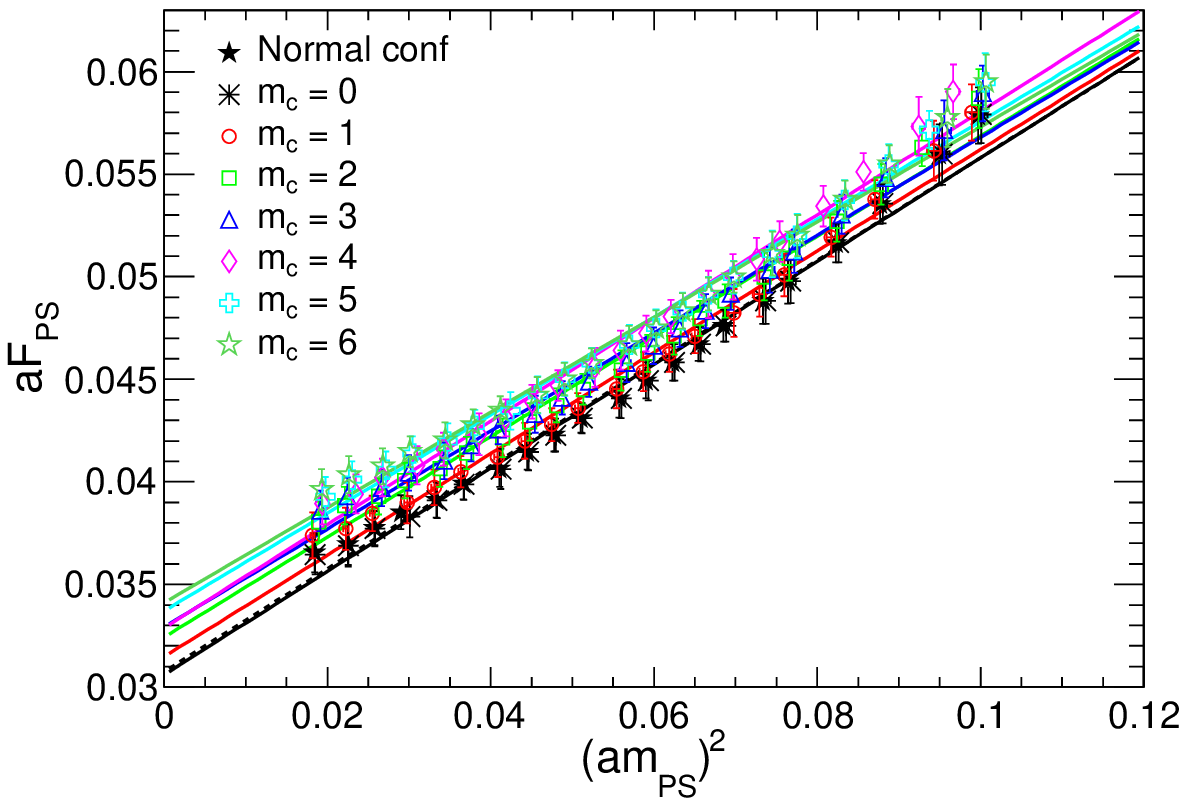}
    \hfill
    \setlength\abovecaptionskip{-3.0pt}
    \caption{The decay constant $aF_{PS}$ versus the square of the pseudoscalar mass $(am_{PS})^{2}$. The colored symbols and lines represent the numerical results and fitting results, respectively. The black dotted line indicates the fitting result of the normal configuration.}\label{fig:afpi_ampi2_chiPt_fit_1}
  \end{minipage}
  \hspace{3mm}
  \begin{minipage}{0.48\hsize}
   \vspace{-2mm}
    \centering
    \includegraphics[keepaspectratio, width=71mm]{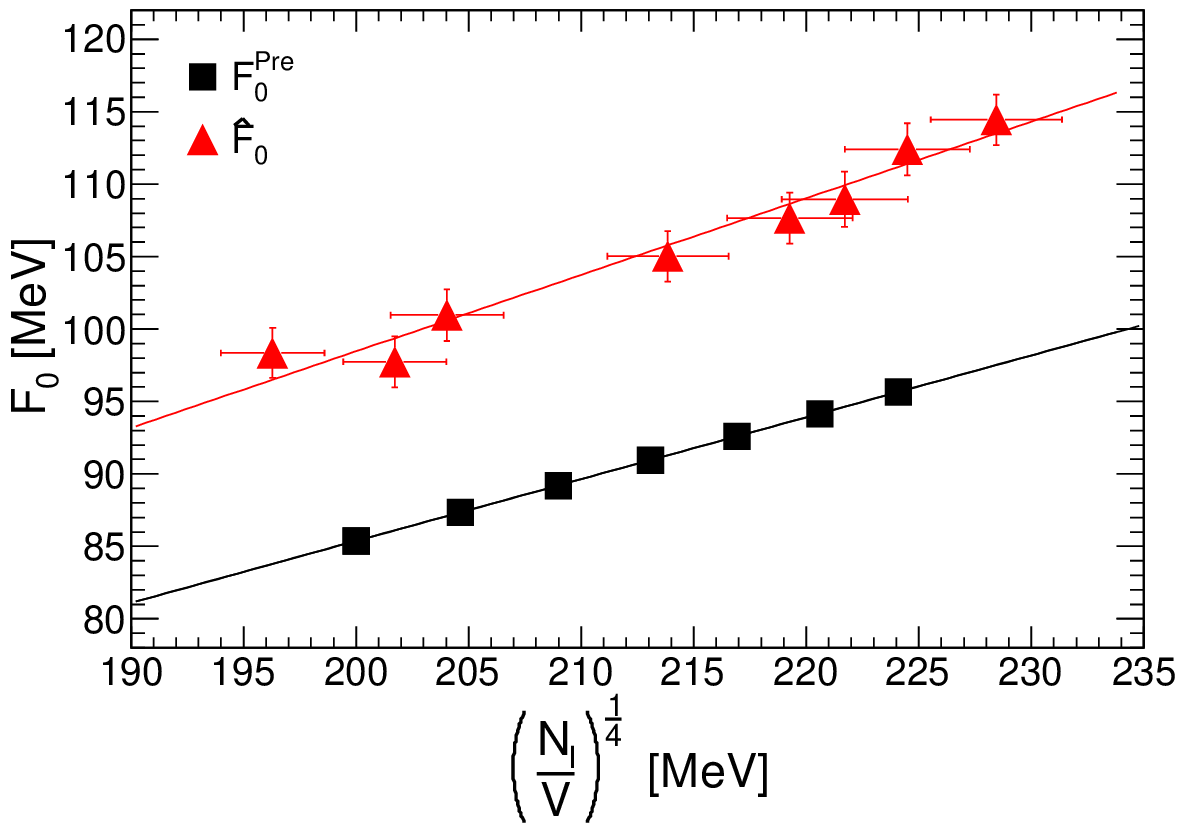}
    \setlength\abovecaptionskip{-3.0pt}
     \caption{Comparisons of the numerical results of $\hat{F}_{0}$ with the prediction $F_{0}^{Pre}$. The solid black and red lines indicate the fitting results of the prediction and the numerical result, respectively.}\label{fig:af0_niv}
  \end{minipage}
\end{figure}
\begin{table}[tbp]
  \small
  \centering
  \begin{tabular}{|c|c|c|c|c|c|c|} \hline
    $m_{c}$  & $aF_{0}$ & $L_{5}^{q}$  & $\hat{F}_{0}$ [MeV] & $\hat{F}_{\pi}$ [MeV] & $FR [(am_{PS})^{2}]$ & $\chi^{2}/d. o. f.$ \\ 
    & $\times10^{-2}$  & $\times10^{-3}$ &  & & $\times10^{-2}$ & \\ \hline
    {\footnotesize Normal conf} & 3.08(5) & 1.93(4)  & 98.4(1.7)  & 101.3(1.7) & 1.8 - 10.0 & 9.4/19.0 \\ \hline
    0 & 3.06(6) & 1.93(4)  & 97.7(1.8)  & 100.7(1.7) & 1.8 - 10.0 & 8.7/19.0 \\\hline
    1 & 3.15(6) & 1.95(5)  & 101.0(1.8) & 103.8(1.7) & 1.8 - 10.0 & 9.5/19.0 \\\hline
    2 & 3.24(5) & 1.98(5)  & 105.0(1.7) & 107.9(1.7) & 1.8 - 10.0 & 9.7/19.0 \\\hline
    3 & 3.29(5) & 1.97(5)  & 107.7(1.8) & 110.5(1.7) & 1.9 - 10.1 & 9.7/19.0 \\\hline
    4 & 3.29(6) & 2.07(5)  & 109.0(1.9) & 112.0(1.9) & 1.9 - 9.7 & 7.6/19.0  \\\hline
    5 & 3.37(5) & 2.01(5)  & 112.4(1.8) & 115.3(1.7) & 1.9 - 10.1 & 8.4/19.0 \\\hline
    6 & 3.41(5) & 1.98(5)  & 114.4(1.7) & 117.3(1.7) & 1.9 - 10.1 & 9.9/19.0 \\\hline           
  \end{tabular}
  \caption{The fitting results of $aF_{0}$ and $L_{5}^{q}$. The computed results of the renormalized decay constants $\hat{F}_{0}$ and $\hat{F}_{\pi}$.}\label{tb:afpi_ampi2_chipt1}
\end{table}

Figure~\ref{fig:afpi_ampi2_chiPt_fit_1} shows the correlation between the decay constant $aF_{PS}$ of the pseudoscalar and the square of the pseudoscalar mass $(am_{PS})^{2}$. This demonstrates that the logarithmic divergence does not appear near the chiral limit and that the decay constant $aF_{PS}$ linearly increases with increasing square mass $(am_{PS})^{2}$. These behaviors correspond to the features that are analogized from the SU(2) Lagrangian in the quenched chiral perturbation theory~\cite{Colangelo1}. 
In the studies of the overlap Dirac operator in quenched QCD, these features have already been mentioned by other scholars~\cite{Giusti2, Giusti1}. 
We fit the following formula derived from the quenched chiral perturbation theory~\cite{Colangelo1} to the numerical results:
\begin{equation}
aF_{PS} = aF_{0} \left[1 + 4L_{5}^{q}\frac{(am_{PS})^{2}}{(aF_{0})^{2}}\right]\label{eq:chipt_f0_fit}.
\end{equation}
 The factor $L_{5}^{q}$ is similar to a low-energy constant in the quenched chiral perturbation theory~\cite{Colangelo1}. We suppose that the PCAC relation holds. The decay constant $F_{PS}$ in the chiral limit $ \bar{m}_{q} \rightarrow 0$ corresponds to $F_{0}$.

The results of $aF_{0}$ and $L_{5}^{q}$ obtained by fitting formula~(\ref{eq:chipt_f0_fit}) are listed in table~\ref{tb:afpi_ampi2_chipt1}. The fitting results of $L_{5}^{q}$ are approximately 2.5 times larger than the result of another group~\cite{Heitger1}. This has been explained in the study of~\cite{Giusti2}. The fitting results demonstrate that the intercept $aF_{0}$ increases with increasing magnetic charge $m_{c}$; however, the slope $L_{5}^{q}$ does not vary.

To quantitatively demonstrate the reason for increasing the decay constants with increasing magnetic charge $m_{c}$, we calculate the renormalized decay constants $\hat{F}_{0}$ and $\hat{F}_{\pi}$. The renormalized decay constant of the pseudoscalar is $\hat{F}_{PS}  = Z_{A}F_{PS}$. The renormalization constants $Z_{A}$ are shown in table~\ref{tb:rho_1}.

The renormalized decay constant $\hat{F}_{0}$ of the normal configurations is
\begin{equation}
 \hat{F}_{0} =  98.4(1.7) \ \  [\mbox{MeV}].\label{eq:f0_num_res_1}
\end{equation}
The renormalized decay constants $\hat{F}$ are calculated in the $\epsilon$-regime and the $p$-regime by the other groups~\cite{Giusti1, Giusti2}. Our result is slightly smaller than the results of the other groups because the renormalization constant $Z_{A}$ is smaller than that of other groups, as mentioned in subsection~\ref{sec:zs_za}. The computed results of the renormalized decay constants $\hat{F}_{0}$ are listed in table~\ref{tb:afpi_ampi2_chipt1}.

Now, we compare the prediction $F_{0}^{Pre}$ with the numerical results of the renormalized decay constant $\hat{F}_{0}$, as shown in figure~\ref{fig:af0_niv}. The results of the prediction $F_{0}^{Pre}$ are given in table~\ref{tb:instantons_chiral_fpi}. To quantitatively compare the renormalized decay constant of the numerical result with the prediction~(\ref{eq:pion_decay}), we fit the following function:
\begin{align}
  F_{PS} = A_{F}\left(\frac{N_{I}}{V}\right)^{\frac{1}{4}} + B.\label{eq:fit_func1}
\end{align}
The fitting results of $\hat{F}_{0}$ are $A_{F}  = 0.53 (7)$, $B = -7 (15)$ [MeV], and $\chi^{2}/d. o. f. = 2.2/6.0$. The intercept $B$ is zero. The slope of the prediction $F_{0}^{Pre}$ is $A_{F}^{Pre}$  = 0.4268.

These results clearly show that the decay constant $\hat{F}_{0}$ increases in direct proportion to the one-fourth root of the number density of the instantons and anti-instantons. The slope of the numerical calculations is consistent with the slope of the prediction~(\ref{eq:pion_decay}). However, the error of the slope $A_{F}$ obtained by fitting is more than 13$\%$. Moreover, the numerical result~(\ref{eq:f0_num_res_1}) is larger than the result of the chiral perturbation theory~(\ref{eq:chi_pt_f0}) and the prediction~(\ref{eq:f0_pred}). Accordingly, we improve the computations in the next section.

Next, we substitute the fitting results of $aF_{0}$, $L_{5}^{q}$ and the experimental result of the pion mass~(\ref{eq:exp_mpi}) for formula~(\ref{eq:chipt_f0_fit}). We estimate the renormalized pion decay constant $\hat{F}_{\pi}$ at the physical pion mass. The renormalized pion decay constant $\hat{F}_{\pi}$ that is calculated using the normal configurations is $\hat{F}_{\pi} =  101.3(1.7) \ [\mbox{MeV}]$. This result is consistent with the result of the phenomenological model~\cite{Dyakonov2} $F_{\pi} =$ 98.82 [MeV], which is computed with the values~(\ref{eq:inst_dens_1}) and~(\ref{eq:ins_size}); however, this value is approximately 10$\%$ larger than the experimental result~\cite{PDG_2017}. We list the computed results of the renormalized decay constants $\hat{F}_{\pi}$ in table~\ref{tb:afpi_ampi2_chipt1}. We fit the function~(\ref{eq:fit_func1}) to the numerical results of $\hat{F}_{\pi}$. The fitting results are $A_{F}  = 0.53 (7)$, $B = -4 (15)$ [MeV], and $\chi^{2}/d. o. f. = 2.2/6.0$.

These numerical results suggest that the renormalized decay constants $\hat{F}_{0}$ and $\hat{F}_{\pi}$ increase in direct proportion to the one-fourth root of the number density of the instantons and anti-instantons.

\subsection{The chiral condensate}\label{sec:chi_cond_1}

The chiral condensate is derived from the GMOR relation~(\ref{eq:gmor_res_1}) and formula~(\ref{eq:numerical_fpi_1}) as follows:
\begin{align}
  a^{3}\langle \bar{\psi}\psi \rangle^{GMOR} = - \lim_{(a\bar{m}_{PS})^{2} \rightarrow 0} \frac{2 a\bar{m}_{q}a^{4}G_{PS-SS}}{(am_{PS})^{2}}\label{eq:gmor_2}
\end{align}
We substitute the fitting results of $a^{4}G_{PS-SS}$ and $am_{PS}$ at the bare quark mass $a\bar{m}_{q}$ for the expression~(\ref{eq:gmor_2}) and calculate the chiral condensate $a^{3}\langle \bar{\psi}\psi \rangle^{GMOR}$. We list the computed results in tables~\ref{tb:numerical_results_or_mc0},~\ref{tb:numerical_results_mc1_2},~\ref{tb:numerical_results_mc3_4}, and~\ref{tb:numerical_results_mc5_6} in appendix~\ref{sec:corre_func_fit_res_1}.
\begin{figure}[tbp]
  \begin{minipage}{0.48\hsize}
    \centering
    \includegraphics[keepaspectratio, width=70mm]{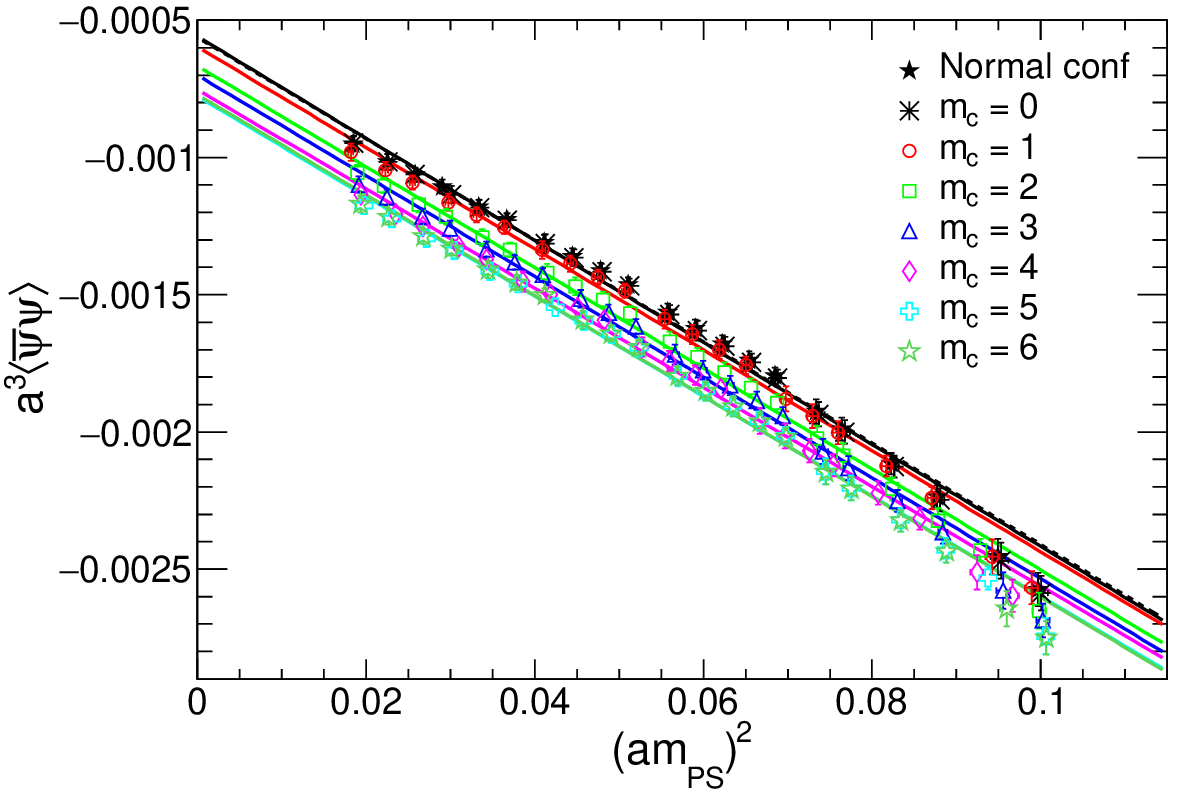}
    \hfill
    \setlength\abovecaptionskip{-3.0pt}
    \caption{The chiral condensate $a^{3}\langle \bar{\psi}\psi \rangle$ versus the square of the pseudoscalar mass $(a\bar{m}_{PS})^{2}$. The colored symbols and lines represent the numerical results and the fitting results, respectively. The dotted line indicates the fitting results of the normal configuration.}\label{fig:chi_condens_amq_1}
  \end{minipage}
  \hspace{3mm}
  \begin{minipage}{0.48\hsize}
    \centering
    \includegraphics[keepaspectratio, width=71mm]{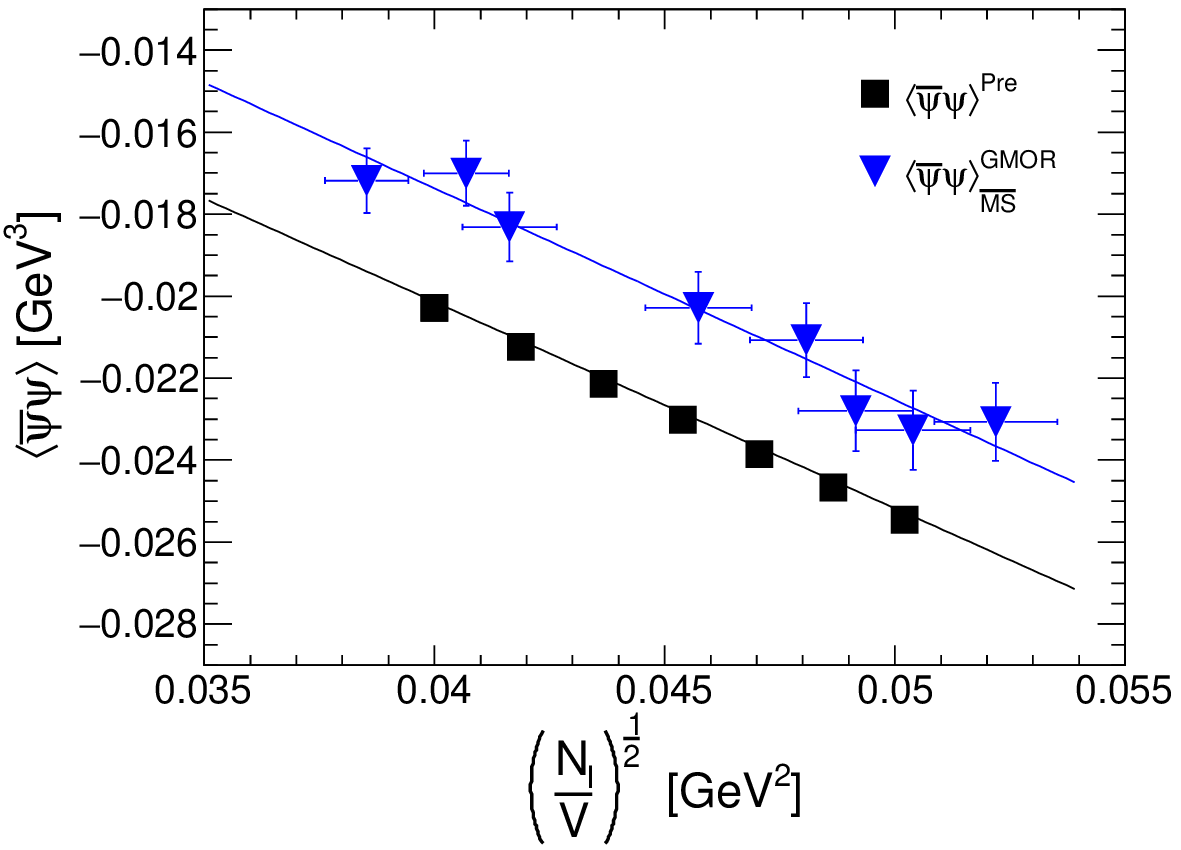}
    \setlength\abovecaptionskip{-3.0pt}
     \caption{Comparisons of the renormalized chiral condensate $\langle \bar{\psi}\psi \rangle_{\overline{MS}}^{GMOR}$ with the prediction $\langle \bar{\psi}\psi \rangle^{Pre}$. The solid blue and black lines indicate the fitting results of the numerical results and prediction, respectively.}\label{fig:chi_niv_1}
  \end{minipage}
\end{figure}

Figure~\ref{fig:chi_condens_amq_1} shows that there are no logarithmic divergences near the chiral limit and that the values of the chiral condensate $a^{3}\langle \bar{\psi}\psi \rangle^{GMOR}$ linearly decrease with the increasing square of the pseudoscalar mass $(am_{PS})^{2}$. Therefore, we interpolate the values of the chiral condensate in the chiral limit $(am_{PS})^{2} \rightarrow 0$ by fitting the linear function
\begin{equation}
a^{3}\langle \bar{\psi}\psi \rangle = aA(am_{PS})^{2} + a^{3}B\label{eq:cond_fit}
\end{equation}
to the computed results. The fitting results of the slope $aA$, intercept $a^{3}B$, and values of $\chi^{2}/d. o. f.$ are given in table~\ref{tb:chi_new_1}. All data points are included in the fitting ranges, and the values of $\chi^{2}/d. o. f.$ range from 0.6 to 1.5; accordingly, we can properly fit the linear function to the computed results. Table~\ref{tb:chi_new_1} indicates that if we increase the magnetic charge $m_{c}$, the values of the intercept $a^{3}B$ decrease, whereas the values of the slope $aA$ do not vary.
\begin{table}[tbp]
  \small
  \centering
  \begin{tabular}{|c|c|c|c|c|c|}\hline
    $m_{c}$ & $aA$ & $a^{3}B$ & {\footnotesize$\langle\bar{\psi}\psi\rangle_{\overline{MS}}^{GMOR}$} & $FR [(am_{PS})^{2}]$ & $\chi^{2}/d. o. f.$ \\
    & &  & {\footnotesize[GeV$^{3}$]} & $\times10^{-2}$ &  \\ \hline
    {\footnotesize Normal conf}& -1.85(3)$\times10^{-2}$  & -5.62(18)$\times10^{-4}$ &  -1.72(8)$\times10^{-2}$ &  1.8 - 10.0 & 29.0/19.0 \\\hline
    0 & -1.86(4)$\times10^{-2}$  & -5.59(18)$\times10^{-4}$ &  -1.70(8)$\times10^{-2}$ & 1.8 - 11.0 & 28.0/19.0 \\\hline
    1 & -1.84(4)$\times10^{-2}$  & -5.97(19)$\times10^{-4}$ &  -1.83(8)$\times10^{-2}$ & 1.8 - 9.9 & 24.9/19.0 \\\hline
    2 & -1.84(4)$\times10^{-2}$  & -6.67(19)$\times10^{-4}$ &  -2.03(9)$\times10^{-2}$ & 1.8 - 10.0 & 19.9/19.0 \\\hline
    3 & -1.83(4)$\times10^{-2}$  & -7.00(19)$\times10^{-4}$ &  -2.11(9)$\times10^{-2}$ & 1.9 - 11.0 & 22.2/19.0 \\\hline
    4 & -1.81(4)$\times10^{-2}$  & -7.5(2)$\times10^{-4}$    & -2.28(10)$\times10^{-2}$ & 1.9 - 9.7 & 10.7/19.0 \\\hline
    5 & -1.82(4)$\times10^{-2}$  & -7.8(2)$\times10^{-4}$    & -2.33(10)$\times10^{-2}$ & 1.9 - 11.0 & 15.2/19.0 \\\hline
    6 & -1.83(4)$\times10^{-2}$  & -7.71(19)$\times10^{-4}$& -2.31(10)$\times10^{-2}$ & 1.9 - 11.0 & 20.1/19.0 \\\hline
  \end{tabular}
  \caption{The results of the slope $aA$ and intercept $a^{3}B$ obtained by fitting the function~(\ref{eq:cond_fit}). The results of the renormalized chiral condensate $\langle\bar{\psi}\psi\rangle_{\overline{MS}}^{GMOR}$ into $\overline{MS}$-scheme at 2 [GeV].}\label{tb:chi_new_1}
\end{table}

We define the renormalized chiral condensate into the $\overline{MS}$-scheme at 2 [GeV] as follows:
\begin{equation}
  \langle \bar{\psi}\psi \rangle_{\overline{MS}}^{GMOR} \ \ \ (2 \ \ [\mbox{GeV}]) \equiv \frac{Z_{S} Z_{A}^{2}}{0.72076}\langle \bar{\psi}\psi \rangle^{GMOR}\label{eq:chiral_def1}
\end{equation}
We use the value $\bar{m}_{\overline{MS}}(\mu)/M = 0.72076$ ($\mu$ = 2 [GeV]) in reference~\cite{ALPHA1}, the computed results of the renormalization constant $\hat{Z}_{S}$ in table~\ref{tb:pcac_1_3}, and the renormalization constant $Z_{A} = 1.3822(5)$ of the normal configuration. We list the computed results of the renormalized chiral condensate $\langle\bar{\psi}\psi\rangle_{\overline{MS}}^{GMOR}$ into $\overline{MS}$-scheme at 2 [GeV] in table~\ref{tb:chi_new_1}.

The numerical result of the renormalized chiral condensate in the $\overline{MS}$-scheme at 2 [GeV] that is computed using the normal configurations is
\begin{align}
  \langle \bar{\psi}\psi \rangle_{\overline{MS}}^{GMOR} \ \ \ ( 2 \ \ [\mbox{GeV}] ) & = -1.72 (7) \times10^{-2} \ \ [\mbox{GeV}^{3}] = - ( 258 (4) \ \ [\mbox{MeV}] )^{3}.
\end{align}
This result is reasonably consistent with the results of the phenomenological models~(\ref{eq:chiral_ins1}), the computed result that uses the experimental results~(\ref{eq:gmor_res_1}), and the result of the numerical computations by another group~\cite{Giusti3}. Therefore, we can correctly compute the renormalized chiral condensate.

To quantitatively compare prediction~(\ref{eq:chiral_ins_prediction1}) with the numerical results, we fit the following curve to the computed results of $\langle \bar{\psi}\psi \rangle_{\overline{MS}}^{GMOR}$, as shown in figure~\ref{fig:chi_niv_1}:
\begin{equation}
\langle \bar{\psi}\psi \rangle  = -A_{\chi}\left(\frac{N_{I}}{V}\right)^{\frac{1}{2}} + B.
\end{equation}
The results of the prediction $\langle \bar{\psi}\psi \rangle^{Pre}$ are in table~\ref{tb:instantons_chiral_fpi}. The fitting results are $A_{\chi} = 0.52 (8)$ [GeV], $B = 3 (4)\times10^{-3}$ [GeV$^{3}$], and $\chi^{2}/d. o. f. = 2.0/6.0$. The value of the intercept $B$ is zero. The slope of the prediction~(\ref{eq:chiral_ins_prediction1}) is $A_{\chi}^{Pre} = 0.5070$ [GeV]. The slope $A_{\chi}$ of the numerical result corresponds to the slope $A_{\chi}^{Pre}$ of the prediction. Therefore, the value of the chiral condensate decreases in direct proportion to the square root of the number density of the instantons and anti-instantons. The proportionality constant of the numerical result is consistent with the result of the phenomenological model. The error of the slope $A_{\chi}$, however, is more than 15$\%$. Therefore, we improve the computational method in the next section.


\section{Instanton effects}

We have demonstrated that the decay constant increases and that the chiral condensate decreases when increasing the number density of the instantons and anti-instantons. In this section, we show the instanton effects on the observables. We first determine the normalization factors by matching the numerical results with the experimental results of the pion and kaon. We then re-evaluate the decay constants and the chiral condensate using the normalization factors. Suppose that the light quark masses become heavy by increasing the number density of the instantons and anti-instantons. We evaluate the instanton effects on the masses of the light quarks and mesons and the decay constants of the mesons. Finally, we estimate the catalytic effect on the charged pion.

\subsection{The normalization factors}\label{sssect:normalization_1}

When determining the scale of the lattice~\cite{Alton1, Giusti3} by matching the experimental results with the numerical results, we suppose that it is possible that the final results in physical units are overestimated or underestimated by multiplying or dividing by the surplus factor together with the lattice spacing. Therefore, we improve the calculation method in references~\cite{Alton1, Giusti3}. We set the scale of the lattice so that it is analytically calculated ($a = 8.5274\times10^{-2}$ [fm]). We match the numerical results of the decay constant $aF_{PS}$ and the square of the mass $(am_{PS})^{2}$ with the experimental results of the pion and kaon and determine the normalization factors.
\begin{figure}[tbp]
\centering
    \includegraphics[width=83mm]{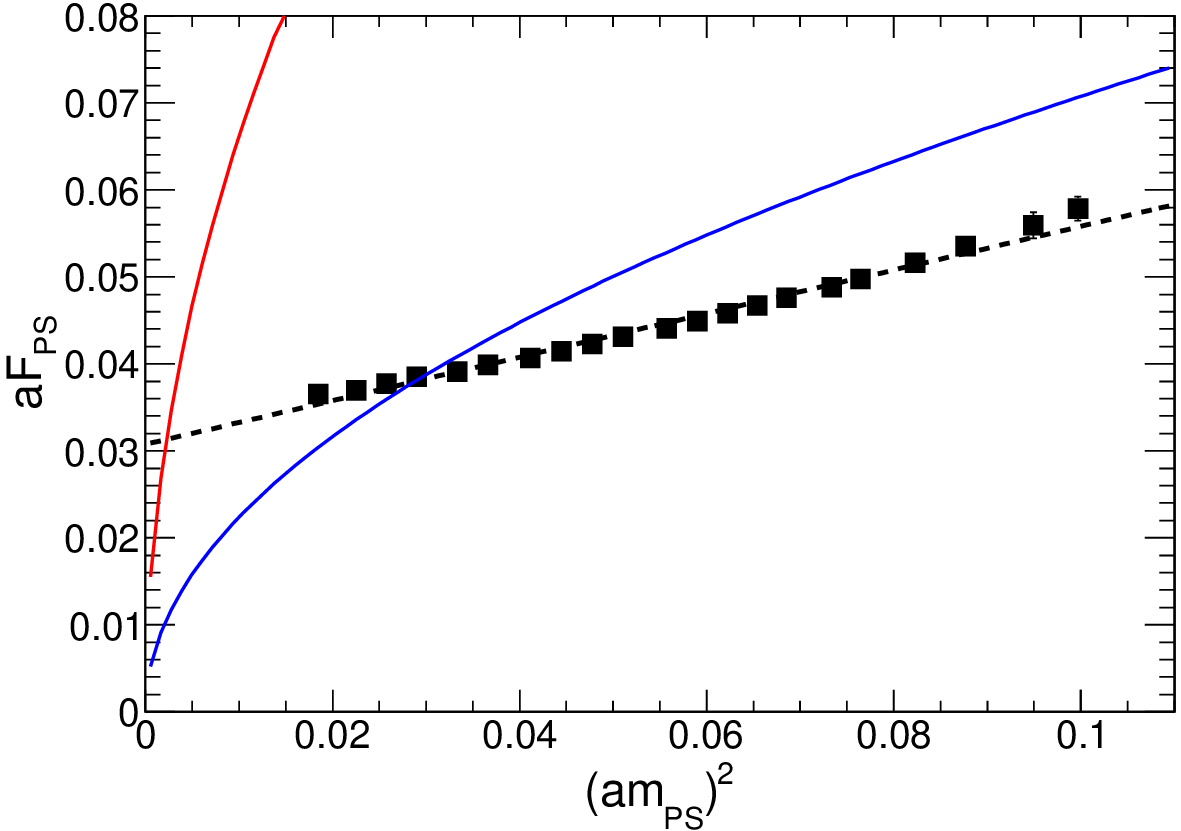}
\setlength\abovecaptionskip{-3.0pt}
\caption{The decay constant $aF_{PS}$ versus the square of the mass $(am_{PS})^{2}$. The black symbols and dotted line are the numerical results and fitting results of the normal configuration, respectively. The solid red and blue curves indicate equations~(\ref{eq:fpi_mpi_exp1}) and~(\ref{eq:fk_mk_exp1}), respectively.}\label{fig:afpi_ampi2_1}
\end{figure}

First, we fit the linear function
\begin{equation}
  aF_{PS} = a^{-1}A(am_{PS})^{2} + aB\label{eq:fit_func_fpi_mps2},
\end{equation}
which is defined without using chiral perturbation theory, to the data points on the planes of $aF_{PS}$ and $(am_{PS})^{2}$, as shown in figure~\ref{fig:afpi_ampi2_1}. The normal configurations are used. The fitting results are $a^{-1}A = 0.251(10), \ aB = 3.08(5)\times10^{-2}$, and $\ \chi^{2}/d. o. f. = 9.4/19.0$ (table~\ref{tb:afpi_ampi2_1}). All data points are included in the fitting range.

We make two equations concerning the pion and kaon using the experimental results~\cite{PDG_2017} as follows:
\begin{align}
  &  aF_{PS} = C_{\pi}^{Exp.}am_{PS}, \ \ C_{\pi}^{Exp.} = \frac{F_{\pi^{-}}^{Exp.}}{\sqrt{2}m_{\pi^{\pm}}^{Exp.}} = \frac{92.277}{139.57061}\label{eq:fpi_mpi_exp1}\\
  & aF_{PS} = C_{K}^{Exp.}am_{PS}, \ \ C_{K}^{Exp.} = \frac{F_{K^{-}}^{Exp.}}{\sqrt{2}m_{K^{\pm}}^{Exp.}} = \frac{110.11}{493.677}\label{eq:fk_mk_exp1}
\end{align}
We do not consider the errors of the experimental results. We plot these curves in figure~\ref{fig:afpi_ampi2_1}. We then analytically compute the intersections between the linear function obtained by fitting equations~(\ref{eq:fpi_mpi_exp1}) and~(\ref{eq:fk_mk_exp1}). The computed results of the intersections for the pion are ($aF_{PS}^{\pi}$, $am_{PS}^{\pi}$) = (3.13(6), 4.74(8)) and for the kaon are ($aF_{PS}^{K}$, $am_{PS}^{K}$) = (3.80(10), 0.171(4)) (table~\ref{tb:int_sec_1}). The normalization factors $Z_{\pi}$ for the pion and $Z_{K}$ for the kaon of the normal configuration are estimated using these intersections as follows:
\begin{align}
  Z_{\pi} & = \frac{F_{\pi^{-}}^{Exp.}}{\sqrt{2}F_{PS}^{\pi}} = \frac{m_{\pi^{\pm}}^{Exp.}}{m_{PS}^{\pi}} = 1.27(2)\label{eq:zpi}\\
  Z_{K} & =  \frac{F_{K^{-}}^{Exp.}}{\sqrt{2}F_{PS}^{K}} = \frac{m_{K^{\pm}}^{Exp.}}{m_{PS}^{K}} = 1.25(3)\label{eq:zpk}
\end{align}
The scale is the Sommer scale $r_{0} = 0.5$ [fm]. These normalization factors are consistent within the errors.

We suppose that the normalization factors do not vary even if we vary the values of the magnetic charge because we numerically confirm that the renormalization constants do not vary. Therefore, we apply the normalization factors of the normal configuration to the results calculated using the configurations with the additional monopoles and anti-monopoles.

\subsection{The instanton effects on the decay constant $F_{0}$}

We use the results of $aF_{0}$ in table~\ref{tb:afpi_ampi2_chipt1} obtained by fitting the function of chiral perturbation theory and re-evaluate the decay constant in the chiral limit using the normalization factor $Z_{\pi}$ as follows:
\begin{equation}
  F_{0}^{Z} = Z_{\pi}F_{0}.
\end{equation}
The result of the normal configuration is $F_{0}^{Z} = 91(2)$. This value is 7$\%$ larger than our predicted value~(\ref{eq:f0_pred}). We list the computed results of $F_{0}^{Z}$ using the normal configurations and the configurations with the additional monopoles and anti-monopoles in table~\ref{tb:mpi_mk_fpi_fk}.

In the analysis of the decay constant $\hat{F}_{0}$ and $\hat{F}_{\pi}$ in subsection~\ref{sec:fps_comp}, we find that the decay constant increases in direct proportion to the one-fourth root of the instanton density. Therefore, we fit the following curve to the numerical result of the decay constant $F_{0}^{Z}$, as shown in figure~\ref{fig:af0_ampi_res_1}: $F_{0} = A_{F}\left(\frac{N_{I}}{V}\right)^{\frac{1}{4}}\label{eq:fit_func2}.$ The fitting result is $A_{F} = 0.446(4)$, which is reasonably consistent with the slope $A_{F}^{Pre}$ = 0.4268 of prediction~(\ref{eq:pion_decay}). The value of $\chi^{2}/d.o.f.$ is 3.0/7.0. These results indicate that the decay constant increases in direct proportion to the one-fourth root of the number density of the instantons and anti-instantons. The increase is consistent with the prediction.

\subsection{The instanton effects on the chiral condensate}

Next, we redefine the chiral condensate derived using the slope $aA$ of the PCAC relation and the decay constant $F_{0}^{Z}$ as follows:
\begin{align}
  a^{3}\langle \bar{\psi}\psi \rangle^{Z} & = - \lim_{a\bar{m}_{q} \rightarrow 0} \frac{(Z_{\pi}am_{PS})^{2}(Z_{\pi}aF_{PS})^{2}}{2 a\bar{m}_{q}^{Z}} = - \frac{aA}{2}{(aF_{0}^{Z})^{2}}\label{eq:gmor_new2}
\end{align}
Here, we suppose the PCAC relation, and we use the following equation:
\begin{equation}
a\bar{m}_{q}^{Z} = \frac{(Z_{\pi}am_{PS})^{2}}{aA} = Z_{\pi}^{2}a\bar{m}_{q}.
\end{equation}
We calculate the chiral condensate $a^{3}\langle \bar{\psi}\psi \rangle^{Z}$ by substituting the fitting results of the slope $aA^{(2)}$ in table~\ref{tb:pcac_1_3}, the results of the decay constant $aF_{0}$ in table~\ref{tb:afpi_ampi2_chipt1}, and the normalization factor~(\ref{eq:zpi}) for formula~(\ref{eq:gmor_new2}). The renormalized chiral condensates in the $\overline{MS}$-scheme at 2 [GeV] are evaluated as follows:
\begin{equation}
  \langle \bar{\psi}\psi \rangle_{\overline{MS}}^{Z} = \frac{Z_{S}}{0.72076}\langle \bar{\psi}\psi \rangle^{Z}
\end{equation}
We use the renormalization constant for the scalar density $Z_{S} = 0.93(3)$ of the normal configuration. The renormalized chiral condensate $\langle \bar{\psi}\psi \rangle_{\overline{MS}}^{Z}$ in the $\overline{MS}$-scheme at 2 [GeV] that is calculated using the normal configurations is
\begin{equation}
 \langle \bar{\psi}\psi \rangle_{\overline{MS}}^{Z}  \  (2 \ [\mbox{GeV}]) = -1.96(12)\times10^{-2} \ [\mbox{GeV}^{3}] = -(269 (5) \ [\mbox{MeV}])^{3}\nonumber.
\end{equation}

Incidentally, we need to confirm the discretization effects on the results computed by formula~(\ref{eq:gmor_new2}) because we separate the lattice spacing and normalization factor and evaluate the chiral condensate. To analyze the effects of the discretization, we generate the configurations by setting the physical volume to $V_{phys}$ = 9.8582 [fm$^{4}$] ($V = 16^{3}\times32$, $\beta = 6.0000$) and varying the lattice spacing and lattice volume. We estimate the chiral condensate in the continuum limit by interpolation. The result in the continuum limit of the renormalized chiral condensate in the $\overline{MS}$-scheme at 2 [GeV] is
\begin{equation}
  \langle \bar{\psi}\psi \rangle_{\overline{MS}}^{Z} \ (2 \ [\mbox{GeV}]) = -1.95(5)\times10^{-2} \ [\mbox{GeV}^{3}] = -(269(2) \ [\mbox{MeV}])^{3}\nonumber.
\end{equation}
This result shows that there are no effects of discretization. We will report this result in the future~\cite{Hasegawa5}.

These results correspond to the findings of the analytic computation~(\ref{eq:gmor_res_1}) and the prediction of the normal configuration in table~\ref{tb:instantons_chiral_fpi}. Moreover, these results are consistent with the outcomes of other groups~\cite{Giusti3} and~\cite{Wennekers1}, which are calculated with the overlap Dirac operator, and the findings of the studies that use the $N_{f} = 2$ and $N_{f} = 2 + 1$ dynamical fermions~\cite{Aoki1}.
\begin{figure}[tbp]
  \begin{minipage}{0.48\hsize}
    \centering
    \includegraphics[keepaspectratio, width=71mm]{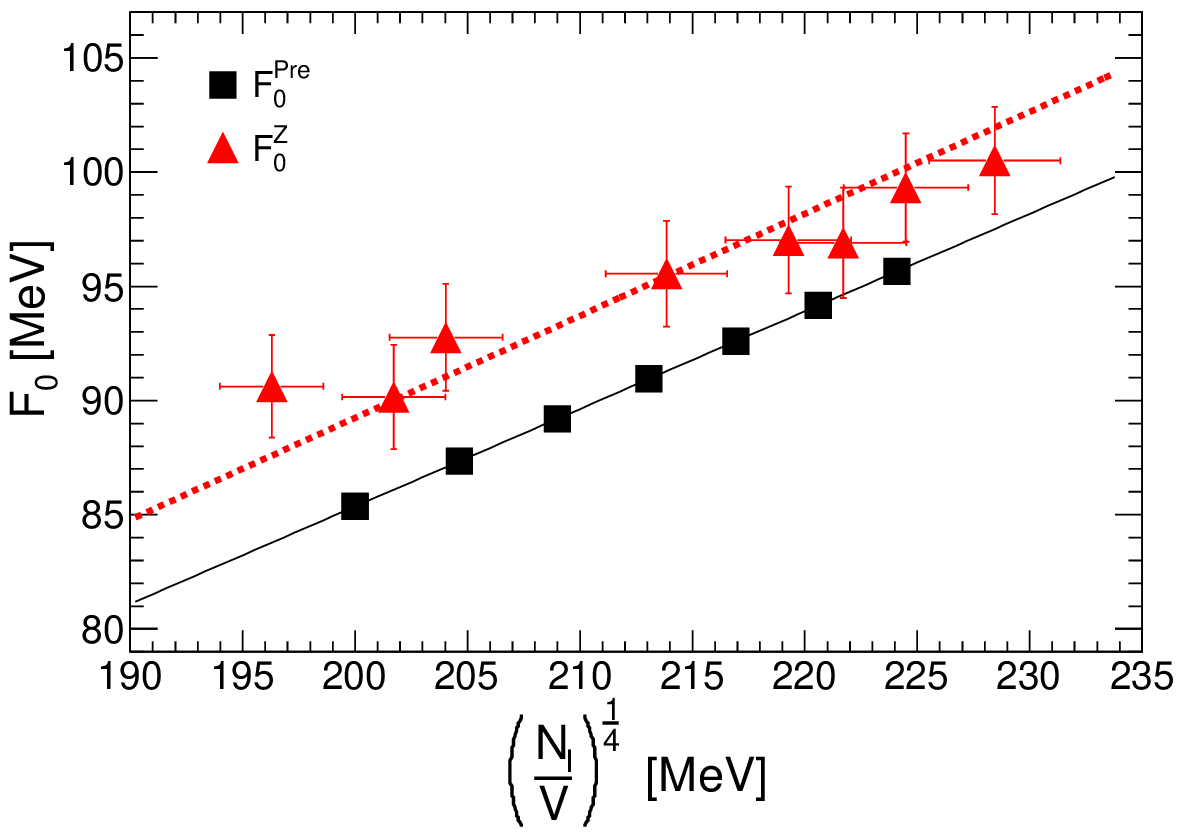}
    \hfill
    \setlength\abovecaptionskip{-3.0pt}
    \caption{Comparisons of the decay constant $F_{0}^{Z}$ with the prediction $F_{0}^{Pre}$. The solid black line and dashed red line represent the fitting results.}\label{fig:af0_ampi_res_1}
  \end{minipage}
  \hspace{3mm}
  \begin{minipage}{0.48\hsize}
    \centering
    \includegraphics[keepaspectratio, width=71mm]{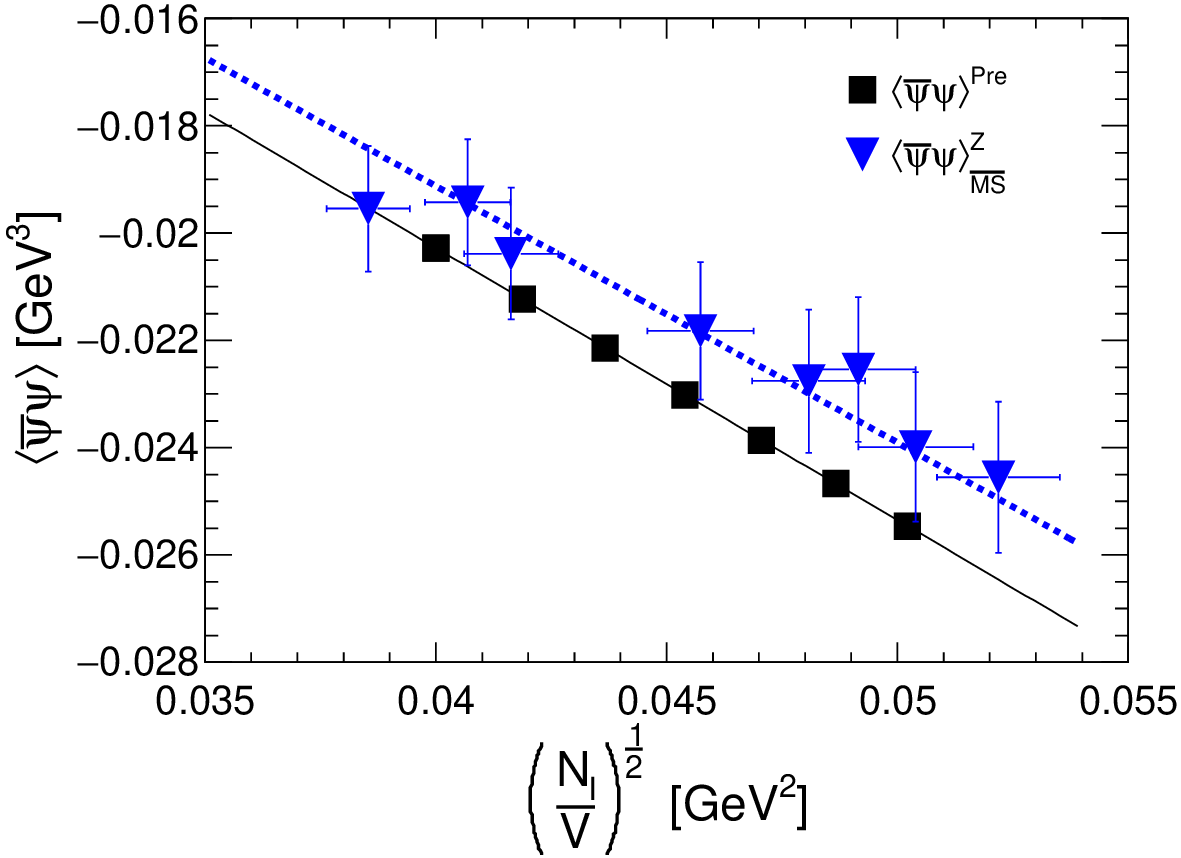}
    \setlength\abovecaptionskip{-3.0pt}
     \caption{Comparisons of the chiral condensate $\langle \bar{\psi}\psi \rangle_{\overline{MS}}^{Z}$ with the prediction $\langle \bar{\psi}\psi \rangle^{Pre}$. The solid black line and dashed blue line represent the fitting results.}\label{fig:chi_cond_ins_dens}
  \end{minipage}
\end{figure}
\begin{table}[tbp]
  \small
  \centering
  \begin{tabular}{|c|c|c|c|c|}\hline
    $m_{c}$ & {\footnotesize$\langle \bar{\psi}\psi\rangle_{\overline{MS}}^{Z}$  [GeV$^{3}$]}   & $R^{Pre}$ & $R_{\chi}^{Z}$ & $R_{\chi}^{\Sigma}$\\ \hline
    {\footnotesize Normal conf} & -1.96(12)$\times10^{-2}$ & - & - & -  \\ \hline
    0  & -1.94(12)$\times10^{-2}$ & 1.0000   & 0.99(4) &  0.99(4) \\ \hline
    1  & -2.04(12)$\times10^{-2}$ & 1.0469   & 1.04(4) &  1.06(4) \\\hline
    2  & -2.18(13)$\times10^{-2}$ & 1.0918   & 1.12(4) &  1.16(4) \\\hline
    3  & -2.28(13)$\times10^{-2}$ & 1.1349   & 1.16(5) &  1.25(5) \\\hline
    4  & -2.25(14)$\times10^{-2}$ & 1.1765   & 1.15(5) &  1.31(5) \\\hline
    5  & -2.40(14)$\times10^{-2}$ & 1.2166   & 1.23(5) &  1.33(5) \\\hline
    6  & -2.46(14)$\times10^{-2}$ & 1.2555   & 1.26(5) &  1.33(5) \\\hline
  \end{tabular}
  \caption{The renormalized chiral condensate $\langle \bar{\psi}\psi \rangle_{\overline{MS}}^{Z}$ and the ratios of the prediction $R^{Pre}$ and chiral condensates $R_{\chi}^{Z}$ and $R_{\chi}^{\Sigma}$.}\label{tb:chi_rat_1}
\end{table}

We list the calculated results of the renormalized chiral condensates in the $\overline{MS}$-scheme at 2 [GeV] in table~\ref{tb:chi_rat_1}.

In subsection~\ref{sec:chi_cond_1}, we confirm that the values of the chiral condensate decrease in direct proportion to the square root of the number density of the instantons and anti-instantons. We re-evaluate the decreases in the chiral condensate by fitting the following function, as shown in figure~\ref{fig:chi_cond_ins_dens}: $\langle \bar{\psi}\psi \rangle = -A_{\chi}\left(\frac{N_{I}}{V}\right)^{\frac{1}{2}}$. The fitting results are $A_{\chi} = 0.478(11) \ [\mbox{GeV}]$ and $\chi^{2}/d.o.f. = 1.5/7.0$. The error of $A_{\chi}$ is approximately 2$\%$. The slope $A_{\chi}$ of the numerical result is reasonably consistent with the slope $A_{\chi}^{Pre}$ = 0.5070 [GeV] of the prediction~(\ref{eq:chiral_ins_prediction1}).

In the phenomenological models of instantons~\cite{Dyakonov3, Shuryak2}, the average size of the instanton~(\ref{eq:ins_size}) is a free parameter, and it cannot be determined in the models. Therefore, there is a great need to confirm it via numerical calculations. We estimate it from the fitting result of the slope $A_{\chi} = 0.478(11)$. The inverse of the average size of the instanton is
\begin{equation}
\frac{1}{\bar{\rho}} = 5.66(13) \times10^{2} \ \  [\mbox{MeV}].
\end{equation}
This result is remarkably consistent with the values in the models~\cite{Shuryak1_1}.

These results demonstrate that the renormalized chiral condensate in the $\overline{MS}$-scheme at 2 [GeV] decreases in direct proportion to the square root of the number density of the instantons and anti-instantons. The slope $A_{\chi}$ and the average size of the instanton closely correspond to the results of the phenomenological models~\cite{Dyakonov3, Shuryak2}.
\begin{figure}[tbp]
\centering
    \includegraphics[width=83mm]{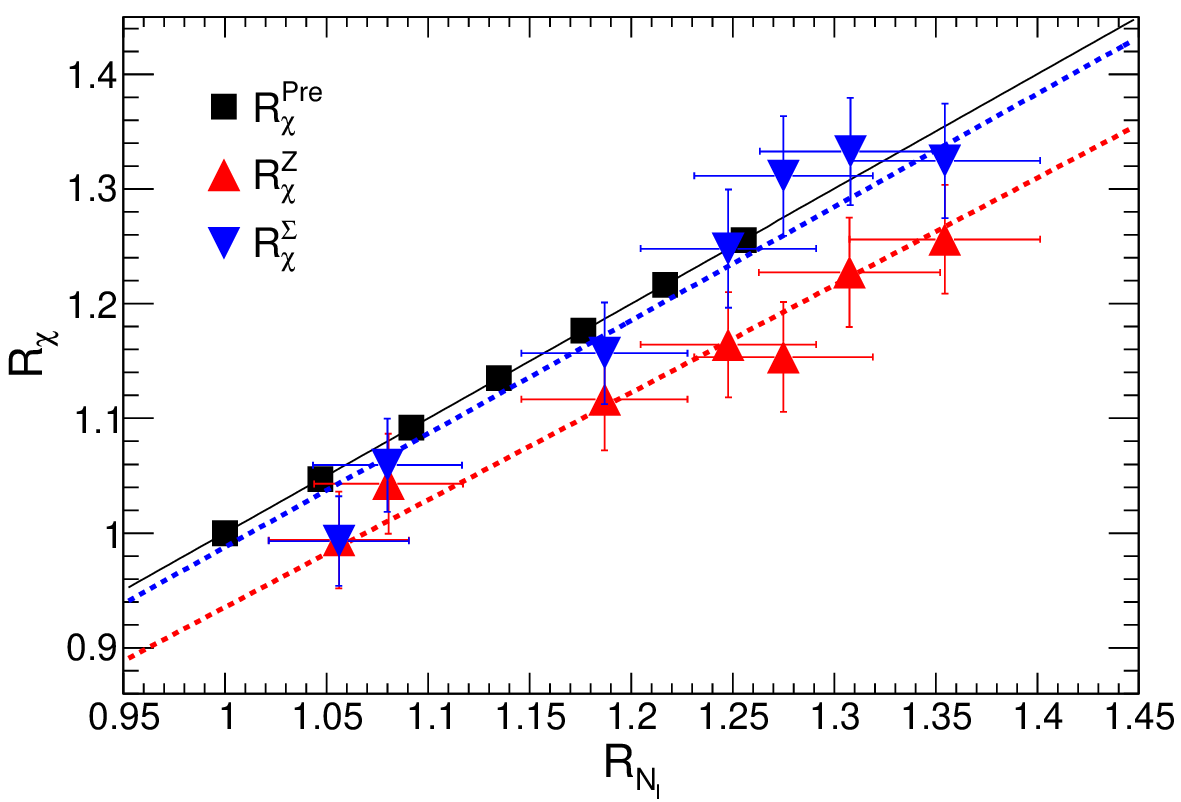}
\setlength\abovecaptionskip{-3.0pt}
\caption{Comparisons of the numerical results of $R_{\chi}^{Z}$ and $R_{\chi}^{\Sigma}$ with the prediction $R^{Pre}$. The black, red, and blue lines indicate the fitting results of $R^{Pre}$, $R_{\chi}^{Z}$, and $R_{\chi}^{\Sigma}$, respectively.}\label{fig:chi_cond_ratio}
\end{figure}
\begin{table}[tbp]
  \small
  \centering
  \begin{tabular}{|c|c|c|c|c|c|}\hline
    & $R^{Pre}$ & $R_{\chi}^{Z}$ & $R_{\chi}^{\Sigma}$ & $R_{\bar{m}_{ud}}$ &  $R_{\bar{m}_{uds}}$ \\ \hline          
    $A_{R}$ & 1.000 & 0.935(19) & 0.988(19) & 0.92(2) & 0.93(3) \\ \hline 
    $\chi^{2}/d. o. f.$ & 0.0/6.0 & 0.8/6.0 & 2.1/6.0 & 1.8/6.0 & 0.6/6.0 \\ \hline
  \end{tabular}
  \caption{The fitting results of the slope $A_{R}$.}\label{tb:fitting_ratios}
\end{table}

To remove the uncertainty that comes from the renormalization constant and the normalization factor and to clearly show the decreases in the chiral condensate, we calculate the ratio $R_{\chi}$ between the chiral condensate of the normal configuration $\langle{\bar\psi} \psi \rangle^{nor}$ and the chiral condensate of the configurations with the additional monopoles and anti-monopoles $\langle{\bar\psi} \psi\rangle^{add}$ as follows:
\begin{equation}
  R_{\chi} = \frac{\langle{\bar\psi} \psi \rangle^{add}}{\langle{\bar\psi} \psi \rangle^{nor}} = \left(\frac{N_{I}^{add}}{N_{I}^{nor}}\right)^{\frac{1}{2}}, \ R_{N_{I}} =  \left(\frac{N_{I}^{add}}{N_{I}^{nor}}\right)^{\frac{1}{2}}\label{eq:pred_chi_cond_ratio}
\end{equation}
This ratio is derived from prediction~(\ref{eq:chiral_ins_prediction1}). Moreover, we compute the ratio $R_{N_{I}}$ between the total number of instantons and anti-instantons of the normal configuration $N_{I}^{nor}$ and the total number of instantons and anti-instantons of the configurations with the additional monopoles and anti-monopoles $N_{I}^{add}$.

We derive the prediction of the ratio $R^{Pre}$ using the result~(\ref{eq:inst_dens_2}) and relation~(\ref{eq:inst_dens_pre_1}). We compute the ratios of $R_{\chi}^{Z}$ and $R_{\chi}^{\Sigma}$ using the numerical results of the chiral condensate $\langle\bar{\psi}\psi\rangle_{\overline{MS}}^{Z}$ in table~\ref{tb:chi_rat_1} and the scale parameter $\Sigma^{a}$ in table~\ref{tb:sigma_all_a}, respectively. The computed results are given in table~\ref{tb:chi_rat_1}. Figure~\ref{fig:chi_cond_ratio} clearly shows that the increases in the ratios $R_{\chi}^{Z}$ and $R_{\chi}^{\Sigma}$ correspond to the prediction $R^{Pre}$. To clearly show the consistency, we fit the following function, which is shown in the same figure:
\begin{equation}
  R_{\chi} = A_{R}R_{N_{I}}.
\end{equation}
All data points are included in the fitting range. The slopes of the numerical results correspond to the slope of the prediction as shown in table~\ref{tb:fitting_ratios}.

Finally, these results demonstrate that chiral symmetry breaking is induced by increasing the number of instantons and anti-instantons, which are created by the additional monopoles and anti-monopoles.

\subsection{The decay constants and masses of the pion and kaon}

\begin{table}[tbp]
  \small
  \centering
  \begin{tabular}{|c|c|c|c|c|}\hline
    $m_{c}$ & $a^{-1}A$ & $aB$ $\times10^{-2}$ & $FR [(am_{PS})^{2}]$ $\times10^{-2}$ & $\chi^{2}/d. o. f.$ \\ \hline
    {\footnotesize Normal conf} & 0.251(10)  & 3.08(5) & 1.8 - 10.0 & 9.4/19.0 \\\hline
    0 & 0.252(10)  & 3.06(6) & 1.8 - 10.1 & 8.7/19.0 \\\hline
    1 & 0.247(10)  & 3.15(6) & 1.8 - 9.9 & 9.5/19.0 \\\hline
    2 & 0.244(9)   & 3.24(5) & 1.8 - 10.0 & 9.7/19.0 \\\hline
    3 & 0.239(9)   & 3.29(5) & 1.9 - 10.1 & 9.7/19.0 \\\hline
    4 & 0.252(10)  & 3.29(6) & 1.9 - 9.7 & 7.6/19.0 \\\hline
    5 & 0.239(9)   & 3.37(5) & 1.9 - 10.1 & 8.4/19.0 \\\hline
    6 & 0.232(9)   & 3.41(5) & 1.9 - 10.1 & 9.9/19.0 \\\hline
  \end{tabular}
  \caption{The fitting results of the slope $a^{-1}A$ and intercept $aB$ obtained by fitting function~(\ref{eq:fit_func_fpi_mps2}).}\label{tb:afpi_ampi2_1}
\end{table}
To estimate the decay constants and masses of the pion and kaon, we first obtain the linear functions by fitting the function~(\ref{eq:fit_func_fpi_mps2}) to the computed results of $aF_{PS}$ and $(am_{PS})^{2}$ using the configurations with the additional monopoles and anti-monopoles. The fitting results are shown in table~\ref{tb:afpi_ampi2_1}. Each fitting range includes all data points of each magnetic charge, and the values of $\chi^{2}/d. o. f.$ are from 0.4 to 0.5. The fitting results of the intercept $aB$ correspond entirely to the fitting results $aF_{0}$ in table~\ref{tb:afpi_ampi2_chipt1}, which are obtained by fitting the function of the chiral perturbation theory.

We then calculate the intersections between the linear functions that are obtained by fitting equations~(\ref{eq:fpi_mpi_exp1}) and~(\ref{eq:fk_mk_exp1}). The computed intersections are in table~\ref{tb:int_sec_1}. The decay constants and the masses of the pion and the kaon are calculated using these intersections and the normalization factors $Z_{\pi}$ and $Z_{K}$.

The computed results of the decay constants of the pion $F_{\pi}^{Z}$ and the kaon $F_{K}^{Z}$ and the ratios $\frac{F_{\pi}^{Z}}{F_{0}^{Z}}$ and $\frac{F_{K}^{Z}}{F_{\pi}^{Z}}$ are given in table~\ref{tb:mpi_mk_fpi_fk}. Similarly, we list the computed results of the masses of the pion $m_{\pi}^{Z}$ and kaon $m_{K}^{Z}$ and the mass ratio $\frac{m_{K}^{Z}}{m_{\pi}^{Z}}$ in the same table~\ref{tb:mpi_mk_fpi_fk}. The table shows that the decay constants $F_{\pi}^{Z}$ and $F_{K}^{Z}$ and the masses $m_{\pi}^{Z}$ and $m_{K}^{Z}$ of the normal configuration are exactly consistent with the experimental results ($\frac{F_{\pi}^{Exp.}}{\sqrt{2}} = 92.23(12)$ [MeV], $\frac{F_{K}^{Exp.}}{\sqrt{2}} = 110.1(6)$ [MeV], $m_{\pi}^{Exp.} = 139.5706(2)$ [MeV], and $m_{K}^{Exp.} = 493.677(16)$ [MeV]~\cite{PDG_2017}). Moreover, the decay constants and the masses increase by increasing the magnetic charges $m_{c}$, whereas the ratios do not vary.
\begin{table}[tbp]
  \small
  \centering
  \begin{tabular}{|c|c|c|c|c|} \hline
    $m_{c}$ & $aF_{PS}^{\pi}$  $\times10^{-2}$ & $am_{PS}^{\pi}$ $\times10^{-2}$ & $aF_{PS}^{K}$ $\times10^{-2}$ & $am_{PS}^{K}$ \\\hline
    {\footnotesize Normal conf} & 3.13(6) & 4.74(8) & 3.80(10) & 0.171(4) \\ \hline
    0 & 3.12(6) & 4.71(9) & 3.78(10) & 0.170(5) \\ \hline
    1 & 3.21(6) & 4.85(9) & 3.91(10) & 0.175(5) \\ \hline
    2 & 3.30(6) & 5.00(8) & 4.05(10) & 0.181(5) \\ \hline
    3 & 3.35(6) & 5.07(8) & 4.10(10) & 0.184(5) \\ \hline
    4 & 3.35(6) & 5.07(9) & 4.17(12) & 0.187(5) \\ \hline
    5 & 3.43(6) & 5.19(8) & 4.23(11) & 0.190(5) \\ \hline
    6 & 3.47(5) & 5.26(8) & 4.26(10) & 0.191(5) \\ \hline
  \end{tabular}
  \caption{The computed results of the intersections. The superscripts $\pi$ and $K$ indicate the interceptions calculated using equations~(\ref{eq:fpi_mpi_exp1}) and~(\ref{eq:fk_mk_exp1}), respectively.}\label{tb:int_sec_1}
\end{table}
\begin{table}[tbp]
  \small
  \centering
  \begin{tabular}{|c|c|c|c|c|c|c|c|c|} \hline
    $m_{c}$ & $F_{0}^{Z}$ & $F_{\pi}^{Z}$& $F_{K}^{Z}$ & $\frac{F_{\pi}^{Z}}{F_{0}^{Z}}$ & $\frac{F_{K}^{Z}}{F_{\pi}^{Z}}$ & $m_{\pi}^{Z}$ & $m_{K}^{Z}$ & $\frac{m_{K}^{Z}}{m_{\pi}^{Z}}$ \\ 
    & {\footnotesize[MeV]} & {\footnotesize[MeV]} & {\footnotesize[MeV]} & & & {\footnotesize[MeV]} & {\footnotesize[MeV]} & \\ \hline
    {\footnotesize Normal conf} &  91(2)  & 92(2)  & 110(4)  &  1.02(3) & 1.19(5)  & 140(4)  & 494(18) & 3.54(16)\\\hline
    0 &  90(2)  & 92(2)  & 110(4)  &  1.02(3) & 1.19(5)  & 139(4)  & 491(18) & 3.54(16)\\ \hline    
    1 &  93(2)  & 95(2)  & 113(4)  &  1.02(3) & 1.20(5)  & 143(4)  & 507(19) & 3.55(16)\\ \hline
    2 &  96(2)  & 97(2)  & 117(4)  &  1.02(2) & 1.20(5)  & 147(4)  & 525(19) & 3.57(16)\\  \hline   
    3 &  97(2)  & 99(2)  & 119(4)  &  1.02(2) & 1.20(5)  & 150(4)  & 532(19) & 3.56(15)\\ \hline    
    4 &  97(2)  & 99(3)  & 121(4)  &  1.02(3) & 1.22(6)  & 149(4)  & 541(20) & 3.62(17)\\ \hline    
    5 &  99(2)  & 101(2) & 122(4)  &  1.02(2) & 1.21(5)  & 153(4)  & 549(20) & 3.58(16)\\  \hline
    6 & 101(2)  & 102(2) & 123(4)  &  1.02(2) & 1.20(5)  & 155(4)  & 552(19) & 3.57(15)\\  \hline
  \end{tabular}
  \caption{The numerical results of $F_{0}^{Z}$, $F_{\pi}^{Z}$, and $F_{K}^{Z}$ and the ratios of these decay constants. The numerical results of $m_{\pi}^{Z}$ and $m_{K}^{Z}$ and the mass ratio $\frac{m_{K}^{Z}}{m_{\pi}^{Z}}$. The decay constant in the chiral perturbation theory is $F_{0}^{\chi PT} = 86.2(5)$ [MeV], and the ratio is $\frac{F_{\pi}}{F_{0}^{\chi PT}}$ = 1.071(6)~\cite{Colangelo3}.}\label{tb:mpi_mk_fpi_fk}
\end{table}

\subsection{The light quark masses}

We evaluate the renormalized average masses of the light quarks $\hat{\bar{m}}_{ud}^{\overline{MS}}$ and $\hat{\bar{m}}_{uds}^{\overline{MS}}$ and the renormalized mass of the strange quark $\hat{m}_{s}^{\overline{MS}}$ in the $\overline{MS}$-scheme at 2 [GeV].

The average mass of the light quarks $\bar{m}_{ud}$~(\ref{eq:light_quark_masses1}) is derived from the PCAC relation concerning the pion as follows:
\begin{equation}
  a\bar{m}_{ud}^{Z} = \frac{(Z_{\pi}am_{PS}^{\pi})^{2}}{aA^{(2)}}
\end{equation}
The average mass of the light quarks $\bar{m}_{sud}$~(\ref{eq:light_quark_masses2}) and the strange quark mass $m_{s}$ are derived from the PCAC relation concerning the kaon as follows:
\begin{align}
  & a\bar{m}_{sud}^{Z} = \frac{am_{s}^{Z} + a\bar{m}_{ud}^{Z}}{2} =  \frac{(Z_{K}am_{PS}^{K})^{2}}{aA^{(2)}}\\
  & am_{s}^{Z} =  \frac{2(Z_{K}am_{PS}^{K})^{2} - (Z_{\pi}am_{PS}^{\pi})^{2}}{aA^{(2)}}\label{eq:ms_comp1}
\end{align}
We use the fitting results of the slope $A^{(2)}$ in table~\ref{tb:pcac_1_3}. The renormalized masses of the light quarks in the $\overline{MS}$-scheme at 2 [GeV] are evaluated by the following formula:
\begin{equation}
  \hat{m}_{q}^{\overline{MS}} = \frac{0.72076}{Z_{S}}m_{q}^{Z}, \ \ (m_{q}^{Z} =  \bar{m}_{ud}^{Z}, \ \bar{m}_{sud}^{Z}, \ m_{s}^{Z}).
\end{equation}
The renormalization constant $Z_{S} = 0.93(3)$ of the normal configurations is used. The renormalized masses of the light quarks in the $\overline{MS}$-scheme at 2 [GeV], which are calculated using the normal configurations, are
\begin{align}
   \hat{\bar{m}}_{ud}^{\overline{MS}} \  (2 \ [\mbox{GeV}]) = 4.1(3) \ \ [\mbox{MeV}], \ \  \hat{m}_{s}^{\overline{MS}} \ (2 \ [\mbox{GeV}]) = 98(8) \ \ [\mbox{MeV}]\label{eq:qmass1}.
\end{align}
\begin{table}[tbp]
 \small
  \centering
      {\renewcommand{\arraystretch}{1.2}
        \begin{tabular}{|c|c|c|c|c|c|c|c|c|c|} \hline
          $m_{c}$ & $\bar{m}_{ud}^{Pre}$ & $\hat{\bar{m}}_{ud}^{\overline{MS}}$ & $\hat{\bar{m}}_{sud}^{\overline{MS}}$ & $m_{s}^{Pre}$ & $\hat{m}_{s}^{\overline{MS}}$ & $\frac{m_{s}^{Z}}{\bar{m}_{ud}^{Z}}$ & $R_{\bar{m}_{ud}}$ & $R_{\bar{m}_{sud}}$ & $R_{m_{s}}$ \\
          &\footnotesize{[MeV]}&\footnotesize{[MeV]}&\footnotesize{[MeV]}&\footnotesize{[MeV]}&\footnotesize{[MeV]}& & & & \\ \hline
          \footnotesize{Normal conf} & 4.1(3) & 3.5$_{-0.3}^{+0.7}$ & 51(4)  & 96$_{-4}^{+8}$ &  98(8) &  24(2) & - & - & - \\  \hline
          0 & 4.0(3) & 3.5$_{-0.3}^{+0.7}$ & 51(4) & 96$_{-4}^{+8}$ & 97(8)   &  24(3)  &  0.99(5) &  0.99(7) &  0.99(11)   \\ \hline
          1 & 4.3(3) & 3.7$_{-0.3}^{+0.7}$ & 54(4) & 101$_{-4}^{+8}$ & 104(9)  &  24(3)  &  1.05(5) &  1.06(8) &  1.06(12)   \\ \hline
          2 & 4.5(3) & 3.8$_{-0.3}^{+0.8}$ & 58(5) & 105$_{-4}^{+9}$ & 111(9)  &  24(3)  &  1.11(5) &  1.13(8) &  1.13(12)   \\ \hline
          3 & 4.6(3) & 4.0$_{-0.3}^{+0.8}$ & 59(5) & 109$_{-5}^{+9}$ & 112(9)  &  24(3)  &  1.13(6) &  1.14(8) &  1.14(12)   \\ \hline
          4 & 4.7(3) & 4.1$_{-0.4}^{+0.8}$ & 61(5) & 113$_{-5}^{+9}$ & 117(10) &  25(3)  &  1.14(6) &  1.19(9) &  1.19(13)   \\ \hline
          5 & 4.8(3) & 4.3$_{-0.4}^{+0.9}$ & 62(5) & 117$_{-5}^{+10}$ & 119(10) &  25(3)  &  1.18(6) &  1.21(9) &  1.21(13)    \\ \hline
          6 & 4.9(3) & 4.4$_{-0.4}^{+0.9}$ & 63(5) & 121$_{-5}^{+10}$ & 121(10) &  24(2)  &  1.21(6) &  1.23(9) &  1.23(13)   \\ \hline
        \end{tabular}
      }
      \caption{The predictions and numerical results of the light quark masses and the mass ratio of the quarks $\frac{m_{s}^{Z}}{\bar{m}_{ud}^{Z}}$. The ratios of the quark masses $R_{\bar{m}_{ud}}$, $R_{\bar{m}_{sud}}$, and $R_{m_{s}}$.}\label{tb:quarks_1}
\end{table}

In this study, we estimate the light quark masses using the normalization factors, which are calculated by matching the numerical results with the experimental results. To analyze the effects of the discretization, we estimate the quark masses in the continuum limit via interpolation. The renormalized average mass of the light quarks $\hat{\bar{m}}_{ud}^{\overline{MS}}$ and the strange quark $\hat{m}_{s}^{\overline{MS}}$ in the $\overline{MS}$-scheme at 2 [GeV] in the continuum limit are
\begin{equation}
\hat{\bar{m}}_{ud}^{\overline{MS}} \ (2 \ [\mbox{GeV}]) =  4.09(10) \ [\mbox{MeV}], \ \ \hat{m}_{s}^{\overline{MS}} \ (2 \ [\mbox{GeV}]) = 98(3) \ [\mbox{MeV}]\nonumber.
\end{equation}
These results are entirely consistent with the computed results of the normal configuration~(\ref{eq:qmass1}). Moreover, these findings are consistent with the experimental results $\bar{m}_{ud}^{Exp.} = 3.5_{-0.3}^{+0.7}$ [MeV] and $m_{s}^{Exp.} = 96_{-4}^{+8}$ [MeV]~\cite{PDG_2017}. Furthermore, the mass ratio of the computed results in the continuum limit is
\begin{equation}
\frac{\hat{m}_{s}^{\overline{MS}}}{\hat{\bar{m}}_{ud}^{\overline{MS}}} \ \ (2 \ [\mbox{GeV}]) = 24.0(9)\nonumber.
\end{equation}
This finding is 12$\%$ smaller than the experimental result $\frac{m_{s}}{\bar{m}_{ud}}$ = 27.3(7)~\cite{PDG_2017}, whereas it is consistent with the outcome of the chiral perturbation theory~\cite{Gasser3}. We obtain these results without using any consequences of the chiral perturbation theory. We adequately calculate the light quark masses. We will report these results in the future~\cite{Hasegawa5}.

We list the computed results of the renormalized masses of the light quarks in the $\overline{MS}$-scheme at 2 [GeV] and the mass ratio $\frac{m_{s}^{Z}}{\bar{m}_{ud}^{Z}}$ in table~\ref{tb:quarks_1}.

\subsection{The instanton effects on the light quark masses, the meson masses, and the decay constants}

\begin{figure}[tbp]
  \begin{minipage}{0.48\hsize}
    \centering
    \includegraphics[keepaspectratio, width=70mm]{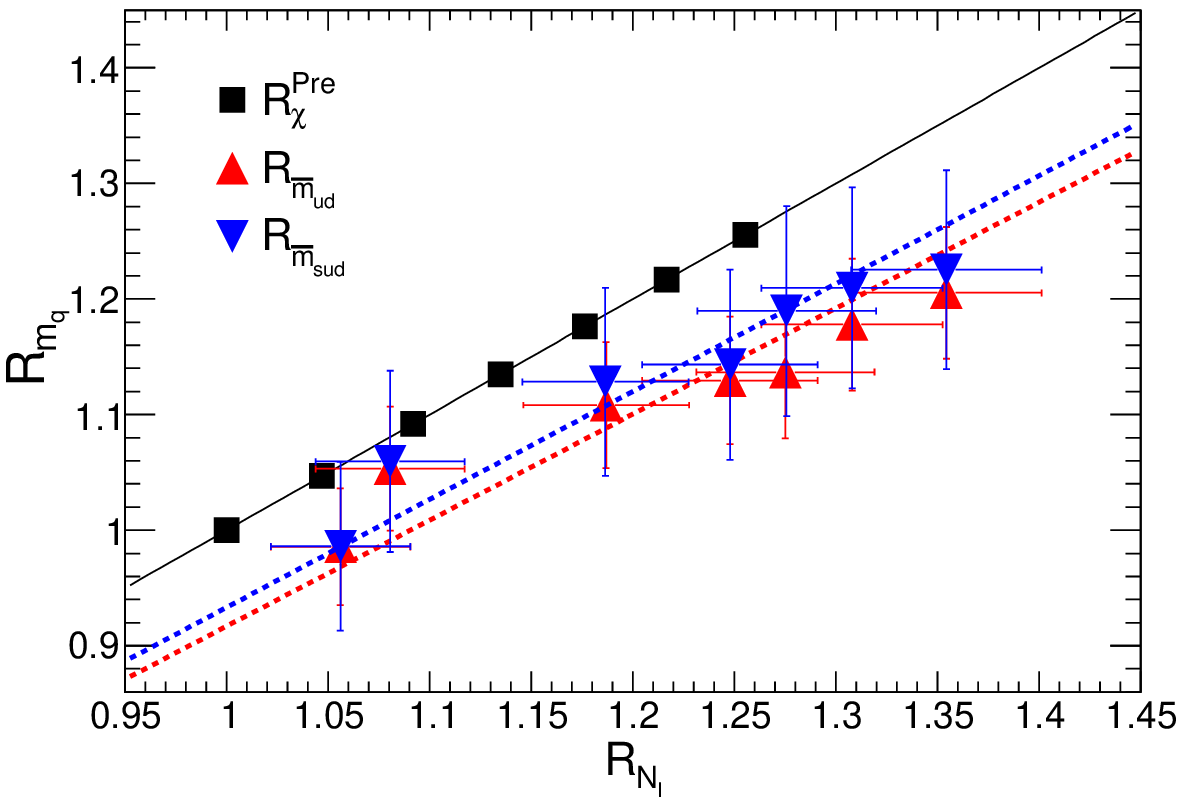}
    \hfill
    \setlength\abovecaptionskip{-3.0pt}
    \caption{Comparisons of $R_{m_{q}}$ with the prediction $R^{Pre}$. The black, red, and blue lines indicate the fitting results.}\label{fig:ratios_mas_red_mc_1}
  \end{minipage}
  \hspace{3mm}
  \begin{minipage}{0.48\hsize}
    \vspace{-4mm}
    \centering
    \includegraphics[keepaspectratio, width=70mm]{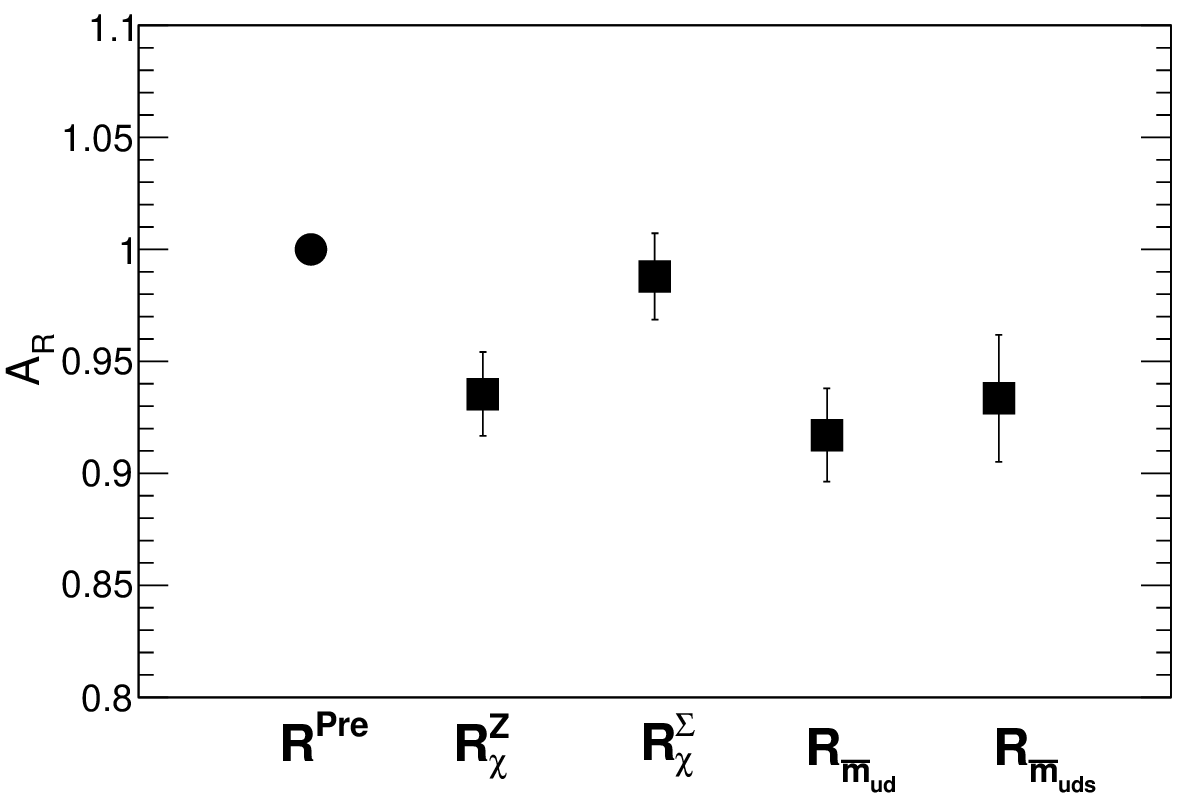}
    \setlength\abovecaptionskip{-3.0pt}
    \caption{Comparisons of the fitting results of slope $A_{R}$ with the prediction $A_{R}^{Pre} = 1.000$.}\label{fig:comparisons_slopes_ratios}
  \end{minipage}
\end{figure}
\begin{figure}[tbp]
\centering
    \includegraphics[width=150mm]{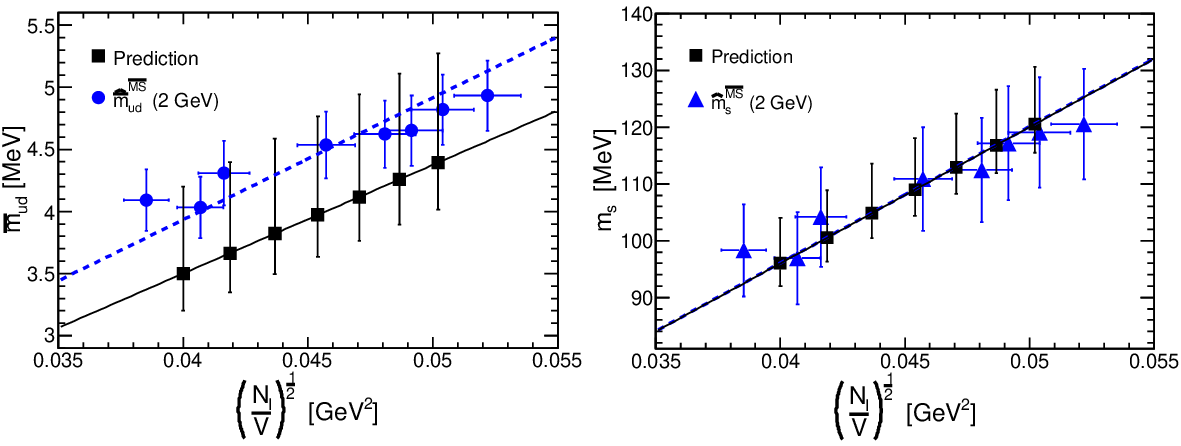}
\setlength\abovecaptionskip{-3.0pt}
\caption{Comparisons of the light quark masses $\bar{m}_{ud}^{\overline{MS}}$ (left) and $m_{s}^{\overline{MS}}$ (right) in the $\overline{MS}$-scheme at 2 [GeV] with the predictions. The black and blue lines indicate the fitting results of the predictions and numerical results, respectively.}\label{fig:mud_ms_ins_fit}
\end{figure}

We suppose that the increases in the light quark masses by increasing the number density of the instantons and anti-instantons correspond to the increase in the ratio of the chiral condensates $R^{Pre}$. This assumption comes from the Nambu-Jona-Lasinio model~\cite{Nambu1, Nambu2, Goldstone1}.

To clearly show the increases in the light quark masses, we evaluate the mass ratios $R_{m_{q}} = \frac{m_{q}^{add}}{m_{q}^{nor}}$, ($m_{q} = \bar{m}_{ud}$, $\bar{m}_{sud}$, and $m_{s}$). The quark masses $m_{q}^{nor}$ are calculated using the normal configurations. The quark masses $m_{q}^{add}$ are computed using the configurations with the additional monopoles and anti-monopoles. The results are in table~\ref{tb:quarks_1}. The errors of the ratio $R_{m_{s}}$ are large because the normalization factors $Z_{\pi}$ and $Z_{K}$ in formula~(\ref{eq:ms_comp1}) do not cancel out.

Figure~\ref{fig:ratios_mas_red_mc_1} clearly shows that the increases in the ratios $R_{\bar{m}_{ud}}$ and $R_{\bar{m}_{sud}}$ correspond to the increase in the prediction $R^{Pre}$. Similar to the evaluations of the rations of the chiral condensate, we fit the function $R_{m_{q}} = A_{R}R_{N_{I}}$ and compare the fitting results of the slope $A_{R}$ with the prediction $A_{R}^{Pre} = 1.000$, which are in table~\ref{tb:fitting_ratios}, as shown in figure~\ref{fig:comparisons_slopes_ratios}. The slopes of $R_{\bar{m}_{ud}}$ and $R_{\bar{m}_{uds}}$ correspond closely to the slopes of $R_{\chi}^{Z}$ of the chiral condensate and $R_{\chi}^{\Sigma}$ of the scale parameter.

Now, we derive the prediction regarding the light quark masses $\bar{m}_{ud}^{Pre}$ and $m_{s}^{Pre}$ from the experimental results and the ratio $R^{Pre}$ in table~\ref{tb:chi_rat_1}, and we list them in table~\ref{tb:quarks_1}. We compare them with the numerical results, as shown in figure~\ref{fig:mud_ms_ins_fit}. The computed results of the square root of the instanton density $\left(\frac{N_{I}^{Pre}}{V}\right)^{\frac{1}{2}}$ in table~\ref{tb:instantons_chiral_fpi} are used for the predictions.
\begin{table}[tbp]
  \small
  \centering
      {\renewcommand{\arraystretch}{1.0}
        \begin{tabular}{|c|c|c|c|c|} \hline
          &  $m_{ud}^{Pre}$ & $\hat{\bar{m}}_{ud}^{\overline{MS}}$ & $m_{s}^{Pre}$ & $\hat{m}_{s}^{\overline{MS}}$  \\ \hline
          $A_{q}$ {\footnotesize [MeV$^{-1}$]} & 8.8(4)$\times10^{-5}$ & 9.8(2)$\times10^{-5}$ & $2.40(5)\times10^{-3}$ & $2.40(7)\times10^{-3}$ \\ \hline 
          $\chi^{2}/d. o. f.$ & 0.0/6.0 & 3.0/7.0 & 0.0/6.0 & 1.1/7.0 \\ \hline\hline
          & $m_{\pi}^{Pre}$ & $m_{\pi}^{Z}$ & $m_{K}^{Pre}$ & $m_{K}^{Z}$  \\ \hline
          $A_{m}$ & 0.697853(4) & 0.688(7) & 2.46833(3) & 2.45(3) \\  \hline 
          $\chi^{2}/d. o. f.$ & 0.0/6.0 & 2.8/7.0 & 0.0/7.00 & 0.8/7.0 \\  \hline \hline
          & $F_{\pi}^{Pre}$ & $F_{\pi}^{Z}$ & $F_{K}^{Pre}$ & $F_{K}^{Z}$ \\ \hline
          $A_{f}$ & 0.46139(16) & 0.4546(4) & 0.55055(18) & 0.547(7) \\  \hline 
          $\chi^{2}/d. o. f.$  & 0.0/6.0 & 2.8/7.0 & 0.0/6.0 & 0.8/7.0 \\  \hline
        \end{tabular}            
      }
      \caption{Comparisons of the fitting results of the slopes $A_{q}$ (upper), $A_{m}$ (middle), and $A_{f}$ (lower) with the predictions.}\label{tb:fitting_results_aq_am_af}
\end{table}
\begin{table}[tbp]
  \small
  \centering 
  \begin{tabular}{|c|c|c|c|c|} \hline
    $m_{c}$ & $m_{\pi}^{Pre}$ {\footnotesize[MeV]} & $m_{K}^{Pre}$ {\footnotesize[MeV]} & $F_{\pi}^{Pre}$ {\footnotesize[MeV]} & $F_{K}^{Pre}$ {\footnotesize[MeV]}  \\ \hline
    {\footnotesize Normal conf} & 139.5706(2) & 493.666(16)  &  92.28(8)  & 110.1(4)\\ \hline
    0 & 139.5706(2) & 493.666(16)  &  92.28(8)  & 110.1(4)\\ \hline
    1 & 142.8069(2) & 505.113(16)  &  94.42(9)  & 112.7(4)\\ \hline
    2 & 145.8370(2) & 515.831(17)  &  96.42(9)  & 115.1(4)\\ \hline
    3 & 148.6892(2) & 525.919(17)  &  98.31(9)  & 117.3(5)\\ \hline
    4 & 151.3862(2) & 535.458(17)  & 100.09(9)  & 119.4(5)\\ \hline
    5 & 153.9462(2) & 544.513(18)  & 101.78(9)  & 121.5(5)\\ \hline
    6 & 156.3845(2) & 553.138(18)  & 103.39(10) & 123.4(5)\\ \hline
  \end{tabular}
  \caption{The predictions of the masses $m_{\pi}^{Pre}$ and $m_{K}^{Pre}$ and the decay constants $F_{\pi}^{Pre}$ and $F_{K}^{Pre}$. The experimental results are used in~\cite{PDG_2017}.}\label{tb:pre_mpi_mk_fpi_mk}
\end{table}
\begin{figure}[tbp]
\centering
    \includegraphics[width=150mm]{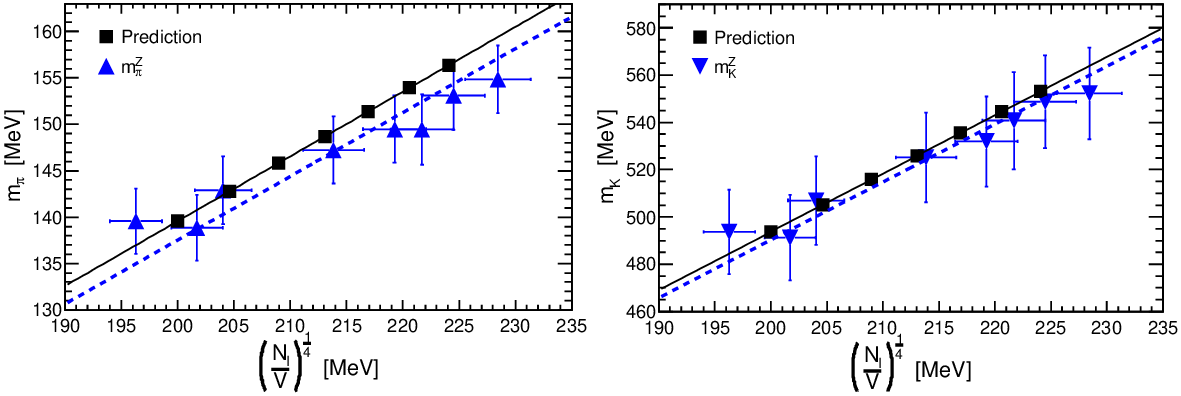}
\setlength\abovecaptionskip{-3.0pt}
\caption{Comparisons of the masses $m_{\pi}$ (left) and $m_{K}$ (right) with the predictions. The black and blue lines indicate the fitting results of the predictions and numerical results, respectively.}\label{fig:mpi_mk_ins_fit}
\end{figure}

To more precisely check their consistency, we fit the following function to the numerical results and predictions: $y = A_{q}\left(\frac{N_{I}}{V}\right)^{\frac{1}{2}}$. All data are included in the fitting ranges. We list the fitting results in table~\ref{tb:fitting_results_aq_am_af}. The fitting results are reasonably consistent with the predictions. Therefore, the light quark masses increase in direct proportion to the square root of the number density of the instantons and anti-instantons.

From the PCAC relation, the pion and kaon masses increase in direct proportion to the one-fourth root of the number density of the instantons and anti-instantons because the light quark masses increase in direct proportion to the square root of the number density of the instantons and anti-instantons. We make the predictions $m_{\pi}^{Pre}$ and $m_{K}^{Pre}$ concerning the pion and kaon masses using the experimental results and the ratio $\left(R^{Pre}\right)^{\frac{1}{2}}$. The predictions are in table~\ref{tb:pre_mpi_mk_fpi_mk}. We compare the predictions with the numerical results, as shown in figure~\ref{fig:mpi_mk_ins_fit}. The computed results of the instanton density $\left(\frac{N_{I}^{Pre}}{V}\right)^{\frac{1}{4}}$ in table~\ref{tb:instantons_chiral_fpi} are used for the predictions. We fit the following function: $y = A_{m}\left(\frac{N_{I}}{V}\right)^{\frac{1}{4}}$. All data are included in the fitting range. We list the fitting results in table~\ref{tb:fitting_results_aq_am_af}. The fitting results agree perfectly with the predictions. Therefore, the pion $m_{\pi}^{Pre}$ and kaon $m_{K}^{Pre}$ masses increase in direct proportion to the one-fourth root of the number density of the instantons and anti-instantons.
\begin{figure}[tbp]
\centering
    \includegraphics[width=150mm]{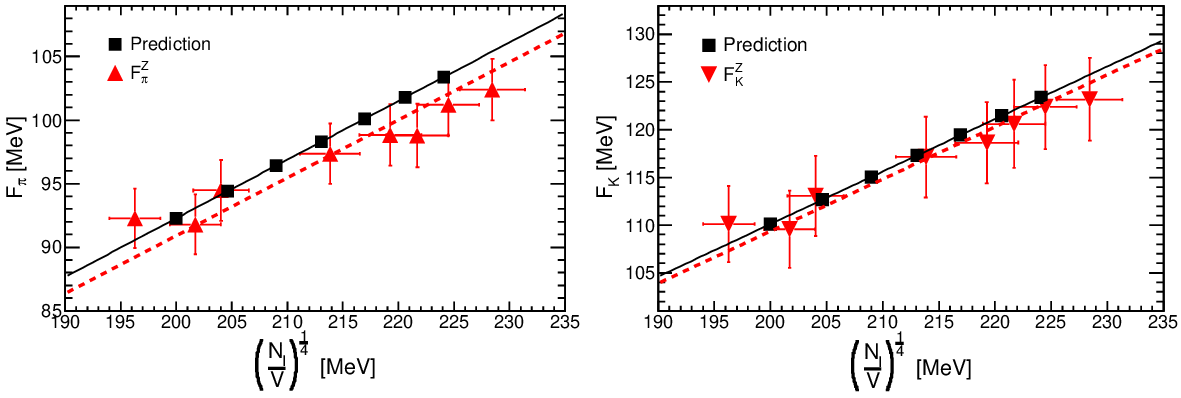}
\setlength\abovecaptionskip{-3.0pt}
\caption{Comparisons of the decay constants $F_{\pi}$ (left) and $F_{K}$ (right) with the predictions. The black and red lines indicate the fitting results of the predictions and numerical results, respectively.}\label{fig:fpi_fk_ins_fit}
\end{figure}

We have confirmed that the formula in the quenched chiral perturbation theory~(\ref{eq:chipt_f0_fit}) holds. Therefore, the decay constants of the pion and kaon are in direct proportion to the one-fourth root of the number density of the instantons and anti-instantons. To confirm this, we make the predictions $F_{\pi}^{Pre}$ and $F_{K}^{Pre}$ regarding the decay constants of the pion and kaon in the same way as the pion and kaon masses and fit the following function to the numerical results of $F_{\pi}^{Z}$ and $F_{K}^{Z}$ and their predictions: $y = A_{f}\left(\frac{N_{I}}{V}\right)^{\frac{1}{4}}$. All results are included in the fitting ranges. Table~\ref{tb:fitting_results_aq_am_af} indicates that the fitting results are remarkably consistent with the predictions. Figure~\ref{fig:fpi_fk_ins_fit} clearly shows that the decay constants of the pion and kaon increase in direct proportion to the one-fourth root of the number density of the instantons and anti-instantons.

\subsection{Catalytic effect on the pion decay}\label{sssec:decay_width1}

Lastly, we estimate the catalytic effect on the charged pion. One charged pion $\pi^{\pm}$ decays to a lepton $l^{\pm}$ (an electron $e$ or a muon $\mu$) and a neutrino $\nu_{l}$ as follows:
\begin{equation}
 \pi^{+} \rightarrow l^{+} + \nu_{l}, \ \  \pi^{-} \rightarrow l^{-} + \bar{\nu}_{l} 
\end{equation}
These decays are induced by the weak interaction, and the decay width of the charged pion is derived~\cite{Kugo1} as follows:
\begin{equation}
  \Gamma (\pi^{-} \rightarrow l + \bar{\nu}_{l}) = \frac{(G_{F}F_{\pi}\cos\theta_{c})^{2}}{4\pi m_{\pi}^{3}}m_{l}^{2}(m_{\pi}^{2} - m_{l}^{2})^{2}\label{eq:decay_width_1} 
\end{equation}
This formula indicates that the decay width is proportional to the mass of the lepton. The mass ratio of these masses is $m_{e}^{Exp.}/m_{\mu}^{Exp.}$ = $4.83633170(11) \times10^{-3}$. Therefore, over 99 \% of the charged pions decay to the muon. Therefore, we estimate the total decay width of the charged pion from the partial decay width, where the charged pion decays to the muon. We suppose that monopoles and instantons do not affect the masses of the leptons.

The decay width of the charged pion, which is estimated by substituting the experimental results for formula~(\ref{eq:decay_width_1}), is $\Gamma (\pi^{-} \rightarrow \mu + \bar{\nu}_{\mu}) = 3.77439 \times 10^{7} \ [\mbox{sec}^{-1}]$. The Dirac constant is $\hbar = 6.582119514 (40) \times10^{-16}$ [eV$\cdot s$]~\cite{PDG_2017}, and the Fermi constant is $G_{F} = 1.1663787(6) \times 10^{-5}$ [GeV$^{-2}$]~\cite{PDG_2017}. Here, we do not consider the errors of the experimental results because they are substantially smaller than the errors of the numerical results.

The lifetime of the charged pion is estimated by the formula $\tau = \frac{1}{\Gamma (\pi^{-} \rightarrow \mu + \bar{\nu}_{\mu})}$ because the branching ratio of the charged pions, which decay to muons, is almost 100$\%$. The lifetime of the charged pion is $\tau = 2.64944 \times 10^{-8} \ [\mbox{sec}]$.

The difference between the experimental lifetime of the charged pion~\cite{PDG_2017} and the result of the theoretical calculations is less than 1.8$\%$. Therefore, we derive the decay width of the charged pion using the formula~(\ref{eq:decay_width_1}) and calculate its lifetime.

The decay width, which is estimated with the numerical results of the pion decay constant $F_{\pi}^{Z}$ and the pion mass $m_{\pi}^{Z}$ of the normal configuration as the input values, is
\begin{equation}
  \Gamma = 3.8(3) \times 10^{7} \ \ [\mbox{sec}^{-1}]\nonumber.
\end{equation}
Similarly, the lifetime is
\begin{equation}
  \tau = 2.6(2) \times 10^{-8} \ \ [\mbox{sec}]\nonumber.
\end{equation}
These results are consistent with the results of the theoretical calculations and experiments. Therefore, we can correctly estimate the decay width and lifetime of the charged pion using formula~(\ref{eq:decay_width_1}) and the numerical results of $F_{\pi}^{Z}$ and $m_{\pi}^{Z}$.
\begin{table}[tbp]
  \small
  \centering
  \begin{tabular}{|c|c|c|c|c|} \hline
    $m_{c}$ & $\Gamma^{Pre}$ [sec$^{-1}$]  & $\Gamma$ [sec$^{-1}$] & $\tau^{Pre}$ [sec]  & $\tau$ [sec] \\ \hline
    {\footnotesize Normal conf} & 3.774(7)$\times10^{7}$   & 3.8(3)$\times10^{7}$ & 2.649(5)$\times10^{-8}$ & 2.6(2)$\times10^{-8}$  \\ \hline
    0 & 3.774(7)$\times10^{7}$   & 3.6(3)$\times10^{7}$ & 2.649(5)$\times10^{-8}$ & 2.8(3)$\times10^{-8}$   \\ \hline
    1 & 4.544(7)$\times10^{7}$   & 4.6(4)$\times10^{7}$ & 2.201(4)$\times10^{-8}$ & 2.2(2)$\times10^{-8}$   \\ \hline
    2 & 5.333(10)$\times10^{7}$  & 5.7(5)$\times10^{7}$ & 1.875(3)$\times10^{-8}$ & 1.75(15)$\times10^{-8}$ \\ \hline
    3 & 6.136(11)$\times10^{7}$  & 6.4(6)$\times10^{7}$ & 1.630(3)$\times10^{-8}$ & 1.57(14)$\times10^{-8}$ \\ \hline
    4 & 6.951(13)$\times10^{7}$  & 6.4(6)$\times10^{7}$ & 1.439(3)$\times10^{-8}$ & 1.57(14)$\times10^{-8}$ \\ \hline
    5 & 7.775(14)$\times10^{7}$  & 7.5(7)$\times10^{7}$ & 1.286(2)$\times10^{-8}$ & 1.33(12)$\times10^{-8}$ \\ \hline
    6 & 8.606(16)$\times10^{7}$  & 8.1(7)$\times10^{7}$ & 1.162(2)$\times10^{-8}$ & 1.24(11)$\times10^{-8}$ \\ \hline
              \end{tabular}
  \caption{The decay width $\Gamma$ and lifetime $\tau$ of the charged pion together with their predictions $\Gamma^{Pre}$ and $\tau^{Pre}$.}\label{tb:width_1}
\end{table}

We have shown that the increases in the mass and decay constant of the pion are in direct proportion to the one-fourth root of the number density of the instantons and anti-instantons, which are precisely consistent with the predictions. We substitute the numerical results of $F_{\pi}^{Z}$ and $m_{\pi}^{Z}$ and predictions $F_{\pi}^{Pre}$ and $m_{\pi}^{Pre}$ for formula~(\ref{eq:decay_width_1}) and estimate the catalytic effect on the charged pion. The numerical results of $F_{\pi}^{Z}$ and $m_{\pi}^{Z}$ and the predictions of $F_{\pi}^{Pre}$ and $m_{\pi}^{Pre}$ are given in tables~\ref{tb:mpi_mk_fpi_fk} and~\ref{tb:pre_mpi_mk_fpi_mk}. The computed results of the decay width $\Gamma$ and lifetime $\tau$ of the charged pion together with the predictions $\Gamma^{Pre}$ and $\tau^{Pre}$ are shown in table~\ref{tb:width_1}.
\begin{figure}[tbp]
\centering
    \includegraphics[width=150mm]{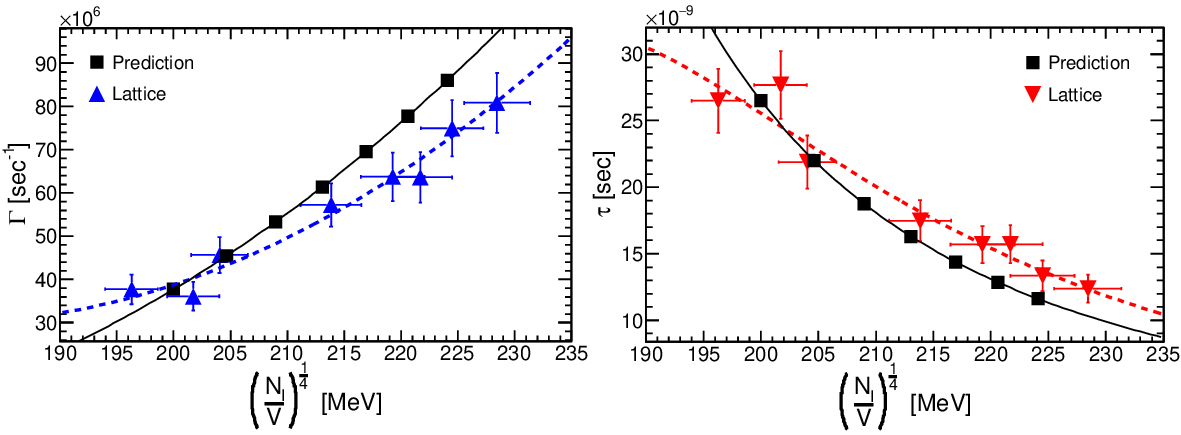}
\setlength\abovecaptionskip{-3.0pt}
\caption{Comparisons of the decay width $\Gamma$ (left) and the lifetime $\tau$ (right) of the charged pion with the predictions. The black symbols and lines indicate the predictions and fitting results, respectively. The blue and red symbols and lines indicate the numerical results and fitting results, respectively.}\label{fig:Gamma_tau_ins_fit}
\end{figure}
\begin{table}[tbp]
  \small
  \centering
      {\renewcommand{\arraystretch}{1.0}
        \begin{tabular}{|c|c|c|c|c|c|} \hline
          & Analytic &  $\Gamma^{Pre}$ & $\Gamma$ & $\tau^{Pre}$ & $\tau$  \\ \hline
          $p_{1}$ {\footnotesize[Sec$^{-1}\cdot$MeV$^{-3}$]}& 25.89 & 25.9(9) & 29(16) & 25.91(4) & 29(11) \\ \hline
          $p_{2}$ {\footnotesize[Sec$^{-1}\cdot$MeV$^{-1}$]}& 1.187 & $1.19(8)\times10^{6}$ & $1.8(1.4)\times10^{6}$ & $1.189(3)\times10^{6}$ & $1.8(9)\times10^{6}$ \\ \hline
          $p_{3}$ {\footnotesize[Sec$^{-1}\cdot$MeV]}& 1.360 & $1.36(17)\times10^{10}$ & $3(3)\times10^{10}$ & $1.364(8)\times10^{10}$ & $3(2)\times10^{10}$ \\ \hline 
          $\chi^{2}/d. o. f.$ & - & 0.0/4.0 & 2.2/5.0 & 0.0/4.0 & 2.1/5.0 \\ \hline 
        \end{tabular}
          }
      \caption{The analytic computations and fitting results of the parameters $p_{1}$, $p_{2}$, and $p_{3}$.}\label{tb:fitting_results_gamma_tau}
\end{table}

We then make the following function of $\left(\frac{N_{I}}{V}\right)^{\frac{1}{4}}$, which is derived from formula~(\ref{eq:decay_width_1}):
\begin{equation}
  y(x) = p_{1}x^{3} - p_{2}x + \frac{p_{3}}{x}, \ x = \left(\frac{N_{I}}{V}\right)^{\frac{1}{4}}.
\end{equation}
We fit this curve $y(x)$ to the results of the decay width $\Gamma$ and the curve $y(x)^{-1}$ to the results of the lifetime $\tau$, as shown in figure~\ref{fig:Gamma_tau_ins_fit}, and we determine the free parameters $p_{1}$,  $p_{2}$, and $p_{3}$. All data are included in the fitting ranges. The fitting results are presented in table~\ref{tb:fitting_results_gamma_tau}. The errors of the fitting results are large. However, the values of $\chi^{2}/d. o. f.$ are sufficiently small. Finally, figure~\ref{fig:Gamma_tau_ins_fit} clearly demonstrates that the decay width of the charged pion becomes wider by increasing the number density of the instantons and anti-instantons. Similarly, the lifetime of the charged pion becomes shorter by increasing the number density of the instantons and anti-instantons. This is the catalytic effect on the charged pion.


\section{Summary and conclusions}

We performed numerical computations to inspect the monopole and instanton effects in QCD on observables. To carefully check the monopole and instanton effects, in this research, we added monopoles and anti-monopoles to the configurations with larger lattice volumes and finer lattice spacings than in our previous study. We prepared normal configurations and configurations in which the monopoles and anti-monopoles were added; then, we first observed the effects of the monopoles by calculating the physical quantities with these configurations.

First, we have shown that the additional monopole and anti-monopole do not affect the scale of the lattice when calculating the lattice spacing. We then calculated the monopole density and measured the length of the monopole loops. We have shown that the monopole density increases and that the physical length of the monopole loops becomes linearly extended when increasing the values of the magnetic charges. These results indicate that the monopole creation operator makes only the long monopole loops, which are the crucial elements for the mechanism of color confinement.

Second, we calculated the eigenvalues and eigenvectors of the overlap Dirac operator using these configurations. We analytically estimated the total number of instantons and anti-instantons from the values of the topological charges. We quantitatively showed that the monopole with magnetic charge $m_{c} = 1$ and the anti-monopole with magnetic charge $m_{c} = -1$ produce one instanton or one anti-instanton. Moreover, we showed that the monopole creation operator creates the topological charges without affecting the vacuum structure by comparing the distributions of the topological charges with the predictions of the distribution functions.

Third, we confirmed that the distributions of the nearest-neighbor spacing and the spectral rigidity correspond perfectly with the results of the GUE in the GRMT, even if we add the monopoles and anti-monopoles to the configurations. Moreover, the ratios of the low-lying eigenvalues and the distributions of the first eigenvalues of each topological sector agree with the results of the GUE in the chRMT. We found that the additional monopoles and anti-monopoles do not affect the eigenvalues and change only the scale parameter $\Sigma$ of the eigenvalue distribution. The scale parameter linearly increases when increasing the magnetic charges.

These results are consistent with the results obtained in previous research~\cite{DiGH3,DiGHP1}.

In previous research~\cite{DiGH2, DiGHP1, DiGH5}, we have already shown that the values of the chiral condensate decrease and that the decay constants slightly increase with increasing magnetic charge; however, we have not explained why. In this research, we made predictions to quantitatively explain the decrease in the values of the chiral condensate and the increase in the decay constants.

We evaluated the renormalized decay constants and the renormalized chiral condensate by calculating the correlation functions of the scalar density and pseudoscalar density. We directly compared these numerical results with the predictions. We found that the values of the chiral condensate decrease in direct proportion to the square root of the number density of the instantons and anti-instantons. Moreover, the decay constant of the pseudoscalar increases in direct proportion to the one-fourth root of the number density of the instantons and anti-instantons. These results correspond to our predictions and the consequences of the phenomenological models of instantons.

The purpose of this research is to clearly show the effects of the monopoles and instantons in QCD on physical quantities, which are measured experimentally. However, it is difficult to directly determine the decay constants of the pion and kaon or their masses only through numerical calculations in quenched QCD without using the results of the chiral perturbation theory or the experimental results. Therefore, we matched the numerical results of the decay constant and the square of the pseudoscalar mass with the experimental results of the pion and kaon and determined the normalization factors. We recomputed the physical quantities using these normalization factors and evaluated the instanton effects.

We confirmed that the increases in the decay constant in the chiral limit and the decreases in the renormalized chiral condensate are consistent with the predictions. We clearly showed that the decay constants of the pion and kaon are larger than the experimental results and that the masses of the pion, kaon, and light quarks become heavier than when the number density of the instantons and anti-instantons are increased.

To quantitatively evaluate the decreases and increases in the physical quantities, we calculated the ratios of the computed results of the configuration with the additional monopoles and anti-monopoles to the computed results of the normal configurations. We demonstrated that the increase in the ratio of the chiral condensates $R_{\chi}$ when increasing the number density of the instantons and anti-instantons accords with the following relation: $R_{\chi} = \left(\frac{N_{I}^{add}}{N_{I}^{nor}}\right)^{\frac{1}{2}}$. 

We found that the mass ratios $R_{m_{q}}$ of the light quarks are consistent with this ratio $R_{\chi}$; thus, the light quark masses increase in direct proportion to the square root of the number density of the instantons and anti-instantons. Additionally, the masses and decay constants of the pion and kaon increase in direct proportion to the one-fourth root of the number density of the instantons and anti-instantons. 


Finally, we estimated the decay width and lifetime of the charged pion using the numerical results of the pion decay constant and the pion mass as the input values. We demonstrated that the decay width of the charged pion becomes wider than the experimental result and that the lifetime of the charged pion becomes shorter by increasing the number density of the instantons and anti-instantons. This is the catalytic effect on the charged pion.

These are the monopole and instanton effects of the Adriano monopole.


\appendix

\section{The definitions of the massless Wilson Dirac operator}

The massless Wilson Dirac operator $D_{W}$ is defined as follows:
\begin{align}
 & D_{W} = \frac{1}{2}\left[ \gamma_{\mu}(\nabla_{\mu}^{*} + \nabla_{\mu}) - a\nabla_{\mu}^{*}\nabla_{\mu}\right]\label{eq:def_wilson1}\\
 & [\nabla_{\mu}\psi](n) = \frac{1}{a}\left[ U_{\mu}(n)\psi(n+\hat{\mu}) - \psi(n) \right], \  [\nabla_{\mu}^{*}\psi](n) = \frac{1}{a}\left[ \psi(n) - U_{\mu}(n-\hat{\mu})^{\dagger}\psi(n-\hat{\mu}) \right]\nonumber
 \end{align}

\section{The prediction of the number of zero modes $N_{Z}^{Pre}$}\label{sec:number_zero}

We analytically calculate the number of zero modes $N_{Z}^{Pre}$ using the result of $N_{I}$~(\ref{eq:inst_dens_2}). Here, we use the notation in reference~\cite{DiGH3}. The topological charge of the normal configurations is given by $\delta$, and the total number of instantons and anti-instantons is $N$ in the expressions below.

\noindent For $m_{c} = 5$,
\begin{align}
  N_{Zero}^{Pre} & = \frac{1}{2^{5}}\left[ \langle |\delta + 5| \rangle + \langle |\delta - 5| \rangle\right] + \frac{5}{2^{5}}\left[\langle |\delta + 3| \rangle + \langle |\delta - 3| \rangle \right] \nonumber\\
  & +\frac{10}{2^{5}}\left[\langle |\delta + 1| \rangle + \langle |\delta - 1| \rangle \right] \nonumber\\
  & = \frac{1}{2^{5}}\left( \frac{4N}{\sqrt{2\pi N}}e^{-\frac{25}{2N}} + \frac{10}{\sqrt{2\pi N}}\int_{-5}^{5}e^{-\frac{\delta^{2}}{2N}} d\delta  \right) + \frac{5}{2^{5}}\left( \frac{4N}{\sqrt{2\pi N}}e^{-\frac{9}{2N}} + \frac{6}{\sqrt{2\pi N}}\int_{-3}^{3}e^{-\frac{\delta^{2}}{2N}} d\delta  \right) \nonumber \\
  & + \frac{10}{2^{5}}\left( \frac{4N}{\sqrt{2\pi N}}e^{-\frac{1}{2N}} + \frac{2}{\sqrt{2\pi N}}\int_{-1}^{1}e^{-\frac{\delta^{2}}{2N}} d\delta   \right).
\end{align}

\noindent For $m_{c} = 6$,
\begin{align}
  N_{Zero}^{Pre} & = \frac{1}{2^{6}}\left[ \langle |\delta + 6| \rangle + \langle |\delta - 6| \rangle \right] + \frac{6}{2^{6}}\left[\langle |\delta + 4| \rangle + \langle |\delta - 4| \rangle \right]\nonumber\\
  & + \frac{15}{2^{6}}\left[\langle |\delta + 2| \rangle + \langle |\delta - 2| \rangle \right] + \frac{20}{2^{6}} \langle |\delta| \rangle \nonumber\\
& = \frac{1}{2^{6}}\left(\frac{4N}{\sqrt{2\pi N}}e^{-\frac{18}{N}} + \frac{12}{\sqrt{2\pi N}}\int_{-6}^{6}e^{-\frac{\delta^{2}}{2N}} d\delta \right)  + \frac{6}{2^{6}}\left(\frac{4N}{\sqrt{2\pi N}}e^{-\frac{8}{N}} + \frac{8}{\sqrt{2\pi N}}\int_{-4}^{4}e^{-\frac{\delta^{2}}{2N}} d\delta \right)\nonumber \\ 
& + \frac{15}{2^{6}}\left(\frac{4N}{\sqrt{2\pi N}}e^{-\frac{2}{N}} + \frac{4}{\sqrt{2\pi N}}\int_{-2}^{2}e^{-\frac{\delta^{2}}{2N}} d\delta \right) + \frac{5}{8}\sqrt{\frac{N}{2\pi}}.
\end{align}

\section{The distribution functions of the topological charges $P(Q + m_{c})$}\label{sec:dis_top_func}

Here, we briefly derive the distribution functions of the topological charges $P(Q + m_{c})$~\cite{DiGH3}. We define the following distribution function for the magnetic charge $k$ as
\begin{equation}
  p_{1}(Q + k) \equiv p_{0}(Q + k)  +  p_{0}(Q - k)
\end{equation}
The distribution functions $p_{0}(Q \pm k)$ are defined by the Gaussian distribution functions as follows:
\begin{equation}
  p_{0}(Q \pm k) = \frac{\mathrm{e}^{-\frac{(Q \pm k)^{2}}{2 \langle \delta^{2}\rangle}}}{\sqrt{2\pi \langle \delta^{2}\rangle}}
\end{equation}
The distribution function for $m_{c} = 5$ is
\begin{align}
  P(Q + 5)  = \left[\frac{1}{2^{5}}p_{1}(Q + 5) + \frac{5}{2^{5}}p_{1}(Q + 3) + \frac{10}{2^{5}}p_{1}(Q + 1)\right] \left[ 1 + \mathcal{O}(V^{-1})\right]\label{eq:dis_func_mc5}.
\end{align}
\noindent For $m_{c} = 6$,
\begin{align}
 P(Q + 6)  =  \left[\frac{1}{2^{6}}p_{1}(Q + 6) + \frac{6}{2^{6}}p_{1}(Q + 4) + \frac{15}{2^{6}}p_{1}(Q + 2) + \frac{20}{2^{6}}p_{0}(Q)\right]\left[ 1 + \mathcal{O}(V^{-1})\right]\label{eq:dis_func_mc6}.
\end{align}
\clearpage

\section{The fitting results of $a^{4}G_{PS-SS}$, $am_{PS}$, and $a\rho$}\label{sec:corre_func_fit_res_1}
\vspace{-7mm}
\begin{table*}[h]
  \begin{footnotesize}
  \centering
  \begin{tabular}{|c|c|c|c|c|c|c|c|c|}\hline
    \multicolumn{9}{|c|}{Normal Conf}  \\ \hline
    $\bar{m}_{q}$ & $a\bar{m}_{q}$ & $a^{4}G_{PS-SS}$ & $am_{PS}$  & $(am_{PS})^{2}$ & $aF_{PS}$ & $a^{3}\langle \bar{\psi}\psi\rangle$ & $ FR (t/a) $ & $\chi^{2}/ d. o. f.$ \\
                {\footnotesize[MeV]} & $\times10^{-2}$ &  $\times10^{-3}$ &   & $\times10^{-2}$ & $\times10^{-2}$ & $\times10^{-3}$ & &  \\ \hline
                            30 &    1.2964 &  0.677(13) &  0.1358(10) & 1.85(3) & 3.65(10) & -0.95(3) &  7 - 25 & 15.1/17.0 \\ \hline
                            35 &    1.5125 &  0.757(16) &  0.1501(11) & 2.25(3) & 3.70(10) & -1.02(3) &  8 - 24 & 8.3/15.0 \\ \hline
                            40 &    1.7286 &  0.792(14) &  0.1606(9) & 2.58(3) & 3.77(8) & -1.06(2) &  8 - 24 & 14.2/15.0  \\ \hline
                            45 &    1.9447 &  0.825(12) &  0.1703(8) & 2.90(3) & 3.85(8) & -1.11(3) &  8 - 24 & 23.1/15.0  \\ \hline
                            50 &    2.1607 &  0.911(16) &  0.1826(10) & 3.34(4) & 3.91(9) & -1.18(3) &  9 - 23 & 9.6/13.0 \\ \hline
                            55 &    2.3768 &  0.946(15) &  0.1914(9) & 3.66(3) & 3.99(8) & -1.23(2) &  9 - 23 & 14.8/13.0 \\ \hline
                            60 &    2.5929 &  1.04(2) &  0.2027(11) & 4.11(4) & 4.06(9) & -1.31(3) & 10 - 22 & 4.9/11.0 \\ \hline
                            65 &    2.8090 &  1.077(19) &  0.2109(10) & 4.45(4) & 4.15(9) & -1.36(3) & 10 - 22 & 7.3/11.0 \\ \hline
                            70 &    3.0250 &  1.115(17) &  0.2186(9) & 4.78(4) & 4.23(8) & -1.41(3) & 10 - 22 & 10.5/11.0  \\ \hline
                            75 &    3.2411 &  1.152(16) &  0.2259(8) & 5.10(4) & 4.31(7) & -1.46(3) & 10 - 22 & 14.9/11.0 \\ \hline
                            80 &    3.4572 &  1.26(2) &  0.2361(11) & 5.57(5) & 4.47(9) & -1.57(3) & 11 - 21 & 3.8/9.0 \\ \hline
                            85 &    3.6732 &  1.30(2) &  0.2430(10) & 5.90(5) & 4.49(9) & -1.62(3) & 11 - 21 & 5.3/9.0 \\ \hline
                            90 &    3.8893 &  1.35(2) &  0.2495(9) & 6.23(5) & 4.58(9) & -1.68(3) & 11 - 21 & 7.2/9.0 \\ \hline
                            95 &    4.1054 &  1.39(2) &  0.2558(9) & 6.54(4) & 4.67(8) & -1.74(3) & 11 - 21 & 9.7/9.0 \\ \hline
                            100 &   4.3215 & 1.42(2) &  0.2617(8) & 6.85(4) & 4.76(8) & -1.80(3) & 11 - 21 & 12.8/9.0  \\ \hline
                            105 &   4.5375 & 1.56(3) &  0.2708(12) & 7.33(6) & 4.88(11) & -1.93(4) & 12 - 20 & 2.3/7.0 \\ \hline
                            110 &   4.7536 & 1.60(3) &  0.2764(11) & 7.64(6) & 4.98(11) & -1.99(4) & 12 - 20 & 3.0/7.0   \\ \hline
                            120 &   5.1858 & 1.68(3) &  0.2868(10) & 8.23(6) & 5.16(10) & -2.12(4) & 12 - 20 & 4.8/7.0 \\ \hline
                            130 &   5.6179 & 1.75(3) &  0.2961(9) & 8.77(5) & 5.35(10) & -2.24(4) & 12 - 20 & 7.5/7.0  \\ \hline
                            140 &   6.0501 & 1.93(5) &  0.3081(14) & 9.49(8) & 5.59(15) & -2.46(7) & 13 - 19 & 0.9/5.0 \\ \hline
                            150 &   6.4822 & 1.98(5) &  0.3158(12) & 9.97(8) & 5.79(14) & -2.57(6) & 13 - 19 & 1.3/5.0  \\ \hline             
                            \multicolumn{9}{|c|}{$m_{c} = 0$}  \\ \hline 
                            30 &  1.2964 &  0.676(14) &  0.1360(10) & 1.85(3) & 3.64(10) & -0.95(3) &  7 - 25 & 16.2/17.0 \\ \hline
                            35 &  1.5125 &  0.757(16) &  0.1502(11) & 2.26(3) & 3.69(10) & -1.02(3) &  8 - 24 & 8.9/15.0 \\ \hline
                            40 &  1.7286 &  0.793(14) &  0.1607(10) & 2.58(3) & 3.77(9) & -1.06(3) &  8 - 24 & 15.1/15.0 \\ \hline
                              45 &  1.9447 &  0.878(18) &  0.1735(11) & 3.01(4) & 3.83(10) & -1.13(3) &  9 - 23 & 6.5/13.0 \\ \hline
                              50 &  2.1607 &  0.914(16) &  0.1828(10) & 3.34(4) & 3.91(9) & -1.18(3) &  9 - 23 & 10.2/13.0 \\ \hline
                              55 &  2.3768 &  0.949(15) &  0.1916(9) & 3.67(3) & 3.99(8) & -1.23(2) &  9 - 23 & 15.6/13.0  \\ \hline
                              60 &  2.5929 &  1.04(2) &  0.2031(11) & 4.13(4) & 4.06(10) & -1.31(3) & 10 - 22 & 5.2/11.0 \\ \hline
                              65 &  2.8090 &  1.084(19) &  0.2112(10) & 4.46(4) & 4.14(9) & -1.36(3) & 10 - 22 & 7.7/11.0  \\ \hline
                              70 &  3.0250 &  1.122(18) &  0.2190(9) & 4.79(4) & 4.23(8) & -1.42(3) & 10 - 22 & 11.1/11.0 \\ \hline
                              75 &  3.2411 &  1.160(17) &  0.2263(8) & 5.12(4) & 4.31(7) & -1.47(3) & 10 - 22 & 15.7/11.0  \\ \hline
                              80 &  3.4572 &  1.27(3)  &  0.2366(11) & 5.60(5) & 4.41(10) & -1.57(3) & 11 - 21 & 4./9.0  \\ \hline
                              85 &  3.6732 &  1.32(2) &  0.2435(10) & 5.93(5) & 4.49(10) & -1.63(4) & 11 - 21 & 5.5/9.0 \\ \hline
                              90 &  3.8893 &  1.36(2) &  0.2501(9) & 6.25(5) & 4.58(9) & -1.69(3) & 11 - 21 & 7.5/9.0 \\ \hline
                              95 &  4.1054 &  1.40(2) &  0.2563(9) & 6.58(4) & 4.67(8) & -1.75(3) & 11 - 21 & 10.1/9.0 \\ \hline
                              100 & 4.3215 & 1.44(2) &  0.2623(8) & 6.88(4) & 4.76(8) & -1.80(3) & 11 - 21 & 13.3/9.0  \\ \hline
                              105 & 4.5375 & 1.57(3) &  0.2715(12) & 7.37(6) & 4.88(11) & -1.94(4) & 12 - 20 & 2.4/7.0  \\ \hline
                              110 & 4.7536 & 1.62(3) &  0.2771(11) & 7.68(6) & 4.98(11) & -2.00(4) & 12 - 20 & 3.1/7.0 \\ \hline
                              120 & 5.1858 & 1.69(3) &  0.2874(10) & 8.27(6) & 5.17(10) & -2.13(4) & 12 - 20 & 5.0/7.0  \\ \hline
                              130 & 5.6179 & 1.76(3) &  0.2967(9) & 8.80(5) & 5.36(10) & -2.25(4) & 12 - 20 & 7.7/7.0  \\ \hline
                              140 & 6.0501 & 1.95(5) &  0.3087(14) & 9.53(9) & 5.60(15) & -2.47(7) & 13 - 19 & 0.9/5.0   \\ \hline
                              150 & 6.4822 & 2.00(5) &  0.3163(13) & 10.01(8) & 5.79(14) & -2.59(6) & 13 - 19 & 1.30/5.00 \\ \hline
              \end{tabular}
  \end{footnotesize}
  \caption{The fitting results of $a^{4}G_{PS-SS}$ and $am_{PS}$ together with the analytic results of the square of the pseudoscalar mass $(am_{PS})^{2}$, decay constant $aF_{PS}$, and chiral condensate $a^{3}\langle \bar{\psi}\psi\rangle$. The configurations are the normal configuration and the configuration of $m_{c} = 0$.}\label{tb:numerical_results_or_mc0}
\end{table*}

\begin{table*}[tbp]
  \begin{footnotesize}
  \centering
  \begin{tabular}{|c|c|c|c|c|c|c|c|c|} \hline
                \multicolumn{9}{|c|}{$m_{c} = 1$}  \\ \hline
                $\bar{m}_{q}$ & $a\bar{m}_{q}$ & $a^{4}G_{PS-SS}$ & $am_{PS}$  & $(am_{PS})^{2}$ & $aF_{PS}$ & $a^{3}\langle \bar{\psi}\psi\rangle$ & $ FR (t/a) $ & $\chi^{2}/ d. o. f.$ \\
                                  {\footnotesize[MeV]} & $\times10^{-2}$ &  $\times10^{-3}$ &   & $\times10^{-2}$ & $\times10^{-2}$ & $\times10^{-3}$ & &  \\ \hline
                            30  &    1.2964 & 0.687(13) &  0.1348(10) & 1.82(3) & 3.74(11) & -0.98(3) &  7 - 25 & 16.8/17.0 \\ \hline   
                            35  &    1.5125 & 0.770(16) &  0.1492(11) & 2.22(3) & 3.77(10) & -1.05(3) &  8 - 24 & 9.5/15.0 \\ \hline    
                            40  &    1.7286 & 0.805(14) &  0.1597(9) & 2.55(3) & 3.85(9) & -1.09(3) &  8 - 24 & 16.1/15.0 \\ \hline   
                            45  &    1.9447 & 0.890(18) &  0.1725(11) & 2.98(4) & 3.90(10) & -1.16(3) &  9 - 23 & 7.0/13.0 \\ \hline    
                            50  &    2.1607 & 0.925(16) &  0.1819(9) & 3.31(3) & 3.97(9) & -1.21(3) &  9 - 23 & 11.0/13.0 \\ \hline   
                            55  &    2.3768 & 0.959(15) &  0.1907(8) & 3.64(3) & 4.05(8) & -1.25(2) &  9 - 23 & 16.7/13.0  \\ \hline  
                            60  &    2.5929 & 1.05(2) &  0.2021(10) & 4.08(4) & 4.12(10) & -1.34(3) & 10 - 22 & 5.6/11.0 \\ \hline    
                            65  &    2.8090 & 1.090(19) &  0.2102(9) & 4.42(4) & 4.20(9) & -1.39(3) & 10 - 22 & 8.2/11.0 \\ \hline    
                            70  &    3.0250 & 1.127(17) &  0.2179(9) & 4.75(4) & 4.28(8) & -1.44(3) & 10 - 22 & 11.8/11.0 \\ \hline   
                            75  &    3.2411 & 1.162(16) &  0.2252(8) & 5.07(4) & 4.36(7) & -1.49(3) & 10 - 22 & 16.6/11.0 \\ \hline   
                            80  &    3.4572 & 1.27(2) &  0.2354(11) & 5.54(5) & 4.45(9) & -1.59(3) & 11 - 21 & 4.3/9.0  \\ \hline    
                            85  &    3.6732 & 1.31(2) &  0.2422(10) & 5.87(5) & 4.54(9) & -1.64(4) & 11 - 21 & 5.9/9.0 \\ \hline     
                            90  &    3.8893 & 1.35(2) &  0.2488(9) & 6.19(5) & 4.62(9) & -1.70(3) & 11 - 21 & 8.0/9.0 \\ \hline     
                            95  &    4.1054 & 1.39(2) &  0.2550(8) & 6.50(4) & 4.71(8) & -1.76(3) & 11 - 21 & 10.7/9.0 \\ \hline    
                            100 &   4.3215 & 1.52(3) &  0.2642(12) & 6.98(6) & 4.82(12) & -1.88(5) & 12 - 20 & 1.9/7.0  \\ \hline    
                            105 &   4.5375 & 1.56(3) &  0.2700(11) & 7.29(6) & 4.92(11) & -1.94(4) & 12 - 20 & 2.5/7.0  \\ \hline    
                            110 &   4.7536 & 1.60(3) &  0.2756(11) & 7.59(6) & 5.01(10) & -2.00(4) & 12 - 20 & 3.2/7.0 \\ \hline     
                            120 &   5.1858 & 1.67(3) &  0.2858(10) & 8.17(5) & 5.19(9) & -2.12(4) & 12 - 20 & 5.3/7.0  \\ \hline    
                            130 &   5.6179 & 1.74(3) &  0.2951(9) & 8.71(5) & 5.38(9) & -2.24(4) & 12 - 20 & 8.1/7.0 \\ \hline     
                            140 &   6.0501 & 1.91(5) &  0.3070(14) & 9.43(8) & 5.61(15) & -2.46(6) & 13 - 19 & 0.9/5.0 \\ \hline     
                            150 &   6.4822 & 1.96(4) &  0.3145(12) & 9.89(8) & 5.80(14) & -2.57(6) & 13 - 19 & 1.4/5.0 \\ \hline 
                            \multicolumn{9}{|c|}{$m_{c} = 2$}  \\ \hline 
                                        30 &  1.2964 & 0.771(15) &  0.1376(10) & 1.89(3) & 3.80(10) & -1.06(3) &  7 - 25 & 14.0/17.0 \\ \hline          
                                        35 &  1.5125 & 0.805(13) &  0.1487(9) & 2.21(3) & 3.88(9) & -1.10(3) &  7 - 25 & 25.3/17.0  \\ \hline         
                                        40 &  1.7286 & 0.890(16) &  0.1620(9) & 2.62(3) & 3.93(9) & -1.17(3) &  8 - 24 & 13.3/15.0  \\ \hline         
                                        45 &  1.9447 & 0.921(14) &  0.1717(8) & 2.95(3) & 4.01(9) & -1.22(3) &  8 - 24 & 21.45/15.0 \\ \hline         
                                        50 &  2.1607 & 1.009(18) &  0.1838(10) & 3.38(4) & 4.06(9) & -1.29(3) &  9 - 23 & 9.1/13.0 \\ \hline           
                                        55 &  2.3768 & 1.042(16) &  0.1925(9) & 3.70(3) & 4.14(8) & -1.34(3) &  9 - 23 & 13.9/13.0  \\ \hline         
                                        60 &  2.5929 & 1.14(2) &  0.2036(11) & 4.15(4) & 4.21(9) & -1.42(3) & 10 - 22 & 4.8/11.0\\ \hline            
                                        65 &  2.8090 & 1.17(2) &  0.2116(10) & 4.48(4) & 4.29(9) & -1.47(3) & 10 - 22 & 7.0/11.0  \\ \hline          
                                        70 &  3.0250 & 1.207(19) &  0.2193(9) & 4.81(4) & 4.37(8) & -1.52(3) & 10 - 22 & 10.1/11.0 \\ \hline          
                                        75 &  3.2411 & 1.242(18) &  0.2266(8) & 5.13(4) & 4.45(8) & -1.57(3) & 10 - 22 & 14.2/11.0 \\ \hline          
                                        80 &  3.4572 & 1.35(3) &  0.2366(11) & 5.60(5) & 4.54(10) & -1.67(4) & 11 - 21 & 3.67/9.0  \\ \hline          
                                        85 &  3.6732 & 1.39(3) &  0.2434(10) & 5.93(5) & 4.63(9) & -1.73(3) & 11 - 21 & 5.1/9.0  \\ \hline           
                                        90 &  3.8893 & 1.43(2) &  0.2499(9) & 6.25(5) & 4.71(9) & -1.78(4) & 11 - 21 & 6.9/9.0  \\ \hline           
                                        95 &  4.1054 & 1.47(2) &  0.2562(9) & 6.56(4) & 4.80(9) & -1.84(3) & 11 - 21 & 9.2/9.0 \\ \hline            
                                        100 & 4.3215 & 1.51(2) &  0.2621(8) & 6.87(4) & 4.88(8) & -1.89(3) & 11 - 21 & 12.1/9.0 \\ \hline           
                                        105 & 4.5375 & 1.64(4) &  0.2711(12) & 7.35(6) & 5.00(12) & -2.02(5) & 12 - 20 & 2.2/7.0 \\ \hline            
                                        110 & 4.7536 & 1.68(3) &  0.2767(11) & 7.66(6) & 5.09(11) & -2.09(5) & 12 - 20 & 2.8/7.0  \\ \hline           
                                        120 & 5.1858 & 1.75(3) &  0.2870(10) & 8.24(6) & 5.27(10) & -2.21(4) & 12 - 20 & 4.6/7.0 \\ \hline            
                                        130 & 5.6179 & 1.82(3) &  0.2963(9) & 8.78(5) & 5.45(10) & -2.32(4) & 12 - 20 & 7.0/7.0 \\ \hline            
                                        140 & 6.0501 & 1.87(3) &  0.3047(8) & 9.28(5) & 5.63(9) & -2.43(4) & 12 - 20 & 10.4/7.0\\ \hline            
                                        150 & 6.4822 & 2.04(5) &  0.3159(13) & 9.98(8) & 5.87(14) & -2.65(6) & 13 - 19 & 1.2/5.0\\ \hline 
  \end{tabular}
  \end{footnotesize}
  \caption{The fitting results of $a^{4}G_{PS-SS}$ and $am_{PS}$ together with the analytic results of the square of the pseudoscalar mass $(am_{PS})^{2}$, decay constant $aF_{PS}$, and chiral condensate $a^{3}\langle \bar{\psi}\psi\rangle$. The magnetic charges of the configurations are $m_{c} = 1$ and $m_{c} = 2$.}\label{tb:numerical_results_mc1_2}
\end{table*}

\begin{table*}[tbp]
  \begin{footnotesize}
    \centering
    \begin{tabular}{|c|c|c|c|c|c|c|c|c|}  \hline
      \multicolumn{9}{|c|}{$m_{c} = 3$}  \\  \hline
      $\bar{m}_{q}$ & $a\bar{m}_{q}$ & $a^{4}G_{PS-SS}$ & $am_{PS}$  & $(am_{PS})^{2}$ & $aF_{PS}$ & $a^{3}\langle \bar{\psi}\psi\rangle$ & $ FR (t/a) $ & $\chi^{2}/ d. o. f.$ \\
                  {\footnotesize[MeV]} & $\times10^{-2}$ &  $\times10^{-3}$ &   & $\times10^{-2}$ & $\times10^{-2}$ & $\times10^{-3}$ & &  \\ \hline
                            30 &    1.2964 & 0.810(16) &  0.1383(10) & 1.91(3) & 3.86(10) & -1.10(3) &  7 - 25 & 11.5/17.0 \\ \hline             
                            35 &    1.5125 & 0.849(13) &  0.1498(9) & 2.24(3) & 3.93(9) & -1.15(3) &  7 - 25 & 22.0/17.0  \\ \hline            
                            40 &    1.7286 & 0.936(16) &  0.1632(9) & 2.66(3) & 3.97(9) & -1.21(3) &  8 - 24 & 11.8/15.0 \\ \hline             
                            45 &    1.9447 & 0.970(14) &  0.1731(8) & 3.00(3) & 4.04(9) & -1.26(3) &  8 - 24 & 19.7/15.0  \\ \hline            
                            50 &    2.1607 & 1.059(18) &  0.1852(10) & 3.43(4) & 4.10(9) & -1.34(3) &  9 - 23 & 8.4/13.0 \\ \hline              
                            55 &    2.3768 & 1.094(17) &  0.1940(9) & 3.76(3) & 4.18(8) & -1.38(3) &  9 - 23 & 13.1/13.0 \\ \hline             
                            60 &    2.5929 & 1.126(15) &  0.2023(8) & 4.09(3) & 4.25(7) & -1.43(2) &  9 - 23 & 19.8/13.0 \\ \hline             
                            65 &    2.8090 & 1.23(2) &  0.2131(9) & 4.54(4) & 4.33(9) & -1.52(3) & 10 - 22 & 6.8/11.0 \\ \hline              
                            70 &    3.0250 & 1.262(19) &  0.2208(9) & 4.87(4) & 4.41(8) & -1.57(3) & 10 - 22 & 9.9/11.0 \\ \hline              
                            75 &    3.2411 & 1.297(18) &  0.2281(8) & 5.20(4) & 4.49(7) & -1.62(3) & 10 - 22 & 14.1/11.0 \\ \hline             
                            80 &    3.4572 & 1.41(3) &  0.2380(10) & 5.66(5) & 4.58(10) & -1.72(4) & 11 - 21 & 3.7/9.0 \\ \hline               
                            85 &    3.6732 & 1.45(3) &  0.2448(10) & 5.99(5) & 4.66(9) & -1.77(3) & 11 - 21 & 5.2/9.0 \\ \hline               
                            90 &    3.8893 & 1.49(2) &  0.2513(9) & 6.32(5) & 4.75(9) & -1.83(4) & 11 - 21 & 7.1/9.0 \\ \hline               
                            95 &    4.1054 & 1.52(2) &  0.2575(8) & 6.63(4) & 4.83(8) & -1.89(3) & 11 - 21 & 9.6/9.0  \\ \hline              
                            100 &   4.3215 & 1.56(2) &  0.2634(8) & 6.94(4) & 4.92(8) & -1.94(3) & 11 - 21 & 12.7/9.0  \\ \hline             
                            105 &   4.5375 & 1.59(2) &  0.2690(7) & 7.42(6) & 5.03(11) & -2.07(5) & 12 - 20 & 2.3/7.0 \\ \hline               
                            110 &   4.7536 & 1.73(3) &  0.2779(11) & 7.72(6) & 5.12(11) & -2.13(4) & 12 - 20 & 3.0/7.0 \\ \hline               
                            120 &   5.1858 & 1.80(3) &  0.2881(9) & 8.30(5) & 5.30(10) & -2.25(4) & 12 - 20 & 5.0/7.0  \\ \hline              
                            130 &   5.6179 & 1.86(3) &  0.2973(8) & 8.84(5) & 5.48(9) & -2.36(4) & 12 - 20 & 7.7/7.0 \\ \hline               
                            140 &   6.0501 & 2.04(5) &  0.3091(13) & 9.55(8) & 5.71(15) & -2.58(7) & 13 - 19 & 0.9/5.0 \\ \hline               
                            150 &   6.4822 & 2.08(5) &  0.3166(12) & 10.03(8) & 5.89(14) & -2.69(6) & 13 - 19 & 1.3/5.0  \\ \hline
                            \multicolumn{9}{|c|}{$m_{c} = 4$}  \\ \hline 
                            30 &  1.2964 & 0.849(15) &  0.1393(9) & 1.94(3) & 3.89(10) & -1.13(3) &  7 - 25 & 18.3/17.0  \\ \hline            
                                        35 &  1.5125 & 0.936(17) &  0.1532(10) & 2.35(3) & 3.94(10) & -1.21(3) &  8 - 24 & 10.9/15.0 \\ \hline             
                                        40 &  1.7286 & 0.968(15) &  0.1635(8) & 2.67(3) & 4.02(9) & -1.25(3) &  8 - 24 & 19.0/15.0 \\ \hline             
                                        45 &  1.9447 & 1.056(19) &  0.1760(10) & 3.10(3) & 4.08(9) & -1.33(3) &  9 - 23 & 8.8/13.0 \\ \hline              
                                        50 &  2.1607 & 1.086(17) &  0.1850(9) & 3.42(3) & 4.16(8) & -1.37(3) &  9 - 23 & 13.9/13.0 \\ \hline             
                                        55 &  2.3768 & 1.18(2) &  0.1964(10) & 3.86(4) & 4.23(10) & -1.45(4) & 10 - 22 & 5.1/11.0 \\ \hline              
                                        60 &  2.5929 & 1.21(2) &  0.2046(9) & 4.18(4) & 4.31(9) & -1.50(3) & 10 - 22 & 7.6/11.0 \\ \hline              
                                        65 &  2.8090 & 1.241(19) &  0.2123(8) & 4.51(4) & 4.39(8) & -1.55(3) & 10 - 22 & 11.1/11.0 \\ \hline             
                                        70 &  3.0250 & 1.270(18) &  0.2196(8) & 4.82(3) & 4.47(7) & -1.59(3) & 10 - 22 & 15.8/11.0 \\ \hline             
                                        75 &  3.2411 & 1.38(3) &  0.2297(10) & 5.28(5) & 4.56(9) & -1.69(4) & 11 - 21 & 4.3/9.0 \\ \hline               
                                        80 &  3.4572 & 1.41(2) &  0.2365(9) & 5.59(4) & 4.64(10) & -1.74(4) & 11 - 21 & 6.0/9.0 \\ \hline               
                                        85 &  3.6732 & 1.44(2) &  0.2430(9) & 5.90(4) & 4.72(9) & -1.79(3) & 11 - 21 & 8.1/9.0 \\ \hline               
                                        90 &  3.8893 & 1.47(2) &  0.2491(8) & 6.21(4) & 4.81(8) & -1.84(3) & 11 - 21 & 10.9/9.0 \\ \hline              
                                        95 &  4.1054 & 1.59(4) &  0.2582(12) & 6.67(6) & 4.91(12) & -1.96(5) & 12 - 20 & 2.0/7.0 \\ \hline               
                                        100 & 4.3215 & 1.62(3) &  0.2639(11) & 6.97(6) & 5.00(11) & -2.02(4) & 12 - 20 & 2.7/7.0 \\ \hline               
                                        105 & 4.5375 & 1.65(3) &  0.2694(10) & 7.26(6) & 5.09(10) & -2.07(4) & 12 - 20 & 3.4/7.0 \\ \hline               
                                        110 & 4.7536 & 1.68(3) &  0.2746(10) & 7.54(5) & 5.17(10) & -2.12(4) & 12 - 20 & 4.3/7.0 \\ \hline               
                                        120 & 5.1858 & 1.73(3) &  0.2841(9) & 8.07(5) & 5.34(10) & -2.22(4) & 12 - 20 & 6.7/7.0 \\ \hline               
                                        130 & 5.6179 & 1.77(3) &  0.2926(8) & 8.56(5) & 5.51(9) & -2.32(4) & 12 - 20 & 9.9/7.0 \\ \hline               
                                        140 & 6.0501 & 1.92(5) &  0.3040(13) & 9.24(8) & 5.73(14) & -2.51(6) & 13 - 19 & 1.2/5.0 \\ \hline               
                                        150 & 6.4822 & 1.94(4) &  0.3109(12) & 9.66(7) & 5.90(13) & -2.60(6) & 13 - 19 & 1.6/5.0 \\ \hline
              \end{tabular}
  \end{footnotesize} 
  \caption{The fitting results of $a^{4}G_{PS-SS}$ and $am_{PS}$ together with the analytic results of the square of the pseudoscalar mass $(am_{PS})^{2}$, decay constant $aF_{PS}$, and chiral condensate $a^{3}\langle \bar{\psi}\psi\rangle$. The magnetic charges of the configurations are $m_{c} = 3$ and $m_{c} = 4$.}\label{tb:numerical_results_mc3_4}
\end{table*}
\begin{table*}[tbp]
  \begin{footnotesize}
    \centering
    \begin{tabular}{|c|c|c|c|c|c|c|c|c|} \hline
      \multicolumn{9}{|c|}{$m_{c} = 5$}  \\ \hline
      $\bar{m}_{q}$ & $a\bar{m}_{q}$ & $a^{4}G_{PS-SS}$ & $am_{PS}$  & $(am_{PS})^{2}$ & $aF_{PS}$ & $a^{3}\langle \bar{\psi}\psi\rangle$ & $ FR (t/a) $ & $\chi^{2}/ d. o. f.$ \\
                  {\footnotesize[MeV]} & $\times10^{-2}$ &  $\times10^{-3}$ &   & $\times10^{-2}$ & $\times10^{-2}$ & $\times10^{-3}$ & &  \\ \hline
                            30 &    1.2964 & 0.896(17) &  0.1406(10) &  1.98(3) & 3.93(10) & -1.18(3) &  7 - 25 & 13.6/17.0 \\ \hline            
                            35 &    1.5125 & 0.929(14) &  0.1516(8) &  2.30(3) & 4.01(9) & -1.22(3) &  7 - 25 & 24.5/17.0 \\ \hline            
                            40 &    1.7286 & 1.016(17) &  0.1648(9) &  2.71(3) & 4.06(9) & -1.29(3) &  8 - 24 & 13.2/15.0 \\ \hline            
                            45 &    1.9447 & 1.046(15) &  0.1744(8) &  3.04(3) & 4.14(9) & -1.34(3) &  8 - 24 & 21.2/15.0 \\ \hline            
                            50 &    2.1607 & 1.137(19) &  0.1863(9) &  3.47(3) & 4.20(9) & -1.42(3) &  9 - 23 & 9.1/13.0  \\ \hline            
                            55 &    2.3768 & 1.168(17) &  0.1950(8) &  3.80(3) & 4.27(8) & -1.46(3) &  9 - 23 & 13.8/13.0 \\ \hline            
                            60 &    2.5929 & 1.26(2) &  0.2060(10) &  4.24(4) & 4.35(9) & -1.55(3) & 10 - 22 & 4.8/11.0  \\ \hline            
                            65 &    2.8090 & 1.30(2) &  0.2140(9) &  4.58(4) & 4.42(9) & -1.59(3) & 10 - 22 & 7.0/11.0 \\ \hline             
                            70 &    3.0250 & 1.33(2) &  0.2216(8) &  4.91(4) & 4.50(8) & -1.64(3) & 10 - 22 & 9.9/11.0 \\ \hline             
                            75 &    3.2411 & 1.368(19) &  0.2288(8) &  5.24(4) & 4.58(7) & -1.69(3) & 10 - 22 & 14.0/11.0 \\ \hline            
                            80 &    3.4572 & 1.48(3) &  0.2387(10) &  5.70(5) & 4.67(10) & -1.80(4) & 11 - 21 & 3.6/9.0 \\ \hline              
                            85 &    3.6732 & 1.52(3) &  0.2455(10) &  6.03(5) & 4.75(9) & -1.85(4) & 11 - 21 & 5.0/9.0 \\ \hline              
                            90 &    3.8893 & 1.56(3) &  0.2520(9) &  6.35(5) & 4.83(9) & -1.91(4) & 11 - 21 & 6.8/9.0 \\ \hline              
                            95 &    4.1054 & 1.59(2) &  0.2582(8) &  6.67(4) & 4.91(8) & -1.96(4) & 11 - 21 & 9.0/9.0 \\ \hline              
                            100 &   4.3215 & 1.63(2) &  0.2641(8) &  6.97(4) & 5.00(8) & -2.02(3) & 11 - 21 & 11.9/9.0  \\ \hline            
                            105 &   4.5375 & 1.66(2) &  0.2697(7) &  7.45(6) & 5.11(11) & -2.14(5) & 12 - 20 & 2.1/7.0 \\ \hline              
                            110 &   4.7536 & 1.80(4) &  0.2784(11) &  7.75(6) & 5.20(11) & -2.20(5) & 12 - 20 & 2.8/7.0 \\ \hline              
                            120 &   5.1858 & 1.86(3) &  0.2887(10) &  8.33(6) & 5.37(10) & -2.32(4) & 12 - 20 & 4.5/7.0 \\ \hline              
                            130 &   5.6179 & 1.92(3) &  0.2979(9) &  8.88(5) & 5.55(10) & -2.43(4) & 12 - 20 & 6.9/7.0 \\ \hline              
                            140 &   6.0501 & 1.96(3) &  0.3062(8) &  9.37(5) & 5.72(9) & -2.53(4) & 12 - 20 & 10.1/7.0  \\ \hline            
                            150 &   6.4822 & 2.13(5) &  0.3172(12) &  10.06(8) & 5.95(14) & -2.74(7) & 13 - 19 & 1.2/5.0 \\ \hline
                            \multicolumn{9}{|c|}{$m_{c} = 6$}  \\ \hline 
                                        30 &  1.2964 & 0.870(16) &  0.1389(10) &  1.93(3) & 3.96(10) & -1.17(3) &  7 - 25 & 9.6/17.0 \\ \hline             
                                        35 &  1.5125 & 0.910(14) &  0.1504(8) &  2.26(2) & 4.03(9) & -1.21(3) &  7 - 25 & 19.0/17.0 \\ \hline            
                                        40 &  1.7286 & 0.996(17) &  0.1636(9) &  2.68(3) & 4.08(9) & -1.29(3) &  8 - 24 & 10.4/15.0 \\ \hline            
                                        45 &  1.9447 & 1.032(15) &  0.1736(8) &  3.01(3) & 4.15(7) & -1.33(2) &  8 - 24 & 17.8/15.0 \\ \hline            
                                        50 &  2.1607 & 1.122(19) &  0.1856(9) &  3.44(3) & 4.20(9) & -1.41(3) &  9 - 23 & 7.8/13.0 \\ \hline             
                                        55 &  2.3768 & 1.158(17) &  0.1945(8) &  3.78(3) & 4.28(8) & -1.46(3) &  9 - 23 & 12.4/13.0 \\ \hline            
                                        60 &  2.5929 & 1.192(15) &  0.2029(7) &  4.12(3) & 4.35(7) & -1.50(2) &  9 - 23 & 19.1/13.0 \\ \hline            
                                        65 &  2.8090 & 1.29(2) &  0.2137(9) &  4.57(4) & 4.42(9) & -1.59(3) & 10 - 22 & 6.7/11.0 \\ \hline             
                                        70 &  3.0250 & 1.330(19) &  0.2214(8) &  4.90(4) & 4.50(8) & -1.64(3) & 10 - 22 & 9.8/11.0  \\ \hline            
                                        75 &  3.2411 & 1.366(18) &  0.2288(8) &  5.23(3) & 4.58(7) & -1.69(3) & 10 - 22 & 14.1/11.0  \\ \hline           
                                        80 &  3.4572 & 1.48(3) &  0.2386(10) &  5.70(5) & 4.67(9) & -1.80(4) & 11 - 21 & 3.8/9.0 \\ \hline              
                                        85 &  3.6732 & 1.52(3) &  0.2455(9) &  6.03(5) & 4.75(9) & -1.85(3) & 11 - 21 & 5.3/9.0  \\ \hline             
                                        90 &  3.8893 & 1.56(2) &  0.2520(9) &  6.35(4) & 4.83(9) & -1.91(4) & 11 - 21 & 7.3/9.0 \\ \hline              
                                        95 &  4.1054 & 1.59(2) &  0.2583(8) &  7.00(4) & 4.92(8) & -1.96(3) & 11 - 21 & 9.9/9.0 \\ \hline              
                                        100 & 4.3215 & 1.63(2) &  0.2642(7) &  6.98(4) & 5.00(8) & -2.02(3) & 11 - 21 & 13.1/9.0 \\ \hline             
                                        105 & 4.5375 & 1.76(3) &  0.2730(11) &  7.45(6) & 5.11(11) & -2.15(4) & 12 - 20 & 2.4/7.0  \\ \hline             
                                        110 & 4.7536 & 1.80(3) &  0.2785(10) &  7.76(6) & 5.20(10) & -2.21(4) & 12 - 20 & 3.1/7.0 \\ \hline              
                                        120 & 5.1858 & 1.87(3) &  0.2888(9) &  8.34(5) & 5.38(9) & -2.32(4) & 12 - 20 & 5.1/7.0 \\ \hline              
                                        130 & 5.6179 & 1.93(3) &  0.2981(8) &  8.88(5) & 5.60(9) & -2.44(4) & 12 - 20 & 7.9/7.0  \\ \hline             
                                        140 & 6.0501 & 2.10(5) &  0.3098(13) &  9.59(8) & 5.78(14) & -2.64(6) & 13 - 19 & 0.9/5.0 \\ \hline              
                                        150 & 6.4822 & 2.14(5) &  0.3173(12) &  10.07(7) & 5.95(13) & -2.75(6) & 13 - 19 & 1.4/5.0 \\ \hline 
              \end{tabular}
  \end{footnotesize}
  \caption{The fitting results of $a^{4}G_{PS-SS}$ and $am_{PS}$ together with the analytic results of the square of the pseudoscalar mass $(am_{PS})^{2}$, decay constant $aF_{PS}$, and chiral condensate $a^{3}\langle \bar{\psi}\psi\rangle$. The magnetic charges of the configurations are $m_{c} = 5$ and $m_{c} = 6$.}\label{tb:numerical_results_mc5_6}
\end{table*}

\begin{table*}[tbp]
  \begin{footnotesize}
  \centering
  \begin{tabular}{|c|c|c|c|c|c|c|c|c|c|} \hline
    \multicolumn{5}{|c|}{Normal Conf}  & \multicolumn{5}{|c|}{$m_{c} = 3$}  \\ \hline
                $\bar{m}_{q}$ & $a\bar{m}_{q}$ & $a\rho$ & $ FR (t/a) $ & $\chi^{2}/ d. o. f.$  &   $\bar{m}_{q}$ & $a\bar{m}_{q}$ & $a\rho$ & $ FR (t/a) $ & $\chi^{2}/ d. o. f.$\\
                            {\footnotesize[MeV]} & $\times10^{-2}$  & $\times10^{-2}$ &   & &  {\footnotesize[MeV]} & $\times10^{-2}$  & $\times10^{-2}$ &  &  \\ \hline
                            30 &    1.2964 & 0.9243(3) & 13 - 19 &  18.6/6.0     &   30 &    1.2964 &  0.9031(3) & 13 - 19 &  45.7/6.0 \\ \hline            
                            35 &    1.5125 & 1.0801(3) & 13 - 19 &  32.4/6.0     &   35 &    1.5125 &  1.0554(3) & 13 - 19 &  83.3/6.0  \\ \hline              
                            40 &    1.7286 & 1.2363(4) & 13 - 19 &  61.0/6.0     &   40 &    1.7286 &  1.2082(4) & 13 - 19 &  140.2/6.0 \\ \hline              
                            45 &    1.9447 & 1.3928(4) & 13 - 19 &  108.3/6.0    &   45 &    1.9447 &  1.3612(4) & 13 - 19 &  219.4/6.0 \\ \hline              
                            50 &    2.1607 & 1.5495(4) & 13 - 19 &  177.3/6.0    &   50 &    2.1607 &  1.5144(4) & 13 - 19 &  322.6/6.0 \\ \hline              
                            55 &    2.3768 & 1.7061(5) & 13 - 19 &  269.0/6.0    &   55 &    2.3768 &  1.6676(5) & 13 - 19 &  449.9/6.0 \\ \hline              
                            60 &    2.5929 & 1.8625(5) & 13 - 19 &  383.1/6.0    &   60 &    2.5929 &  1.8206(5) & 13 - 19 &  600.3/6.0 \\ \hline              
                            65 &    2.8090 & 2.0185(6) & 13 - 19 &  517.5/6.0    &   65 &    2.8090 &  1.9733(6) & 13 - 19 &  771.6/6.0 \\ \hline              
                            70 &    3.0250 & 2.1739(6) & 13 - 19 &  669.1/6.0    &   70 &    3.0250 &  2.1254(6) & 13 - 19 &  960.5/6.0 \\ \hline              
                            75 &    3.2411 & 2.3284(6) & 13 - 19 &  833.9/6.0    &   75 &    3.2411 &  2.2768(6) & 13 - 19 &  1163.1/6.0 \\ \hline
  \multicolumn{5}{|c|}{$m_{c} = 0$}  & \multicolumn{5}{|c|}{$m_{c} = 4$}  \\ \hline
                                        30 &    1.2964 & 0.9256(3) & 13 - 19 &  22.2/6.0     & 30 &    1.2964 &  0.8926(3) & 13 - 19 &  404.3/6.0 \\             
                                        35 &    1.5125 & 1.0815(3) & 13 - 19 &  39.5/6.0     & 35 &    1.5125 &  1.0430(3) & 13 - 19 &  579.6/6.0 \\ \hline             
                                        40 &    1.7286 & 1.2380(3) & 13 - 19 &  74.4/6.0     & 40 &    1.7286 &  1.1937(3) & 13 - 19 &  800.4/6.0 \\ \hline             
                                        45 &    1.9447 & 1.3947(4) & 13 - 19 &  131.2/6.0    & 45 &    1.9447 &  1.3447(4) & 13 - 19 &  1066.7/6.0 \\ \hline            
                                        50 &    2.1607 & 1.5515(4) & 13 - 19 &  212.8/6.0    & 50 &    2.1607 &  1.4959(4) & 13 - 19 &  1375.5/6.0 \\ \hline            
                                        55 &    2.3768 & 1.7083(4) & 13 - 19 &  319.6/6.0    & 55 &    2.3768 &  1.6470(4) & 13 - 19 &  1721.8/6.0 \\ \hline            
                                        60 &    2.5929 & 1.8649(5) & 13 - 19 &  450.5/6.0    & 60 &    2.5929 &  1.7981(5) & 13 - 19 &  2098.6/6.0 \\ \hline            
                                        65 &    2.8090 & 2.0211(5) & 13 - 19 &  602.2/6.0    & 65 &    2.8090 &  1.9488(5) & 13 - 19 &  2497.3/6.0 \\ \hline            
                                        70 &    3.0250 & 2.1766(6) & 13 - 19 &  770.6/6.0    & 70 &    3.0250 &  2.0990(5) & 13 - 19 &  2908.7/6.0 \\ \hline            
                                        75 &    3.2411 & 2.3314(6) & 13 - 19 &  950.9/6.0    & 75 &    3.2411 &  2.2485(6) & 13 - 19 &  3323.0/6.0 \\ \hline
                                          \multicolumn{5}{|c|}{$m_{c} = 1$}  & \multicolumn{5}{|c|}{$m_{c} = 5$}  \\ \hline
                            30 &    1.2964 & 0.9221(3) & 13 - 19 &  38.2/6.0   &   30 &    1.2964 & 0.8857(3) & 13 - 19 &  71.0/6.0 \\ \hline               
                            35 &    1.5125 & 1.0775(3) & 13 - 19 &  70.6/6.0   &   35 &    1.5125 & 1.0350(3) & 13 - 19 &  126.2/6.0 \\ \hline              
                            40 &    1.7286 & 1.2333(4) & 13 - 19 &  120.4/6.0  &   40 &    1.7286 & 1.1847(3) & 13 - 19 &  209.6/6.0 \\ \hline              
                            45 &    1.9447 & 1.3893(4) & 13 - 19 &  190.1/6.0  &   45 &    1.9447 & 1.3347(3) & 13 - 19 &  326.8/6.0 \\ \hline              
                            50 &    2.1607 & 1.5456(4) & 13 - 19 &  280.7/6.0  &   50 &    2.1607 & 1.4849(4) & 13 - 19 &  482.3/6.0 \\ \hline              
                            55 &    2.3768 & 1.7017(5) & 13 - 19 &  391.8/6.0  &   55 &    2.3768 & 1.6351(4) & 13 - 19 &  678.4/6.0 \\ \hline              
                            60 &    2.5929 & 1.8577(5) & 13 - 19 &  521.8/6.0  &   60 &    2.5929 & 1.7852(4) & 13 - 19 &  915.0/6.0 \\ \hline              
                            65 &    2.8090 & 2.0133(6) & 13 - 19 &  667.9/6.0  &   65 &    2.8090 & 1.9350(5) & 13 - 19 &  1189.4/6.0 \\ \hline             
                            70 &    3.0250 & 2.1683(6) & 13 - 19 &  826.5/6.0  &   70 &    3.0250 & 2.0843(5) & 13 - 19 &  1496.4/6.0 \\ \hline             
                            75 &    3.2411 & 2.3225(7) & 13 - 19 &  993.5/6.0  &   75 &    3.2411 & 2.2329(5) & 13 - 19 &  1828.6/6.0 \\ \hline
                                  \multicolumn{5}{|c|}{$m_{c} = 2$}  & \multicolumn{5}{|c|}{$m_{c} = 6$}  \\ \hline
                                        30 &    1.2964 & 0.9115(3) & 13 - 19 &  35.3/6.0      &    30 &    1.2964 & 0.8801(3) & 13 - 19 &  68.4/6.0 \\ \hline  
                                        35 &    1.5125 & 1.0652(3) & 13 - 19 &  67.4/6.0      &    35 &    1.5125 & 1.0284(3) & 13 - 19 &  117.9/6.0 \\ \hline 
                                        40 &    1.7286 & 1.2194(3) & 13 - 19 &  119.5/6.0     &    40 &    1.7286 & 1.1772(3) & 13 - 19 &  191.6/6.0 \\ \hline 
                                        45 &    1.9447 & 1.3739(4) & 13 - 19 &  196.1/6.0     &    45 &    1.9447 & 1.3262(4) & 13 - 19 &  294.0/6.0 \\ \hline 
                                        50 &    2.1607 & 1.5286(4) & 13 - 19 &  300.1/6.0     &    50 &    2.1607 & 1.4755(4) & 13 - 19 &  428.8/6.0 \\ \hline 
                                        55 &    2.3768 & 1.6833(4) & 13 - 19 &  432.9/6.0     &    55 &    2.3768 & 1.6248(4) & 13 - 19 &  597.7/6.0 \\ \hline 
                                        60 &    2.5929 & 1.8378(5) & 13 - 19 &  593.9/6.0     &    60 &    2.5929 & 1.7740(5) & 13 - 19 &  800.4/6.0 \\ \hline 
                                        65 &    2.8090 & 1.9919(5) & 13 - 19 &  781.1/6.0     &    65 &    2.8090 & 1.9228(5) & 13 - 19 &  1034.7/6.0 \\ \hline
                                        70 &    3.0250 & 2.1455(5) & 13 - 19 &  990.7/6.0     &    70 &    3.0250 & 2.0712(5) & 13 - 19 &  1296.3/6.0 \\ \hline
                                        75 &    3.2411 & 2.2984(6) & 13 - 19 &  1217.9/6.0    &    75 &    3.2411 & 2.2190(6) & 13 - 19 &  1579.4/6.0 \\ \hline
              \end{tabular}
              \end{footnotesize}
  \caption{The fitting results of $a\rho$.}\label{tb:numerical_results_arho_all}
\end{table*}

\clearpage


\acknowledgments
The author started this research project with A. Di Giacomo of the University of Pisa. The author is deeply grateful to him for his help and discussions. The author would like to thank M. D'Elia and F. Pucci for their helpful discussions. The author has received financial support from the Istituto Nazionale di Fisica Nucleare at the University of Pisa and the Joint Institute for Nuclear Research. The author performed simulations using the SX-series, computer clusters, and XC40 at the Research Center for Nuclear Physics and the Cybermedia Center at Osaka University and the Yukawa Institute for Theoretical Physics at Kyoto University. We used storage elements from the Japan Lattice Data Grid at the Research Center for Nuclear Physics at Osaka University. We appreciate the computer resources and technical support that was kindly provided by these facilities.

\bibliographystyle{unsrt}
\bibliography{Monopole_instanton_effects_M.Hasegawa_08072020.bib}

\begin{thebibliography}{100}

\bibitem{Clay1}
{Millennium Problems}.
\newblock {Clay Mathematics Institute, Peterborough, New Hampshire, USA,}
  \url{http://www.claymath.org/millennium-problems} (2000).

\bibitem{tHooft2}
{G. 't Hooft}.
\newblock in Proceedings of the EPS International, edited by A. Zichichi, p.
  1225, (1976).

\bibitem{Mandelstam1}
{S. Mandelstam}.
\newblock {II. Vortices and quark confinement in non-Abelian gauge theories}.
\newblock {\em Phys. Rep.}, 23:245, 1976.

\bibitem{Kronfel1}
{A. S. Kronfeld, G. Schierholz, and U. -J. Wiese}.
\newblock {Topology and dynamics of the confinement mechanism}.
\newblock {\em Nucl. Phys. B}, 293:461, 1987.

\bibitem{Maedan1}
{S. Maedan and T. Suzuki}.
\newblock {An Infrared Effective Theory of Quark Confinement Based on Monopole
  Condensation}.
\newblock {\em Prog. Theor. Phys.}, 81(1):229, 1989.

\bibitem{Brandstaeter1}
{F. Brandstaeter, G. Schierholz, and U. -J. Wiese}.
\newblock {Color comfinement, abelian dominance and the dynamics of magnetic
  monopoles in SU (3) gauge theory}.
\newblock {\em Phys. Lett.}, B272:319, 1991.

\bibitem{Hioki1}
{S. Hioki, S. Kitahara, S. Kiura, Y. Matsubara, O. Miyamura, S. Ohno, and T.
  Suzuki}.
\newblock {Abelian dominance in SU (2) color confinement}.
\newblock {\em Phys. Lett.}, B272:326, 1991.

\bibitem{DiGiacomo1}
{A. Di Giacomo and G. Paffuti }.
\newblock {A disorder parameter for dual superconductivity in gauge theories}.
\newblock {\em Phys. Rev. D}, 56:6816, 1997.

\bibitem{Sasaki1}
{S. Sasaki and O. Miyamura}.
\newblock {Lattice Study of $U_{A}$(1) Anomaly: The Role of QCD-Monopoles}.
\newblock {\em Phys. Lett.}, B443:331, 1998.

\bibitem{Bonati2}
{C. Bonati, G. Cossu, M. D'Elia, and A. Di Giacomo}.
\newblock {The disorder parameter of dual superconductivity in QCD revisited}.
\newblock {\em Phys. Rev. D}, 85:065001, 2012.

\bibitem{Sekido2}
{T. Suzuki, K. Ishiguro, Y. Koma, and T. Sekido}.
\newblock {Gauge-independent Abelian mechanism of color confinement in
  gluodynamics}.
\newblock {\em Phys. Rev. D}, 77:034502, Feb 2008.

\bibitem{tHooft1}
{G. 't Hooft}.
\newblock {Magnetic monopoles in unified gauge theories}.
\newblock {\em Nucl. Phys. B}, 79:276, 1974.

\bibitem{Polyakov1}
{A. M. Polyakov}.
\newblock {Particle Spectrum in the Quantum Field Theory}.
\newblock {\em JETP Lett.}, 20(6):194, 1974.

\bibitem{Rubakov1}
{V. A. Rubakov}.
\newblock {Superheavy magnetic monopoles and decay of the proton}.
\newblock {\em Pis'ma Zh. Eksp. Teor. Fiz.}, 33:658, 1981.

\bibitem{Rubakov2}
{V. A. Rubakov}.
\newblock {Adler-Bell-Jackiw anomaly and fermion-number breaking in the
  presence of a magnetic monopole}.
\newblock {\em Nucl. Phys. B}, 203:311, 1982.

\bibitem{Wu2}
{T. T. Wu}.
\newblock {Interaction of a fermion with a monopole I}.
\newblock {\em Nucl. Phys. B}, 222:411, 1983.

\bibitem{Rubakov3}
{V. A. Rubakov}.
\newblock {Monopole catalysis of proton decay}.
\newblock {\em Rep. Prog. Phys.}, 51:189, 1988.

\bibitem{Romanov1}
{V. N. Romanov, V. A. Fateev, and A. S. Schwarz}.
\newblock {Magnetic Monopoles In The Unified Theories Of The Electromagnetic,
  Weak And Strong Interactions. (in Russian)}.
\newblock {\em Yad. Fiz.}, 32:1138, 1980.

\bibitem{Groom1}
{D. E. Groom}.
\newblock {I}n search of the supermassive magnetic monopole.
\newblock {\em Phys. Rep.}, 140(6):323, 1986.

\bibitem{Ueno1}
{K. Ueno, et. at.}
\newblock {Search for GUT monopoles at Super–Kamiokande}.
\newblock {\em Astro. Phys.}, 36:131, 2012.

\bibitem{Patrizii1}
{L. Patrizii and M. Spurio}.
\newblock {Status of Searches for Magnetic Monopoles}.
\newblock {\em Annu. Rev. Nucl. Part. Sci.}, 65:279, 2015.

\bibitem{Nambu1}
{Y. Nambu}.
\newblock {Quasi-Particles and Gauge Invariance in the Theory of
  Superconductivity}.
\newblock {\em Phys. Rev.}, 117:648, 1960.

\bibitem{Nambu2}
{Y. Nambu and G. Jona-Lasinio}.
\newblock {Dynamical Model of Elementary Particles Based on an Analogy with
  Superconductivity. I.}
\newblock {\em Phys. Rev.}, 122:345, 1961.

\bibitem{Goldstone1}
{J. Goldstone}.
\newblock {Field theories with Superconductor solutions}.
\newblock {\em Nuovo Cim.}, 19:154, 1961.

\bibitem{Goldstone2}
{J. Goldstone, A. Salam, and S. Weinberg}.
\newblock {Broken Symmetries}.
\newblock {\em Phys. Rev.}, 127:965.

\bibitem{Gross1}
{D. J. Gross and A. Neveu}.
\newblock {Dynamical symmetry breaking in asymptotically free field theories}.
\newblock {\em Phys. Rev. D}, 10:3235, 1974.

\bibitem{Kugo1}
T.~Kugo.
\newblock {\em The quantum theory of the gauge field I, II}.
\newblock Baifukan, 2002.
\newblock The textbook written in Japanese.

\bibitem{Weinberg1}
{S. Weinberg}.
\newblock {PION SCATTERING LENGTHS}.
\newblock {\em Phys. Rev. Lett.}, 17:616, 1966.

\bibitem{Belavi1}
{A. A. Belavin, A. M. Polyakov, A. S. Schwartz, and Yu. S. Tyupkin}.
\newblock {Pseudoparticle solutions of the Yang-Mills equations}.
\newblock {\em Phys. Lett.}, B59:85, 1975.

\bibitem{Dyakonov6}
D.~Diakonov.
\newblock Instantons at work.
\newblock {\em Prog. Particle and Nuclear Physics}, 51:173, 2003.

\bibitem{Shuryak2}
{T. Sch\"{a}fer and E. V. Shuryak}.
\newblock Instantons in {QCD}.
\newblock {\em Rev. Mod. Phys.}, 70(2):323, 1998.

\bibitem{Dyakonov1}
D.~I. Dyakonov and V.~Yu. Petrov.
\newblock {CHIRAL CONDENSATE IN THE INSTANTON VACUUM}.
\newblock {\em Phys. Lett.}, 147B(4, 5):351, 1984.

\bibitem{Dyakonov2}
D.~I. D'yakonov and V.~Yu. Petrov.
\newblock Meson-current correlation function in instanton vacuum.
\newblock {\em Sov. Phys. JETP}, 62(3):431, 1985.

\bibitem{Dyakonov3}
D.~I. D'yakonov and V.~Yu. Petrov.
\newblock Quark propagator and chiral condensate in an instanton vacuum.
\newblock {\em Sov. Phys. JETP}, 62(2):204, 1985.

\bibitem{Dyakonov4}
D.~I. Dyakonov and V.~Yu. Petrov.
\newblock A theory of light quarks in the instanton vacuum.
\newblock {\em Nucl. Phys. B}, 272:457, 1986.

\bibitem{Ray1}
{M. W. Ray, E. Ruokokoski, S. Kandel, M. M\"{o}tt\"{o}nen, and D. S. Hall}.
\newblock {Observation of Dirac monopoles in a synthetic magnetic field}.
\newblock {\em Nature}, 505:657, 2014.

\bibitem{Ray2}
{M. W. Ray, E. Ruokokoski, K. Tiurev, M. M\"{o}tt\"{o}nen, and D. S. Hall}.
\newblock {Observation of isolated monopoles in a quantum field}.
\newblock {\em Science}, 348(6234):544, 2015.

\bibitem{Moedal1}
{B. Acharya, and et. al.}
\newblock {Search for magnetic monopoles with the MoEDAL prototype trapping
  detector in 8 TeV proton-proton collisions at the LHC}.
\newblock {\em J. of High Energy Phys.}, 08:067, 2016.

\bibitem{Moedal2}
{B. Acharya, and et. al.}
\newblock {Search for Magnetic Monopoles with the MoEDAL Forward Trapping
  Detector in 13 TeV Proton-Proton Collisions at the LHC}.
\newblock {\em Phys. Rev. Lett}, 118:061801, 2017.

\bibitem{Ilgenfritz1}
{E. -M. Ilgenfritz, M. L. laursen, M. M\"{u}ller-Preu\ss ker, G. Schierholz,
  and H. Schiller}.
\newblock {First evidence for the existence of instantons in the quantized
  SU(2) lattice vacuum}.
\newblock {\em Nucl. Phys. B}, 268:693, 1986.

\bibitem{Hart1}
{A. Hart and M. Teper}.
\newblock {Instantons and monopoles in the maximally Abelian gauge}.
\newblock {\em Phys. Lett.}, B371:261, 1996.

\bibitem{Kitahara2}
{S. Kitahara, O. Miyamura, T. Okude, F. Shoji, and T. Suzuki}.
\newblock {Monopoles and hadron spectrum in quenched QCD}.
\newblock {\em Nucl. Phys. B}, 533:576, 1998.

\bibitem{Chernodub1}
{M. N. Chernodub and V. I. Zakharov}.
\newblock {Fermionic signature of the lattice monopoles}.
\newblock {\em Phys. Rev. D}, 65:094020, 2002.

\bibitem{HAoki1}
{H. Aoki, S. Iso, and K. Nagao}.
\newblock {Ginsparg–Wilson relation and 't Hooft–Polyakov monopole on fuzzy
  2-sphere}.
\newblock {\em Nucl. Phys. B}, 684:162, 2004.

\bibitem{DiGH3}
A.~Di Giacomo and M.~Hasegawa.
\newblock Instantons and monopoles.
\newblock {\em Phys. Rev. D}, 91:054512, 2015.

\bibitem{Ginsparg1}
{P. H. Ginsparg and K. G. Wilson}.
\newblock {A remnant of chiral symmetry on the lattice}.
\newblock {\em Phys. Rev. D}, 25:2649, 1982.

\bibitem{Neuberger1}
N.~Neuberger.
\newblock Exactly massless quarks on the lattice.
\newblock {\em Phys. Lett. B}, 417:141, 1998.

\bibitem{Neuberger2}
N.~Neuberger.
\newblock More about exactly massless quarks on the lattice.
\newblock {\em Phys. Lett. B}, 427:353, 1998.

\bibitem{Lusher1}
{M. L\"{u}scher}.
\newblock {Exact chiral symmetry on the lattice and the Ginsparg-Wilson
  relation}.
\newblock {\em Phys. Lett.}, B428:342, 1998.

\bibitem{Chandrasekharan1}
{S. Chandrasekharan}.
\newblock {Lattice QCD with Ginsparg-Wilson fermions}.
\newblock {\em Phys. Rev. D}, 60:074503, 1999.

\bibitem{DiGH4}
A.~Di Giacomo and M.~Hasegawa.
\newblock Monopoles in maximal abelian gauge, number of zero modes, and
  instantons.
\newblock CYBERMEDIA HPC JOURNAL No 5, 21, Osaka University, Cybermedia Center,
  Osaka, Japan, July 2015.
\newblock {I}SSN 2186-473X.

\bibitem{DiGHP1}
{A. Di Giacomo, M. Hasegawa, and F. Pucci}.
\newblock Chiral symmetry breaking and monopoles.
\newblock {\it Proc. Sci.}, CD15, 127 (2015), [hep-lat/1510.07463].

\bibitem{DiGH5}
A.~Di Giacomo and M.~Hasegawa.
\newblock Chiral symmetry breaking, instantons, and monopoles.
\newblock {\it Proc. Sci.}, Lat2015, 313 (2015), [hep-lat/1512.00359].

\bibitem{Damgaard1}
{S. M. Nishigaki, P. H. Damgaard, and T. Wettig}.
\newblock {Smallest Dirac eigenvalue distribution from random matrix theory}.
\newblock {\em Phys. Rev. D}, 58:087704, 1998.

\bibitem{Damgaard2}
P.~H. Damgaard and S.~M. Nishigaki.
\newblock {Distribution of the $k$-th smallest Dirac operator eigenvalue}.
\newblock {\em Phys. Rev. D}, 63(045012), 2001.
\newblock updated in 2003 [hep-th/0006111].

\bibitem{Edwards1}
{R. G. Edwards, U. M. Heller, J. Kiskis, and R. Narayanan}.
\newblock {Quark Spectra, Topology, and Random Matrix Theory}.
\newblock {\em Phys. Rev. Lett.}, 82:4188, 1999.

\bibitem{Giusti4}
{L. Giusti1, M. L\"{u}scher, P. Weisz, and H. Wittig}.
\newblock {Lattice QCD in the $\epsilon$-regime and random matrix theory}.
\newblock {\em J. High Energy Phys.}, 11:023, 2003.

\bibitem{Giusti6}
{L. Giusti, C. Hoelbling, M. L\"{u}scher, and H. Wittig}.
\newblock {Numerical techniques for lattice QCD in the $\epsilon$-regime}.
\newblock {\em Comp. Phys. Comm.}, 153:31, 2003.

\bibitem{Gimenez1}
{V. Gim\'{e}nez, L. Giusti, F. Rapuano, and M. Talevi}.
\newblock {Lattice quark masses: a non-perturbative measurement}.
\newblock {\em Nucl. Phys. B}, 540:472, 1998.

\bibitem{Giusti3}
{L. Giusti, C. Hoelbling, and C. Rebbi}.
\newblock Light quark masses with overlap fermions in quenched {QCD}.
\newblock {\em Phys. Rev. D}, 64:114508, 2001.
\newblock {E}rratum, Phys. Rev. D {\bf 65}, 079903(E) (2002).

\bibitem{Bochicchio1}
{M. Bochicchio, L. Maiani, G. Martinelli, G. Rossi, and M. Testa}.
\newblock {Chiral symmetry on the lattice with Wilson fermions}.
\newblock {\em Nucl. Phys. B}, 262:331, 1985.

\bibitem{Maiani1}
{L. Maiani and G. Martinelli}.
\newblock {Current algebra and quark masses from a Monte Carlo simulation with
  Wilson fermions}.
\newblock {\em Phys. Lett.}, B178:265, 1986.

\bibitem{Hernandez2}
{P. Hern\'{a}ndez, K. Jansen, L. Lellouch, and H. Wittig}.
\newblock {Non-perturbative renormalization of the quark condensate in
  Ginsparg-Wilson regularizations}.
\newblock {\em J. High Energy Phys.}, 07:018, 2001.

\bibitem{Wennekers1}
J.~Wennekers and H.~Wittig.
\newblock On the renormalized scalar density in quenched {QCD}.
\newblock {\em J. High Energy Phys.}, 09:059, 2005.

\bibitem{Ape1}
{Ape Collaboration, M. Albanese, and et al.}
\newblock {Glueball masses and string tension in lattice QCD}.
\newblock {\em Phys. Lett.}, B 192:163, 1987.

\bibitem{Necco1}
S.~Necco and R.~Sommer.
\newblock {The Nf=0 heavy quark potential from short to intermediate
  distances}.
\newblock {\em Nucl. Phys. B}, 622:328, 2002.

\bibitem{Necco2}
S.~Necco.
\newblock {\em {The static quark potential and scaling behavior of SU(3)
  lattice Yang-Mills theory}}.
\newblock {Ph. D.} thesis, Humboldt Universit\"{a}t zu Berlin, DESY,
  Platanenallee 6, D-15738 Zeuthen, Germany, June 2003.
\newblock [hep-lat/0306005].

\bibitem{tHooft3}
{G. 't Hooft}.
\newblock {TOPOLOGY OF THE GAUGE CONDITION AND NEW CONFINEMENT PHASES IN
  NON-ABELIAN GAUGE THEORIES}.
\newblock {\em Nucl. Phys. B}, 190:455, 1981.

\bibitem{DeGrand1}
{T. A. DeGrand and D. Toussaint}.
\newblock {Topological excitations and Monte Carlo simulation of Abelian gauge
  theory}.
\newblock {\em Phys. Rev. D}, 22:2478, 1980.

\bibitem{Bornyakov4}
{ V. G. Bornyakov, H. Ichie, Y. Koma, Y. Mori, Y. Nakamura, D. Pleiter, M. I.
  Polikarpov, G. Schierholz, T. Streuer, H. St\"{u}ben, and T. Suzuki}.
\newblock {Dynamics of Monopoles and Flux Tubes in Two-Flavor Dynamical QCD}.
\newblock {\em Phys. Rev. D}, 70:074511, 2004.

\bibitem{Bode1}
{A. Bode, T. Lippert, and K. Schilling}.
\newblock {Monopole clusters and critical dynamics in four-dimensional U(1)}.
\newblock {\em Nucl. Phys. B, Proc. Suppl.}, 34:549, 1994.

\bibitem{Hernandez1}
{P. Hern\'{a}ndez, K. Jansen, and M. L\"{u}scher}.
\newblock {Locality properties of Neuberger's lattice Dirac operator}.
\newblock {\em Nucl. Phys. B}, 552:363, 1999.

\bibitem{Edwards3}
{R. G. Edwards, U. M. Heller, J. Kiskis, and R. Narayanan}.
\newblock {Chiral condensate in the deconfined phase of quenched gauge
  theories}.
\newblock {\em Phys. Rev. D}, 61:074504, 2000.

\bibitem{Shuryak1_1}
E.~V. Shuryak.
\newblock {THE ROLE OF INSTANTONS IN QUANTUM CHROMODYNAMICS (I)}.
\newblock {\em Nucl. Phys. B}, 203:93, 1982.

\bibitem{DeDebbio1}
{L. Del Debbio, L. Giusti, and C. Pica}.
\newblock {Topological Susceptibility in SU(3) Gauge Theory}.
\newblock {\em Phys. Rev. Lett.}, 94:032003, 2005.

\bibitem{Guhr1}
{T. Guhr, J.-Z. Ma, S. Meyer, and T. Wilke}.
\newblock {Statistical analysis and the equivalent of a Thouless energy in
  lattice QCD Dirac spectra}.
\newblock {\em Phys. Rev. D}, 59:054501, 1999.

\bibitem{Capitani}
{S. Capitani, C. G\"{o}ckeler, R. Horsley, P. E. L. Rakow, and G. Schierholz}.
\newblock {Operator improvement for Ginsparg–Wilson fermions}.
\newblock {\em Phys. Lett. B}, 468:150, 1999.

\bibitem{Wigner1}
{E. P. Wigner}.
\newblock {Group Theory and its Application to the Quantum Mechanics of Atomic
  Spectra}.
\newblock {\em Academic press, New York}, 1959.

\bibitem{Dyson1}
F.~J. Dyson.
\newblock {Statistical Theory of the Energy Levels of Complex Systems. I}.
\newblock {\em J. Math. Phys.}, 3:140, 1962.

\bibitem{Guhr2}
{T. Guhr, A. M\"{u}ller–Groeling, and H. A. Weidenmüller}.
\newblock {Random-matrix theories in quantum physics: common concepts}.
\newblock {\em Physics Reports}, 299:189--428, 1998.

\bibitem{Dyson2}
F.~J. Dyson and M.~L. Mehta.
\newblock {Statistical Theory of the Energy Levels of Complex Systems. IV}.
\newblock {\em J. Math. Phys.}, 4:701, 1963.

\bibitem{Bohigas1}
{O. Bohigas and M. J. Giannoni}.
\newblock {Level density fluctuations and random matrix theory}.
\newblock {\em Annals of Physics}, 89:393--422, 1974.

\bibitem{Nishigaki1}
P.~H.~Damgaard S.~M.~Nishigaki and T.~Wettig.
\newblock {Smallest Dirac eigenvalue distribution from random matrix theory}.
\newblock {\em Phys. Rev. D}, 58:087704, 1998.

\bibitem{Dammgard1}
P.~H. Damgaard and S.~M. Nishigaki.
\newblock {Distribution of the kth smallest Dirac operator eigenvalue}.
\newblock {\em Phys. Rev. D}, 63:045012 (hep--th/0006111), 2001.

\bibitem{DiGH2}
A.~Di Giacomo and M.~Hasegawa.
\newblock Zero modes, instantons, and monopoles.
\newblock In Victor~Kim Alexander~Andrianov, Nora~Brambilla and Sergei
  Kolevatov, editors, {\em XITH CONFERENCE ON QUARK CONFINEMENT AND HADRON
  SPECTRUM}, volume 1701, page 100008. AIP Conf. Proc., 2016.
\newblock [hep-lat/1412.2704].

\bibitem{Gellmann1}
{Murray Gell-Mann, R. J. Oakes, and B. Renner}.
\newblock {Behavior of Current Divergences under $SU_{3}\times SU_{3}$}.
\newblock {\em Phys. Rev.}, 175:2195, 1968.

\bibitem{Colangelo3}
{G. Colangelo and S. D\"urr}.
\newblock {The pion mass in finite volume}.
\newblock {\em Eur. Phys. J. C}, 33:543, 2004.

\bibitem{PDG_2017}
{C. Patrignani et al., (Particle Data Group)}.
\newblock {REVIEW OF PARTICLE PHYSICS}.
\newblock {\em Chin. Phys. C}, 40:100001, 2016.
\newblock updated in 2017.

\bibitem{Giusti2}
{L. Giusti, P. Hern\'{a}ndez, M. laine, P. Weisz, and H. Wittig}.
\newblock Low-energy couplings of {QCD} from current correlations near the
  chiral limit.
\newblock {\em J. High Energy Phys.}, 04:013, 2004.

\bibitem{DeGrand2}
T.~DeGrand and S.~Schaefer.
\newblock Improving meson two-point functions in lattice {QCD}.
\newblock {\em Comp. Phys. Commun.}, 159:185, 2004.

\bibitem{Niedermayer1}
F.~Niedermayer.
\newblock Exact chiral symmetry, topological charge and related topics.
\newblock {N}ucl. Phys. B (Proc. Suppl.), {\bf 73}, 105 (1999),
  [hep-lat/9810026].

\bibitem{Blum1}
{T. Blum, P. Chen, N. Christ, C. Cristian, C. Dawson, G. Fleming, A. Kaehler,
  X. Liao, G. Liu, C. Malureanu, R. Mawhinney, S. Ohta, G. Siegert, A. Soni, C.
  Sui, P. Vranas, M. Wingate, L. Wu, and Y. Zhestkov}.
\newblock {Quenched lattice QCD with domain wall fermions and the chiral
  limit}.
\newblock {\em Phys. Rev. D}, 69:074502, 2004.

\bibitem{Gasser1}
{J. Gasser and H. Leutwyler}.
\newblock {Chiral perturbation theory to one loop}.
\newblock {\em Ann. Phys.}, 158:142, 1984.

\bibitem{Alexandrou1}
{C. Alexandrou, E. Follana, H. Panagopoulos, and E. Vicari}.
\newblock {One-loop renormalization of fermionic currents with the
  overlap-Dirac operator}.
\newblock {\em Nucl. Phys. B}, 580:394, 2000.

\bibitem{Giusti3_2}
{L. Giusti, C. Hoelbling, and C. Rebbi}.
\newblock Quenched results for the light quark physics with overlap fermions.
\newblock {\em {Nucl. Phys. B (Proc. Suppl.)}}, 106:739, 2002.

\bibitem{Colangelo1}
{G. Colangelo and E. Pallante}.
\newblock {Quenched chiral perturbation theory to one loop}.
\newblock {\em Nucl. Phys. B}, 520:433, 1998.

\bibitem{Giusti1}
{L. Giusti, P. Hern\'{a}ndez, S. Necco, C. Pena, J. Wennerkers, and H. Wittig}.
\newblock Testing chiral effective theory with quenched lattice {QCD}.
\newblock {\em J. High Energy Phys.}, 05:024, 2008.

\bibitem{Heitger1}
{ALPHA Collaboration, J. Heitger, R. Sommer, and H. Wittig}.
\newblock {Effective chiral Lagrangians and lattice QCD}.
\newblock {\em Nucl. Phys. B}, 588:377, 2000.

\bibitem{ALPHA1}
{ALPHA and UKQCD Collaborations, J. Garden, J. Heitger, R. Sommer, H. Wittig}.
\newblock {Precision computation of the strange quark's mass in quenched QCD}.
\newblock {\em Nucl. Phys. B}, 571:237, 2000.

\bibitem{Alton1}
{C. R. Allton, V. Giménez, L. Giusti, and F. Rapuano}.
\newblock {Light quenched hadron spectrum and decay constants on different
  lattices}.
\newblock {\em Nucl. Phys. B}, 489:427, 1997.

\bibitem{Hasegawa5}
M.~Hasegawa.
\newblock {Monopole and instanton effects in the continuum limit on the pion}.
\newblock {in preparation}.

\bibitem{Aoki1}
{S. Aoki, et. al.}
\newblock {Review of lattice results concerning low-energy particle physics},
  2016.

\bibitem{Gasser3}
{J. Gasser and H. Leutwyler}.
\newblock {Chiral perturbation theory: Expansions in the mass of the strange
  quark}.
\newblock {\em Nucl. Phys. B}, 250:465, 1985.

\end{thebibliography}

\end{document}